\documentclass[11pt, a4paper]{article}
\usepackage[margin=1in]{geometry} 
\usepackage{amsmath}
\usepackage{float}
\usepackage{amssymb}
\usepackage{bm}
\usepackage{physics}
\usepackage{algpseudocode}
\usepackage{mathtools}
\usepackage{empheq}
\usepackage{algorithm}
\usepackage{booktabs}
\usepackage{enumitem}
\usepackage{graphicx}
\usepackage{pifont}
\usepackage{subcaption}
\usepackage{float}
\usepackage{tikz}
\usepackage{makecell}
\usepackage{authblk}
\usepackage[authoryear,round,sort]{natbib}
\bibpunct{(}{)}{;}{a}{}{,}
\usepackage[hidelinks, unicode]{hyperref}

\newcommand{\pderiv}[3]{\left(\frac{\partial #1}{\partial #2}\right)_{#3}}
\newcommand{\totalderiv}[2]{\frac{\mathrm{d} #1}{\mathrm{d} #2}}
\newcommand{\Dt}[1]{\frac{\mathrm D #1}{\mathrm Dt}}
 
\def\ge{\geqslant}\def\le{\leqslant}
\def\geq{\geqslant}

\hypersetup{
    colorlinks=true,
    linkcolor=blue,
    citecolor=blue,
    urlcolor=blue
}

\title{\Large\bfseries Flash evaporation Riemann Problem: Thermodynamic Relaxation and Oscillation-Free Real Fluid Finite Volume Method}

\author[1,2]{Haotong Bai}
\author[3]{Ping Yi\thanks{Corresponding author: yipingshjt@sjtu.edu.cn}}
\author[1,2]{Yixin Yang\thanks{Corresponding author: yangyixin@nudt.edu.cn}}
\author[1,2]{Guoyan Zhao}
\author[2]{Wenjia Xie}
\author[1,2]{Mingbo Sun\thanks{Corresponding author: sunmingbo@nudt.edu.cn}}

\affil[1]{Advanced Propulsion Technology Laboratory, National University of Defense Technology, Kaifu District, Changsha, Hunan 410073, China}
\affil[2]{College of Aerospace Science and Engineering, National University of Defense Technology, Kaifu District, Changsha, Hunan 410073, China}
\affil[3]{State Key Laboratory of Ocean Engineering, Shanghai Jiao Tong University,, Shanghai, 200240, China}

\date{}

\begin{document}
\maketitle

\begin{abstract}
The Flash evaporation Riemann problem (FeRP) \citep{Bai2026FeRP} formulated the wave structures of flash evaporation in the vapor-liquid homogeneous equilibrium limit. However, physical flash evaporation involves finite-rate interphase heat and mass transfer, inducing the wave structure to deviate from the equilibrium state. To solve such a non-equilibrium Riemann problem, this paper first derives the governing equation of mechanical equilibrium evolution in a relaxation system, based on which an oscillation-free Homogeneous Relaxation Real Fluid (HRRF) finite volume method is developed. In addition to the Euler equations with mass relaxation, the HRRF method evolves advection equations for the internal energy--pressure affine parameters and the vapor volume fraction, in order to eliminate numerical pressure oscillations inherent in conservative methods while preserving phase-transition-induced pressure changes, and consistently degenerate into a completely pressure-equilibrium method in supercritical flows. The HRRF method is then employed to comprehensively evaluate how thermodynamic relaxation affects the FeRP solution. Results indicate that the non-classical expansion shock of the FeRP loses its traveling-wave characteristics with increasing relaxation time, continuously degenerating into a self-similar rarefaction wave. Compared with the equilibrium limit, an increased mass transfer relaxation time smooths the expansion branch of the RP solution, lowers the intermediate pressure and velocity, and induces a pseudo-splitting of the rarefaction wave as well as a pseudo-advance of the phase transition onset. Meanwhile, the heat transfer relaxation time primarily dictates the temperature difference in the vapor-liquid mixture, thereby modulating the liquid superheat and in turn influencing the mass transfer rate.
\\
\textbf{Keywords~~~}Riemann problem, Real fluid, Thermodynamic relaxation, Finite volume method
\end{abstract}

\section{Introduction}

\begin{figure}[h]
\centering 
\includegraphics[width=0.75\textwidth]{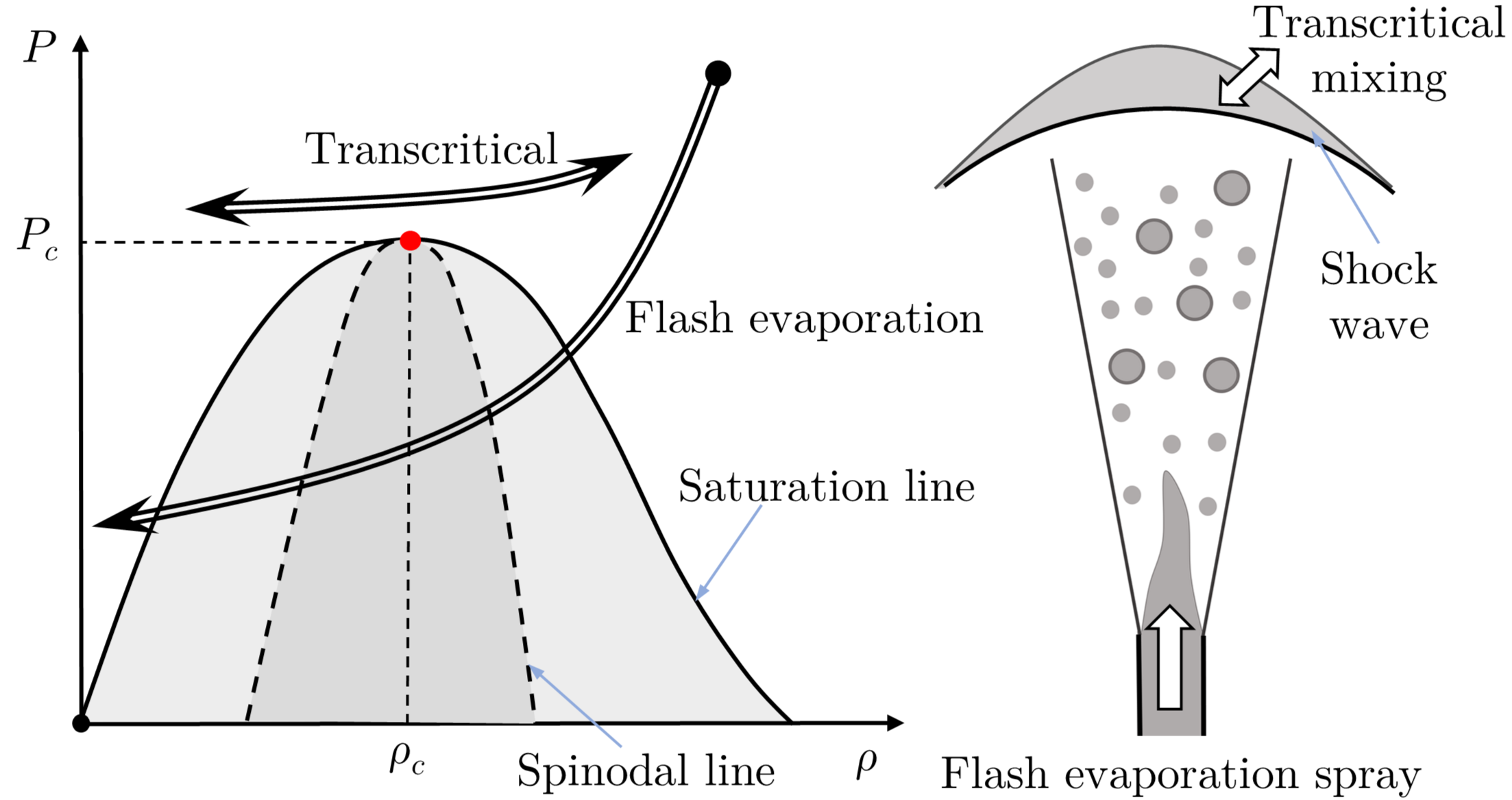} 
\caption{Diagram of flash evaporation and thermodynamic transcritical mixing}
\label{fig1}
\end{figure}

Flash evaporation is prevalent in propulsion and power systems in aerospace and energy systems, such as flash sprays of heated fuel in scramjets \citep{Kim2024} and internal combustion engines \citep{Yang2018}, as well as cavitation in rocket engine propellant feedlines \citep{Caze2024}. Understanding its flow mechanisms through numerical simulation is essential for the design and evaluation of these power systems. However, unlike gradual evaporation driven by temperature, the flash evaporation process is typically accompanied by extreme flow phenomena such as shock discontinuities and transcritical mixing (as shown in Figure \ref{fig1}). How to reliably model and simulate such complex processes remains a challenge in the field of compressible two-phase flows.

Grounded in the physical characteristics of flash evaporation, where phase transition is driven by depressurization with a short characteristic timescale, our previous work proposed the ``Flash evaporation Riemann problem'' (FeRP) \citep{Bai2026FeRP}, which formulates flash evaporation as a Riemann problem, aiming to describe its flow characteristics in the Homogeneous Equilibrium Model (HEM) and Vapor-Liquid Equilibrium (VLE) limits within self-similar wave structures. The expansion branch of the FeRP crosses the saturation line, inducing a discontinuity in the speed of sound and giving rise to rarefaction wave splitting as well as non-classical expansion shocks and composite waves. Based on these wave features, we developed an exact solution method for the FeRP with arbitrary real-fluid equations of state (EoS), providing a benchmark for the design of related CFD algorithms.

However, in real flash evaporation processes, the physical state of the fluid can hardly approach the equilibrium limit within the FeRP framework, especially the thermodynamic equilibrium. Generally, the thermodynamic equilibrium process is much slower than mechanical equilibrium. Based on this physical consideration and numerical convenience, many methods within the HEM framework adopt the thermodynamically frozen speed of sound, namely the mechanical-equilibrium Wood's speed of sound, to replace the thermodynamic equilibrium speed of sound \citep{Yang2019,Yi2019,Yang2020,Saurel2016}. In particular, \citet{Yang2019} employed this approach to numerically simulate flash evaporation jets and obtained results in good agreement with experiments. To examine the rationale of this practice and the mechanisms by which two-phase thermodynamic relaxation affects the Riemann solution, we further extended the FeRP to two-phase non-equilibrium speeds of sound in our previous work, finding that as the degree of non-equilibrium increases, both the intermediate pressure and velocity of the RP solution decrease. Most notably, we also proved theoretically that such a practice of replacing the speed of sound is thermodynamically inconsistent and can even cause a non-physical entropy decrease \citep{Bai2026FeRP}.

This leads to a fundamental question: what is the actual mechanism by which thermodynamic relaxation affects the flash evaporation Riemann solution? To comprehensively answer this question, it is necessary to solve the relaxation system. This calls for developing a physically consistent numerical method within the Homogeneous Relaxation Model (HRM) framework \citep{DownarZapolski1996,Bilicki1990}, under the assumption of mechanical equilibrium and thermodynamic non-equilibrium. However, although exact fluxes can be obtained via the FeRP, the Godunov method \citep{Godunov1959} cannot be directly applied to solve the RPs under real-fluid conditions. This involves a well-known problem in two-phase flows: numerical pressure oscillations \citep{Abgrall1996}. This problem arises because the finite-volume averaging in the Godunov method implicitly assumes instantaneous thermodynamic equilibrium mixing \citep{Saurel2007, Bai2026RFQC2}. In two-phase flows, it manifests as the destruction of pressure equilibrium across phase interfaces; in real fluids, however, such equilibrium is violated whenever fluid elements in different thermodynamic states mix.

Most existing oscillation-free numerical methods for compressible two-phase flows have been proposed within the HEM framework. Early efforts were tailored for two-phase fluids with specific EoS, including non-conservative pressure-equation methods \citep{Karni1994,Karni1996}, double-flux methods \citep{Abgrall2001,Billet2003}, the Kapila and Allaire five-equation model \citep{Kapila2001, Allaire2002}, relaxation-projection methods \citep{Saurel2007,Petitpas2007}, as well as oscillation-free schemes designed for the stiffened gas EoS \citep{Shyue1998,Johnsen2006}, Mie-Gr\"uneisen EoS \citep{Shyue2001}, and van der Waals EoS \citep{Shyue1999,Pantano2017}. Subsequently, to eliminate pressure oscillations in single-phase fluids with arbitrary EoS during thermodynamic transcritical processes, real-fluid pressure-equation methods \citep{terashima2012,Kawai2015,Kitamura2018, Matheis2018}, double-flux methods \citep{Ma2017,yatsuyanagi2022,Zhang2024}, and enthalpy-equation methods \citep{Lacaze2019} were developed. These methods more or less compromise energy conservation in exchange for pressure equilibrium, which is dictated by the mathematical properties of nonlinear EoS in conservative methods \citep{Klein2026}. Although double-flux and pressure-equation methods are theoretically not restricted to single-phase real-fluid EoS, our previous work showed that they may exhibit limitations in FeRP-type configurations, resulting in noticeable deviations in predicted pressure and temperature \citep{Bai2026RFQC1}. To overcome this, we developed a Real Fluid Quasi-Conservative (RFQC) method \citep{Bai2026RFQC1,Bai2026RFQC2}, which evolves the affine parameters of the internal energy--pressure relation to recover an oscillation-free pressure. Its numerical predictions agree well with the exact FeRP solutions, while the introduced energy conservation error remains a second-order small term in smooth regions.

Returning to the flash evaporation problem within the HRM framework: it obviously imposes higher requirements on the numerical method. First, flash evaporation jets are often accompanied by transcritical processes. For instance, when a high-pressure jet is injected into a supercritical environment, the three-dimensional over-expansion effect triggers flash phase transition, followed by an inverse phase transition induced by the intercepting shock \citep{Xu2022}. At the phase transition front, transcritical mixing may occur, as shown in Figure \ref{fig1}. This requires that the relevant HRM method must be based on a real-fluid EoS, and be able to maintain pressure equilibrium during single-phase transcritical mixing, or in other words, consistently degenerate into a single-phase real-fluid oscillation-free method. Secondly, physically speaking, the thermodynamic relaxation during the isochoric-adiabatic mixing of fluid elements in different states should naturally cause a pressure change \citep{Bai2026RFQC2}. This requires the HRM method not only to eliminate the numerical pressure oscillations caused by finite volume averaging, but also to preserve the physical pressure change of the relaxation process.

To the best of our knowledge, various HRM methods already exist in two-phase flows \citep{Saurel2008,James2016,DeLorenzo2018,Faccanoni2021,Pelanti2022} and have achieved good results within the scope of the models they consider. However, constrained by their physical assumptions and numerical constructions, these methods cannot maintain pressure equilibrium in real fluids with complex EoS while simultaneously describing shock waves, thermodynamic transcritical processes, and phase-change relaxation. First among these is the traditional HRM formulation where the vapor phase is assumed to be saturated while the liquid phase can be superheated \citep{DownarZapolski1996,Zheng2018}. Although this assumption is relatively convenient for numerical closure, it limits the scope of application at the model level: considering the advection of superheated vapor and liquid, when they move into the same finite-volume cell, the saturation assumption forces the vapor volume to shrink instantaneously, while the isochoric condition requires the liquid to expand by hundreds of times when its volume fraction is small, which is physically unacceptable. Therefore, our model must require independent vapor and liquid temperatures so that both phases can enter metastable states---which is also closer to physical reality. The HRM methods proposed by \citet{DeLorenzo2019} and \citet{Pelanti2022} are based on such an assumption, and are theoretically not limited to a specific EoS. However, the pressure equilibrium property of such methods can only be maintained in specific simple EoS; when extended to more complex real-fluid EoS (like cubic EoS), pressure oscillations and iterative root-finding problems still exist. Moreover, how to make the HRM method naturally degenerate into an oscillation-free method in single-phase supercritical fluids still urgently needs to be solved. This is precisely the core challenge that this paper aims to overcome.

To investigate the FeRP with thermodynamic relaxation, elucidate the mechanisms of how relaxation affects wave structures, and provide a reliable and physically consistent numerical method for flash evaporation, this paper first proposes an oscillation-free Homogeneous Relaxation Real Fluid (HRRF) finite volume method. In Section 2, we formulate the basic physical assumptions and governing equations of the homogeneous relaxation system. In Section 3, we clarify the origin of numerical pressure oscillations from a physical perspective, derive the mechanical equilibrium evolution equation in a real-fluid homogeneous relaxation system, and outline the strategy for recovering the physical pressure. In Section 4, we further detail the numerical implementation of the HRRF method, the modeling of source terms, and analyze the conservation error introduced by the method. In Section 5, combined with the P-R EoS \citep{Peng1976}, we validate the pressure equilibrium property and modeling accuracy of the HRRF method. Finally, the HRRF method is employed to solve a series of Riemann problems with different heat and mass transfer relaxation times, thereby elucidating the mechanism of how thermodynamic relaxation affects the FeRP solution.

\section{Physical Assumptions and Governing Equations}
\label{sec:basic_eq}
From the Eulerian perspective, the fluid at each point is either single-phase or homogeneously mixed under the HRM assumption. A single-phase fluid is allowed to be in a metastable state; for a homogeneously mixed fluid, the velocity $u$ and pressure $P$ reach equilibrium, while the temperature can take two non-equilibrium states $T_l$ and $T_v$. For a homogeneous mixture, we have:
\begin{equation}
    u_l = u_v = u, \quad P_l = P_v=P.
    \label{eq:HRM_mix}
\end{equation}

\begin{figure}[htbp]
\centering 
\includegraphics[width=0.7\textwidth]{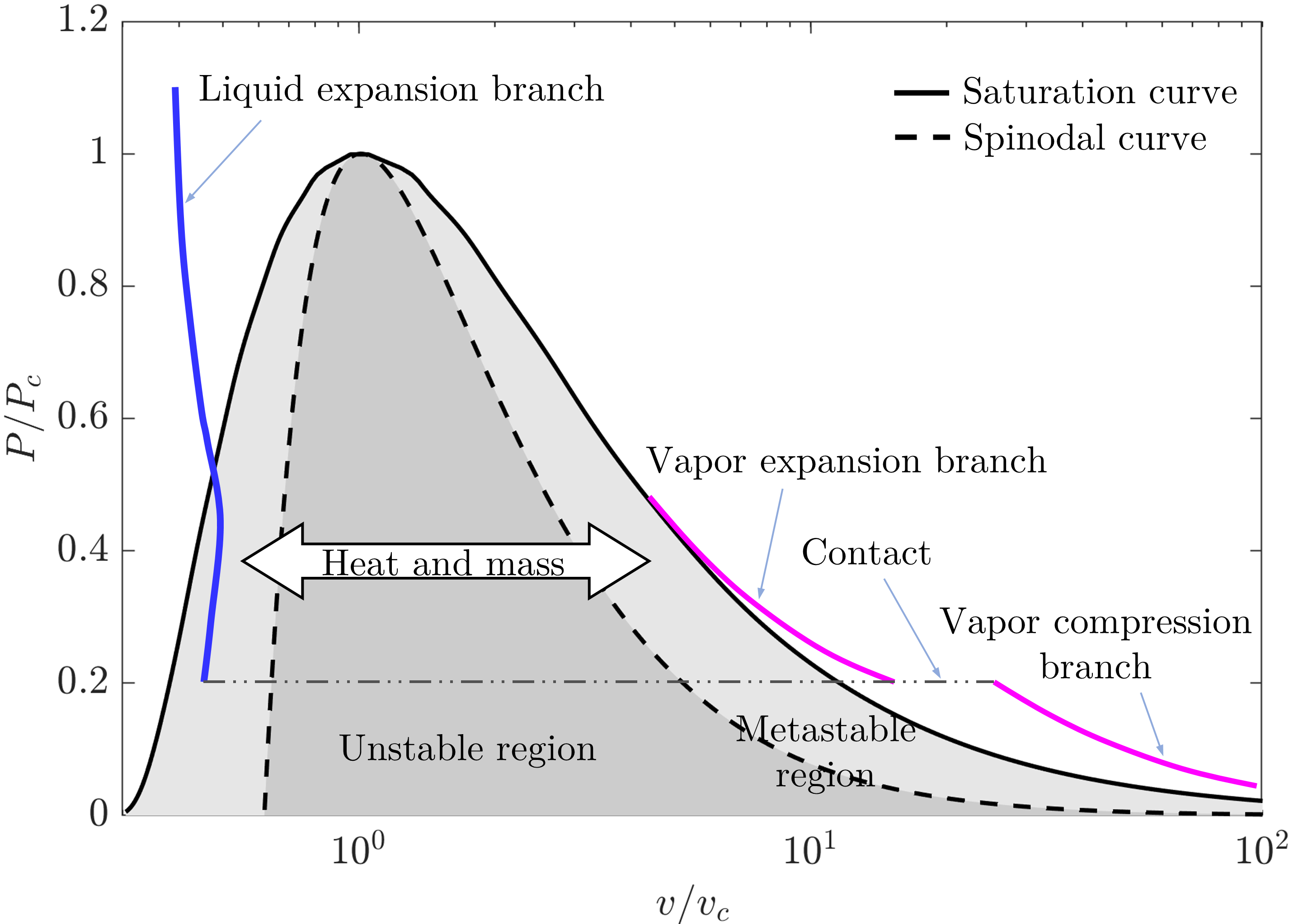} 
\caption{Diagram of the Flash evaporation Riemann Problem with relaxation}
\label{fig2}
\end{figure}

Figure \ref{fig2} shows a $(v,P)$ phase diagram of the FeRP under the HRM assumption, where the liquid phase is initially at high pressure while the vapor phase is at low pressure; driven by the pressure difference, the liquid undergoes flash evaporation. Unlike in the standard FeRP, with the relaxation process considered, the solution curve of the Riemann problem does not always lie on the equilibrium trajectory. The high-pressure liquid initially expands along the isentrope, and initiates phase transition after entering the metastable region, generating the vapor phase and expanding together. The newly generated vapor and the liquid are in mechanical equilibrium, and exchange heat and mass at finite rates, which causes the vapor-liquid state to transition from the frozen state toward thermodynamic equilibrium. In addition, we still assume that the compressed branch does not cross the saturation line in the reverse direction.

The governing equations of such a process are a combination of the one-dimensional Euler equations and the relaxation equation for the vapor mass fraction $\beta$:
\begin{subequations}
\label{eq:system}
\begin{empheq}[left=\empheqlbrace]{align}
&\frac{\partial \mathbf{U}}{\partial t}
+\frac{\partial \mathbf{F}(\mathbf{U})}{\partial x}
=0, \label{eq:euler}\\
&\frac{\partial(\rho\beta)}{\partial t}
+\frac{\partial(\rho\beta u)}{\partial x}
=\rho R. \label{eq:beta}
\end{empheq}
\end{subequations}
where $\mathbf{U}$ is the vector of conservative variables and $\mathbf{F}$ is the flux vector. Their specific expressions are:
\begin{equation}
\mathbf{U}=\left( \begin{matrix}
   \rho   \\
   \rho u  \\
   \rho (e+\frac{1}{2}{{u}^{2}})  \\
\end{matrix} \right),\text{  }\mathbf{F}=\left(\begin{matrix}
   \rho u  \\
   \rho {{u}^{2}}+P  \\
   \rho u(e+\frac{1}{2}{{u}^{2}})+Pu  \\
\end{matrix} \right).
\label{eq:UF}
\end{equation}
$\beta$ is the vapor mass fraction, which satisfies the relation:
\begin{equation}
    \beta=
\left(\frac{1}{\rho}-\frac{1}{\rho_l}\right)
\left(\frac{1}{\rho_v}-\frac{1}{\rho_l}\right)^{-1}
= \frac{v-v_l}{v_v - v_l}.
\end{equation}

In this relaxation system, it is necessary to emphasize the physical quantities of three states: single-phase, two-phase equilibrium, and two-phase relaxation.

\textbf{1. Single-phase fluid}:

Taking $\rho$ and $T$ as independent variables, the pressure $P$ is given by the equation of state:
\begin{equation}
    P=P\left( \rho ,T \right)
    \label{eq:eos}
\end{equation}\par
The internal energy is expressed as:
\begin{equation}
    e(\rho,T)=\int_{{{T}_{0}}}^{T}{{{C}_{v}}\text{d}T}-\int_{{{\rho }_{0}}}^{\rho }{\frac{1}{{{\rho }^{2}}}\left( T\pderiv{P}{T}{\rho}-P \right)}\text{d}\rho +e_0
    \label{eq:e}
\end{equation}
where $C_v$ is the specific heat at constant volume. $C_v$ is also a function of $\rho$ and $T$, which can also be called the second equation of state:
\begin{equation}
    {C_v}={C_v}(\rho ,T)
    \label{eq:cv}
\end{equation}\par

The entropy expression of the single-phase fluid is:
\begin{equation}
    s(\rho,T)=\int_{{{T}_{0}}}^{T}{\frac{{{C}_{v}}}{T}\mathrm{d}T}-\int_{{{\rho }_{0}}}^{\rho }{\frac{1}{{{\rho }^{2}}}{\pderiv{P}{T}{\rho}}\mathrm{d}\rho }+s_0
    \label{eq:s}
\end{equation}\par

The speed of sound is:
\begin{equation}
    {c^2}= \pderiv{P}{\rho}{s}=\pderiv{P}{\rho}{T}+\frac{T}{{{C}_{v}}{{\rho }^{2}}}\pderiv{P}{T}{\rho}^{2}
    \label{eq:c}
\end{equation}

\textbf{2. Saturated vapor-liquid mixture fluid:}

Within the HRM framework, there exists a special scenario: the HEM limit, where the vapor-liquid mixture not only maintains mechanical equilibrium \eqref{eq:HRM_mix}, but also satisfies thermodynamic equilibrium ($T_l = T_v = T_\text{sat}$). Therefore, the vapor mass fraction is uniquely determined by the phase equilibrium condition, the relaxation equation \eqref{eq:beta} no longer serves as an independent evolution equation, and the Euler equations \eqref{eq:euler} are directly closed by the vapor-liquid equilibrium (VLE). This is also the mixing rule in the standard FeRP framework (see Section 2 of \citealt{Bai2026FeRP}). Here, the pressure $P$ is determined solely by the temperature $T$, and their relation is referred to as the saturation curve.

\begin{equation}
    P=P^\text{sat}\left(T \right)
    \label{equ_psat}
\end{equation}\par
Each point $(P,T)$ on the saturation curve corresponds to two density values, namely the saturated vapor density $\rho^\text{sat}_v$ and the saturated liquid density $\rho^\text{sat}_l$, both of which are single-valued functions of $P$ or $T$. The two-phase equilibrium mixture density $\rho$ satisfies the relation:
\begin{equation}
  \rho=\left(\frac{\beta^\text{sat}}{\rho^\text{sat}_v(T)}+\frac{1-\beta^\text{sat}}{\rho^\text{sat}_l(T)}\right)^{-1}
\end{equation}\par
Specifically, the saturated vapor mass fraction here satisfies the two-parameter form under equilibrium mixing:
\begin{equation}
    \beta^\text{sat}(\rho,T)=
\left(\frac{1}{\rho}-\frac{1}{\rho^\text{sat}_l(T)}\right)
\left(\frac{1}{\rho^\text{sat}_v(T)}-\frac{1}{\rho^\text{sat}_l(T)}\right)^{-1}.
\label{eq:betasat}
\end{equation}

The internal energy expression of the saturated two phases is:
\begin{equation}
    e^\text{sat}(\rho,T)=\beta^\text{sat} e^\text{sat}_v(T) +(1-\beta^\text{sat}) e^\text{sat}_l(T)
    \label{eq:mix_esat}
\end{equation}
where $e^\text{sat}_v$ and $e^\text{sat}_l$ are the internal energies at point $(P,T)$ on the vapor and liquid saturation lines, which are also single-valued functions of $P$ or $T$.

The complete equilibrium speed of sound in the two-phase region can be calculated from the ratio of the derivatives of mixture entropy with respect to density and pressure\citep{Saurel2008}, which simplifies to:
\begin{equation}
c_{eq}^2
=\frac{1}{\rho^2} \left(
\frac{\beta}{\rho_v^2(c^\text{sat}_v)^2}
+\frac{1-\beta}{\rho_l^2(c^\text{sat}_l)^2}
+ T\left(
\frac{\beta}{C^\text{sat}_{P,v}}
\left(\frac{\mathrm ds^\text{sat}_v}{\mathrm dP}\right)^2
+\frac{1-\beta}{C^\text{sat}_{P,l}}
\left(\frac{\mathrm ds^\text{sat}_l}{\mathrm dP}\right)^2
\right)\right)^{-1}.
\label{eq:c_eq}
\end{equation}
where $c^\text{sat}_v$, $c^\text{sat}_l$, $C^\text{sat}_{P,v}$, $C^\text{sat}_{P,l}$, and $s^\text{sat}_v, s^\text{sat}_l$ are the speed of sound, specific heat at constant pressure, and entropy at $(P,T)$ on the saturation lines, respectively. They are also single-valued functions of $P$ or $T$. In addition, $C_v$ and $C_P$ satisfy the thermodynamic relation \citep{Sedov1973}:
\begin{equation}
    {C_p}-{C_v}=\frac{T}{{\rho^2}}\frac{\left( {\partial P}/{\partial T}\; \right)_\rho^2}{{\partial P}/{\partial \rho}}.
    \label{eq:CPCv}
\end{equation}

\textbf{3. Thermodynamically relaxing vapor-liquid mixture fluid:}

During two-phase relaxation mixing, the fluid state is jointly determined by the system of equations \eqref{eq:system}. The velocities and pressures of the two phases are in equilibrium, but their temperatures are allowed to be distinct and non-saturated. Here, the vapor mass fraction satisfies the mechanical mixing relation:

\begin{equation}
    \beta=
\left(\frac{1}{\rho}-\frac{1}{\rho_l(P,T_l)}\right)
\left(\frac{1}{\rho_v(P,T_v)}-\frac{1}{\rho_l(P,T_l)}\right)^{-1}.
\label{eq:min_beta}
\end{equation}
\par

The internal energy expression of the two phases is:
\begin{equation}
    e=\beta e(\rho_v, T_v) +(1-\beta) e(\rho_l, T_l).
    \label{eq:mix_e}
\end{equation}

For the Euler part in the relaxation system, its characteristic wave speed is the mechanical mixture speed of sound of the two phases, namely Wood's speed of sound:
\begin{equation}
c_W^2=
\frac{1}{\rho^2} \left(
\frac{\beta}{\rho_v^2c_v^2}
+\frac{1-\beta}{\rho_l^2c_l^2}
\right)
^{-1}.
\label{eq:c_W}
\end{equation}
where $c_v = c(\rho_v, T_v), c_l = c(\rho_l, T_l)$ are the speeds of sound of the two phases.
For the entire system, the finite-rate heat and mass transfer processes further affect its acoustic propagation characteristics, and its dynamic speed of sound is jointly determined by the relaxation times of heat and mass transfer and the perturbation frequency \citep{Radovskii1970}:
\begin{equation}
c_\omega^2
=c_W^2+(c_{\mathrm{eq}}^2-c_W^2)
\left(
\frac{\chi}{1-i\omega\tau_m}
+
\frac{1-\chi}{1-i\omega\tau_t}
\right).
\end{equation}
where $\tau_m, \tau_t$ denote the effective relaxation times of mass and heat transfer, $\omega$ is the perturbation frequency, and $\chi$ denotes the weighting coefficient between the two relaxation modes. It is worth noting that this speed of sound does not enter the actual computation, and only represents the acoustic response of the entire system. Specifically, in the high-frequency and low-frequency limits, the dynamic speed of sound approaches Wood's speed of sound and the complete thermodynamic equilibrium speed of sound, respectively:
\begin{equation}
\begin{aligned}
c_\omega^2\rightarrow c_W^2& \quad (\omega \tau_m\rightarrow\infty, \omega \tau_t\rightarrow\infty)\\
c_\omega^2\rightarrow c_{eq}^2& \quad(\omega \tau_m\rightarrow 0, \omega \tau_t\rightarrow 0) .
\label{eq:c_dynamic}
\end{aligned}
\end{equation}

\section{Pressure Oscillation Problem and Its Remedy for the Homogeneous Relaxation Real Fluid}

In conservative finite volume methods, across vapor-liquid interfaces or during the mixing of nonlinear-EoS fluids in different states, the pressure equilibrium is violated by numerical averaging. Designing an oscillation-free method for the system of equations \eqref{eq:system} is not an easy task: it must not only eliminate the numerical pressure oscillations in two-phase mixtures, but also preserve the pressure fluctuations naturally generated by the two-phase thermodynamic relaxation process, and can consistently degenerate into a completely pressure-equilibrium real-fluid method in the supercritical single-phase state. To overcome this challenge, we must provide a physical explanation for the pressure oscillation problem, and coincidentally, from a physical perspective, it is closely related to the thermodynamic equilibrium behavior of two-phase fluids.

Before proceeding, the concept of pressure equilibrium must be clarified. In the present context, two distinct notions of ``pressure equilibrium'' are involved. The first is that under the homogeneous mixture assumption, different phases mixed at each Eulerian coordinate point are in pressure equilibrium (Equation \eqref{eq:HRM_mix}), with a single-valued pressure $P(x, y)$. For the discrete finite-volume method, this corresponds to a single pressure state $P_{i,j}$ within each grid cell.
The second is that in the Euler equations, there should inherently be no heat and mass exchange between fluid elements; once pressure equilibrium is reached, it should be maintained over time in the absence of external perturbations. The so-called pressure oscillation problem means that this equilibrium property of the Euler equations is violated in the finite-volume method. Meanwhile, if the interphase heat and mass exchange indeed exist, a pressure change should occur. Recovering an oscillation-free pressure means eliminating the pressure oscillations caused by numerical mixing while preserving the physical pressure changes arising from the actual relaxation process.

\subsection{Numerical and Physical Interpretations of Pressure Oscillations}

We examine the energy conservation equation of the one-dimensional Euler equations. At a contact discontinuity where both velocity $u=u_0$ and pressure $p=p_0$ are constant, the energy equation degenerates into a pure advection equation for the volumetric internal energy $\rho e$:
\begin{equation}
    \frac{\partial (\rho e)}{\partial t} + u_0 \frac{\partial (\rho e)}{\partial x} = 0
    \label{eq:rho_e_eq}
\end{equation}
For a first-order finite volume scheme, during the update from time step $n$ to $n+1$, the conservative variable $U$ within a grid cell is the weighted average of the fluid elements entering this grid cell during the evolution. Assuming two fluid elements enter this grid cell, denoted by subscripts $L$ and $R$, and letting $\lambda$ be the weighting coefficient determined by the CFL condition and advection velocity, the discrete update can be expressed as:
\begin{equation}
    \begin{aligned}
       \rho_i^{n+1} &= (1-\lambda)\rho_L + \lambda \rho_R = \bar{\rho} \\
        (\rho e)_i^{n+1} &= (1-\lambda)(\rho e)_L + \lambda (\rho e)_R = \overline{\rho e}
    \end{aligned}
\end{equation}
Numerically speaking, when the equation of state is combined with the two conservative variables $(\rho, \rho e)$ to recover the pressure, we generally have:
\begin{equation}
     p^{n+1} = p\left( \rho^{n+1}, (\rho e)^{n+1} \right) = p\left( \bar{\rho}, \overline{\rho e} \right) \neq p_0
     \label{eq:pneq}
 \end{equation}
This inequality arises because in a nonlinear EoS, the pressure $p$ is not a linear function of the conservative variables $(\rho, \rho e)$. Linear weighting does not commute with nonlinear functions; that is, the pressure of the averaged state does not equal the average of the pressures:
\begin{equation}
p(\bar{\rho}, \overline{\rho e}) \neq (1-\lambda)p\left(\rho_L, (\rho e)_L\right) + \lambda p\left(\rho_R, (\rho e)_R\right) = p_0
\end{equation}

The above explanation is the one adopted in most studies on oscillation-free methods for real fluids and two-phase flows \citep{Shyue1999,Ma2017,Bai2026RFQC1}, and its perspective is based on numerical analysis. However, from a physical perspective, we point out that the so-called pressure oscillation problem is actually caused by the thermodynamic equilibrium assumption implicit in the Godunov method\citep{Bai2026RFQC2}. The operation in Equation \eqref{eq:pneq}:
\begin{equation*}
     p^{n+1} = p\left( \bar{\rho}, \overline{\rho e} \right)
\end{equation*}
essentially applies the two-parameter assumption in continuum mechanics: when a continuum element is in thermodynamic equilibrium, its macroscopic state is completely determined by a thermodynamic potential composed of two independent parameters \citep{Sedov1973}. This implies that recovering pressure from the EoS and the two conservative variables forces two fluid elements in different states into instantaneous thermodynamic equilibrium. For an ideal gas, its isochoric-adiabatic thermodynamic equilibrium process happens to be isobaric; whereas for a real fluid, intermolecular potential energy exists, and the isochoric-adiabatic mixing process within a finite volume cell is not isobaric. If they are forced to reach thermodynamic equilibrium, the pressure will inevitably change \citep{Landau1980}. However, from the perspective of the Euler equations, such an equilibrium process should not occur at all.

Therefore, we can formulate the pressure oscillation problem as follows: the finite-volume method imposes a thermodynamic equilibrium condition on different-state fluid elements within a discrete cell, thereby violating the mechanical equilibrium property of the Euler equations.
Guided by this understanding, how can such a challenge be addressed Physically speaking, thermodynamic equilibrium inherently entails a loss of information; to maintain mechanical equilibrium between fluid elements, additional mechanical equilibrium information must be introduced inside the discrete cell to recover the pressure, i.e., $p^{n+1} = f\left( \bar{\rho}, \overline{\rho e}, x, y, \dots \right)$.

\subsection{Internal Energy--Pressure Affine Relation and Mechanical Equilibrium Recovery at Homogeneous Equilibrium}

The Real Fluid Quasi-Conservative (RFQC) method \citep{Bai2026RFQC1} embodies this physical concept to design an oscillation-free method within the HEM framework. The method extracts mechanical equilibrium information from the internal energy--pressure affine relation and evolves these parameters via advection equations, thereby recovering an oscillation-free pressure. Specifically, according to the isentropic property of the Euler equations, we consider the isentropic affine relation between internal energy and pressure:
\begin{equation}
    (\rho e)_{s=\text{const}} = \xi P + E_0
    \label{eq:rhoe_xi_E0_s}
\end{equation}
where $\xi$ is the slope of the local tangent to the internal energy--pressure isentropic curve, and $E_0$ is the intercept corresponding to this local tangent:
\begin{equation}
    \xi = \pderiv{(\rho e)}{P}{s} =  \frac{e + {p}/{\rho}}{c_{eq}^2} = \frac{h}{c_{eq}^2} , \quad E_0 = (\rho e)_{s=\text{const}} - \xi P
    \label{eq:xi_E0_1}
\end{equation}

The RFQC method freezes $(\xi,E_0)$ within each time step along the particle pathline, and evolves them via the advection equations:
\begin{equation}
    \Dt \xi =0, \quad \Dt {E_0} = 0
    \label{eq:Dxi_DE0_1}
\end{equation}
and recovers the oscillation-free pressure by combining the conservative variables with the evolved thermodynamic parameters:
\begin{equation}
    P = \frac{\overline{\rho e} - E_0}{\xi}
    \label{eq:P_xi_E0}
\end{equation}
Finally, from this recovered pressure and the EoS, the new internal energy $e = e_\text{EoS}(P,\rho)$ and affine parameters $(\xi,E_0)$ are recalculated for pressure recovery in the next time step, which is called thermodynamic re-projection. From a numerical perspective, the RFQC method approximates the isentropic relation between internal energy and pressure via a local tangent, and thus exhibits a second-order curvature deviation from the actual isentropic curve. By recalculating the internal energy, this deviation is converted into an energy error, also termed the re-projection error:
\begin{equation}
    \epsilon =e_\text{EoS}(P,\rho) - \frac{\overline{\rho e}}{\overline{\rho}}.
    \label{eq:epsilon}
\end{equation}
This error is of second order with respect to the time step in smooth regions. Precisely by introducing a controllable error, the RFQC method achieves a balance among computational accuracy, shock capturing, and oscillation-free pressure.

\subsection{Mechanical Equilibrium Evolution Equation for Homogeneous Relaxation System}

Guided by this concept of recovering mechanical equilibrium pressure via affine relations, we now design an oscillation-free finite-volume method for the homogeneous relaxation system. To this end, we first consider the pressure evolution equation. With the mixture density, vapor mass fraction, and phasic specific entropies as independent variables, the pressure is written as:
\begin{equation}
 P=P(\rho,\beta,s_l,s_v).
 \label{eq:p_rho_beta_s}
\end{equation}
Taking the material derivative along the pathline yields:
\begin{equation}
 \begin{aligned}
 \Dt{P}={}&
 \pderiv{P}{\rho}{\beta,s_l,s_v}\Dt{\rho}
 +\pderiv{P}{\beta}{\rho,s_l,s_v}\Dt{\beta}\\
 &+\pderiv{P}{s_l}{\rho,\beta,s_v}\Dt{s_l}
 +\pderiv{P}{s_v}{\rho,\beta,s_l}\Dt{s_v}.
 \end{aligned}
 \label{eq:DPDt_1}
\end{equation}
The first partial derivative fixes the mass fraction and phasic specific entropies, which corresponds to the characteristic speed of sound of the Euler equations \eqref{eq:euler}, namely:
\begin{equation}
 \pderiv{P}{\rho}{\beta,s_l,s_v}=c_W^2.
 \label{eq:wood-entropy-derivative}
\end{equation}
Combining the mass conservation equation and the vapor mass fraction relaxation equation, it is easy to obtain $\mathrm{D}\beta/\mathrm{D}t = R$. The first term of Equation \eqref{eq:DPDt_1} gives the frozen compression response, while the remaining terms constitute the phase-change pressure source term:
\begin{equation}
 \begin{aligned}
 R_P={}
 \pderiv{P}{\beta}{\rho,s_l,s_v}R+\pderiv{P}{s_l}{\rho,\beta,s_v}\Dt{s_l}
 +\pderiv{P}{s_v}{\rho,\beta,s_l}\Dt{s_v}.
 \end{aligned}
 \label{eq:Rp_def}
\end{equation}
$R_P$ represents the contribution from interphase heat and mass exchange to the rate of common pressure change. Therefore, the pressure evolution equation is rewritten as:

\begin{equation}
 \Dt{P}+\rho c_W^2\frac{\partial u}{\partial x}=R_P.
 \label{eq:DPDt_2}
\end{equation}
Specifically, for the hyperbolic part of the relaxation system, $R_P=0$, we have:

\begin{equation}
 \Dt{P}+\rho c_W^2\frac{\partial u}{\partial x}=0. \quad (\text{Hyperbolic part})
 \label{eq:DPDt_hyper}
\end{equation}

Considering a more general internal energy--pressure relation in the relaxation system (without the isentropic assumption), we introduce an affine form for each actual state:
\begin{equation}
\rho e=\xi P+E_0.
\label{eq:rhoe_xi_E0}
\end{equation}
Without specifying the concrete expression of $\xi$ for now, differentiating along the pathline gives:
\begin{equation}
\Dt{(\rho e)}= \xi\Dt{P} +P\Dt{\xi} +\Dt{E_0}.
\label{eq:Drhoe_xi_E0}
\end{equation}
Here, the material derivative of internal energy satisfies:
\begin{equation}
    \Dt{(\rho e)} =e \Dt\rho +\rho \Dt e= e \left( -\rho \frac{\partial u}{\partial x} \right) +  \rho \left( -\frac{p}{\rho}\frac{\partial u}{\partial x} \right) 
    =-\rho h\frac{\partial u}{\partial x},
\label{eq:DrhoeDt_1}
\end{equation}
Substituting Equations \eqref{eq:DPDt_2} and \eqref{eq:DrhoeDt_1} into \eqref{eq:Drhoe_xi_E0} yields:
\begin{equation}
\begin{aligned}
-\rho h\frac{\partial u}{\partial x}
={}&
\xi\left(
-\rho c_W^2\frac{\partial u}{\partial x}+R_P
\right)
+P\Dt{\xi}
+\Dt{E_0}.
\end{aligned}
\end{equation}
Moving the term caused by pressure variation to the left-hand side and rearranging gives:
\begin{equation}
P\Dt{\xi}+\Dt{E_0}=\rho\left(\xi c_W^2-h\right)
\frac{\partial u}{\partial x}-\xi R_P.
\end{equation}

To satisfy the isentropic property of the Euler equations and match the isentropic compression work of pressure with that of internal energy (meaning that mechanical work does not cause entropy change), we set the explicit compression term on the right-hand side to vanish, taking $\xi c_W^2=h$. At this point, $\xi$ is precisely the frozen internal energy--pressure isentropic slope of the hyperbolic part. Using Equations \eqref{eq:DPDt_hyper} and \eqref{eq:DrhoeDt_1}:

\begin{equation}
    \xi = \pderiv{(\rho e)}{P}{\beta,s_l,s_v} = \frac{(\mathrm{D}(\rho e)/\mathrm{D}t)_{\beta,s_l,s_v}}{(\mathrm{D}P/\mathrm{D}t)_{\beta,s_l,s_v}} = \frac{-\rho h {\partial u}/{\partial x}}{-\rho c_W^2 {\partial u}/{\partial x}} = \frac{h}{c_W^2}
    \label{eq:xi_def}
\end{equation}

We have thus derived the governing equation that the internal energy--pressure affine parameters must satisfy to maintain mechanical equilibrium in a homogeneous relaxation system, namely, the mechanical equilibrium evolution equation:
\begin{equation}
P\Dt{\xi} + \Dt{E_0}
=-\xi R_P.
\label{eq:Dxi_DE0_basic}
\end{equation}
Compared with the single-phase supercritical regime or the HEM limit, the mechanical equilibrium evolution equation in the homogeneous relaxation system contains an additional heat and mass transfer source term on the right-hand side.

Now, following the ``freezing'' idea of RFQC, we set the material derivative of $\xi$ to zero and simultaneously remove the paired term in the intercept equation, obtaining:
\begin{equation}
\Dt{\xi}=0,
\qquad
\Dt{E_0}=-\xi R_P.
\label{eq:Dxi_DE0_HRRF}
\end{equation}

Equation \eqref{eq:Dxi_DE0_HRRF} is the thermodynamic parameter equation used to recover the oscillation-free pressure in the proposed method. We update Equation \eqref{eq:Dxi_DE0_HRRF} simultaneously with the system \eqref{eq:system} to obtain the temporary parameters $(\xi_\text{temp}^{n+1},E_{0,\text{temp}}^{n+1})$ after frozen evolution, thereby recovering the oscillation-free pressure $P^{n+1} = (\overline{\rho e} - E_{0,\text{temp}}^{n+1})/\xi_\text{temp}^{n+1}$.

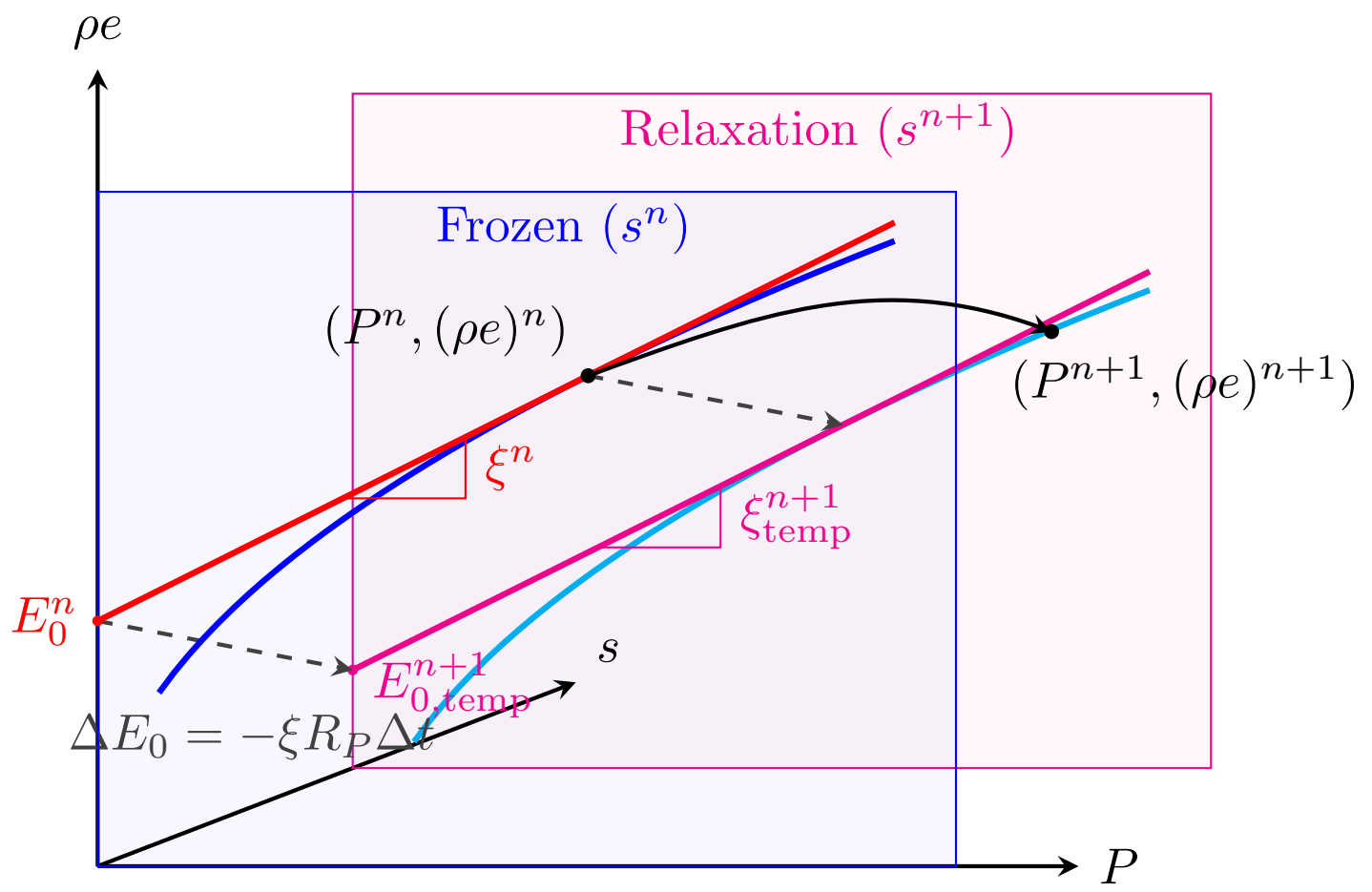
\begin{figure}[htbp]
    \centering
    \begin{tikzpicture}[
        x={(1cm,0cm)}, 
        y={(0.52cm,0.2cm)}, 
        z={(0cm,1cm)}, 
        scale=0.85, 
        >=stealth, 
        font=\small
    ]
        \definecolor{mypink}{RGB}{255, 182, 193}

        \fill[mypink, opacity=0.1] (0,4,0) -- (7,4,0) -- (7,4,5.5) -- (0,4,5.5) -- cycle;
        
        \fill[blue!20, opacity=0.15] (0,0,0) -- (7,0,0) -- (7,0,5.5) -- (0,0,5.5) -- cycle;

        \draw[->, thick] (0,0,0) -- (8,0,0) node[right] {$P$};
        \draw[->, thick] (0,0,0) -- (0,7.5,0) node[above right] {$s$};
        \draw[->, thick] (0,0,0) -- (0,0,6.5) node[above] {$\rho e$};

        \draw[magenta, thin] (0,4,0) -- (7,4,0) -- (7,4,5.5) -- (0,4,5.5) -- cycle;
        \node[magenta] at (3.8, 4, 5.2) {Relaxation ($s^{n+1}$)};
        
        \draw[very thick, cyan, domain=0.5:6.5, samples=50] plot (\x, 4, {2*sqrt(\x) - 1.2});
        
        \draw[very thick, magenta] (0,4,0.8) -- (6.5,4,4.05);
        \filldraw[magenta] (0,4,0.8) circle (1pt) node[right, yshift=-3pt] {$E_{0,\text{temp}}^{n+1}$};
        \draw[magenta, thin] (2,4,1.8) -- (3,4,1.8) -- (3,4,2.3);
        \node[magenta, right] at (3,4,2.05) {$\xi_\text{temp}^{n+1}$};
        
        \coordinate (B) at (5.7, 4.0, 3.56);
        \filldraw[black] (B) circle (1.5pt) node[below right, xshift=-12pt, yshift=-1pt] {$(P^{n+1}, (\rho e)^{n+1})$};

        \draw[->, thick, darkgray, dashed] (0,0,2) -- (0,4,0.8) node[midway, right, xshift=-35pt, yshift=-18pt] {$\Delta E_0 = -\xi R_P \Delta t$};
        \draw[->, thick, darkgray, dashed] (4,0,4) -- (4,4,2.8);

        \draw[blue, thin] (0,0,0) -- (7,0,0) -- (7,0,5.5) -- (0,0,5.5) -- cycle;
        \node[blue] at (3.8, 0, 5.2) {Frozen ($s^n$)};
        
        \draw[very thick, blue, domain=0.5:6.5, samples=50] plot (\x, 0, {2*sqrt(\x)});
        
        \draw[very thick, red] (0,0,2) -- (6.5,0,5.25);
        \filldraw[red] (0,0,2) circle (1pt) node[left] {$E_0^n$};
        \draw[red, thin] (2,0,3) -- (3,0,3) -- (3,0,3.5);
        \node[red, right] at (3,0,3.25) {$\xi^n$};
        
        \coordinate (A) at (4, 0, 4);
        \filldraw[black] (A) circle (1.5pt) node[above left] {$(P^n, (\rho e)^n)$};

        \draw[->, thick, black] (A) to[out=20, in=160] (B);
        
    \end{tikzpicture}
    \caption{Diagram of the affine parameter evolution across isentropic planes. The blue and red lines represent the internal energy-pressure curve and its tangent in the frozen limit, while the light blue and magenta lines indicate the drifted internal energy-pressure curve and its tangent due to thermodynamic relaxation.}
    \label{fig:affine_evolution}
\end{figure}

The physical meaning of the above operations is illustrated in Figure \ref{fig:affine_evolution}. Equation \eqref{eq:Dxi_DE0_HRRF} decomposes the actual pressure evolution process into two steps: the first is frozen evolution, which freezes the slope $\xi$ and approximates the internal energy--pressure evolution curve in the frozen limit with its tangent line; the second is relaxation drift, which shifts the tangent line to approximate the internal energy--pressure evolution curve after thermodynamic relaxation. Since relaxation is a process approaching thermodynamic equilibrium, the two curves before and after relaxation lie on different isentropic planes.

Freezing the slope $\xi$ introduces a second-order energy conservation error related to the curvature of the internal energy--pressure curve, the rigorous proof of which is detailed in Section \ref{sec:eq_reprojetion}.

\subsection{Closure of Equations and Single-Phase Degeneration}
In Section 2, we saw that the governing equations for a single-phase fluid are solely the Euler equations \eqref{eq:euler}, whereas the relaxing two-phase fluid requires the additional vapor transport equation \eqref{eq:beta}. Obviously, when the two-phase mixture transitions to a single phase, the vapor transport equation degenerates into the mass equation or directly vanishes, and the two-phase relaxation system degenerates into a single-phase equilibrium system. This imposes requirements on the design of the method: we must not only recover an oscillation-free pressure in the two-phase region, but also naturally degenerate into a single-phase real-fluid method while satisfying pressure equilibrium.

Now, we consider the implementation scheme of such a method. At this stage, having completed the conservative evolution, relaxation update, and pressure recovery, state variables $(\rho, e, \beta, P)$ are obtained (where $\rho = \overline{\rho}$ and $e = \overline{\rho e}/ \overline{\rho}$). However, the respective densities and temperatures of the two phases, $\rho_v, T_v, \rho_l, T_l$, remain unknown. For these four unknowns, there are exactly four equations to close them, namely the equations of state and the mass and energy mixing rules:
\begin{equation}
\begin{dcases}
    &P = P(\rho_v,T_v);\\
    &P = P(\rho_l,T_l);\\
    &\frac{1}{\rho} = \frac{\beta}{\rho_v}+\frac{1-\beta}{\rho_l};\\
    &e = \beta e(\rho_v,T_v)+(1-\beta)e(\rho_l,T_l).
\end{dcases}
\label{eq:4eq}
\end{equation}
The first two equations ensure that the two phases share the given common pressure, while the last two equations satisfy the constraints of total volume and conservative internal energy, respectively. However, as we are about to see, although such a system of equations theoretically has a solution, solving it undoubtedly requires an iterative method, which increases the computational cost; more unacceptably, the equations fail to converge when approaching the single-phase limit, preventing the current numerical method from degenerating into the pressure-equilibrium RFQC method.

Through the equation of state, we write the temperature as $T(P,\rho_k)$ and denote $e(P,\rho) =e(\rho,T(P,\rho))$; through the mixing rule, we express the liquid density as a function of the vapor density, $\rho_l = \rho_l(\rho_v)$.
Consequently, the system of equations \eqref{eq:4eq} is transformed into a root-finding problem for a nonlinear equation:
\begin{equation}
    F(\rho_v)=\beta e(P,\rho_v) +(1-\beta)e(P,\rho_l(\rho_v))-e=0.
\end{equation}

We point out that this equation has no solution in the single-phase limit. In the single-phase limit, $\rho_v=\rho_l=\rho,\quad T_v=T_l=T$. If solved via Newton iteration, the objective function and its derivative become:
\begin{equation}
    F(\rho)=e_{\mathrm{EOS}}(P,\rho)-e = \epsilon,
    \qquad
    \left. \totalderiv{F}{\rho_v} \right|_{\rho_v=\rho}=0.
\end{equation}
At this point, $F(\rho)$ is precisely the energy deviation \eqref{eq:epsilon} handled by the RFQC re-projection, which is of second order with respect to the time step, but is generally non-zero, while its derivative is zero. Therefore, directly solving for the remaining unknowns via the system \eqref{eq:4eq} cannot guarantee root-finding convergence in the single-phase limit, and the algorithm cannot consistently degenerate into the single-phase pressure-equilibrium method.

To solve this problem, we draw on the ``re-projection'' concept from the RFQC method, namely letting the energy entering the next time step be $e_{\mathrm{EOS}}(P,\rho)$, rather than the internal energy directly calculated from the conservation law. To this end, we drop the last energy mixing equation in \eqref{eq:4eq} as a constraint and introduce the evolution equation for the vapor volume fraction, turning the volume allocation, which originally needed to be solved through energy conservation, into a directly evolved physical quantity.

Again taking the mixture density, vapor mass fraction, and phasic specific entropies as independent variables, the volume fraction is written as:
\begin{equation}
 \alpha=\alpha(\rho,\beta,s_l,s_v).
 \label{eq:alpha_rho_beta_s}
\end{equation}
Taking the material derivative along the pathline and applying the chain rule yields:
\begin{equation}
\begin{aligned}
    \Dt{\alpha}=&
    \pderiv{\alpha}{\rho}{\beta,s_l,s_v}\Dt{\rho}
    +\pderiv{\alpha}{\beta}{\rho,s_l,s_v}\Dt{\beta}\\
    &+\pderiv{\alpha}{s_l}{\rho,\beta,s_v}\Dt{s_l}
    +\pderiv{\alpha}{s_v}{\rho,\beta,s_l}\Dt{s_v}.
    \label{DalphaDt_1}
\end{aligned}
\end{equation}
The first partial derivative fixes the mass fraction and phasic specific entropies, representing the phase volume redistribution caused by pure mechanical compression. Denoting:
\begin{equation}
    \pderiv{\alpha}{\rho}{\beta,s_l,s_v}\Dt{\rho} = -\rho \pderiv{\alpha}{\rho}{\beta,s_l,s_v}\frac{\partial u}{\partial x} = K\frac{\partial u}{\partial x}
\end{equation}
where $K$ characterizes compressibility mismatch of the two phases. The remaining terms in Equation \eqref{DalphaDt_1} constitute the volume fraction source term induced by thermodynamic relaxation, denoted as $R_\alpha$:
\begin{equation}
\begin{aligned}
    R_\alpha={}
    \pderiv{\alpha}{\beta}{\rho,s_l,s_v}R
    +\pderiv{\alpha}{s_l}{\rho,\beta,s_v}\Dt{s_l}
    +\pderiv{\alpha}{s_v}{\rho,\beta,s_l}\Dt{s_v}.
\end{aligned}
\label{eq:Ra_def}
\end{equation}
Finally, the evolution equation for the volume fraction is rewritten as:
\begin{equation}
    \Dt{\alpha}
    =K\frac{\partial u}{\partial x} + R_{\alpha} 
    \label{eq:alpha}
\end{equation}

After obtaining $\alpha$, we can easily calculate the remaining unknowns $\rho_v, T_v, \rho_l, T_l$ through the mixing rules and the EoS, and finally recalculate the internal energy $e_\text{EoS} = \beta e(\rho_v,T_v)+(1-\beta)e(\rho_l,T_l)$. This likewise introduces an energy error $\epsilon = e_\text{EoS} - e$, which is exactly the form of the re-projection error in the RFQC method. In the single-phase limit, the current algorithm construction can consistently degenerate into the RFQC method.

It is easy to notice that Equation \eqref{eq:alpha} is the volume fraction equation in the classic Kapila five-equation model, and $K$ is the interface compressibility term \citep{Kapila2001,tenEikelder2017}. It is worth emphasizing that although we explicitly evolve the volume fraction, from the perspective of algorithm construction, the HRRF method established in this paper is closer to an extension of the RFQC method to the HRM framework. In the Kapila five-equation model, an oscillation-free pressure is directly recovered through the evolution of the $\alpha$ equation, whereas the HRRF method recovers the pressure by evolving the thermodynamic parameters $(\xi, E_0)$, and further achieves the closure of the system and degeneration into the single-phase pressure-equilibrium method through the $\alpha$ equation.

\section{Oscillation-Free Homogeneous Relaxation Real Fluid Finite Volume Method}
Combining the Euler equations \eqref{eq:euler}, the mass fraction relaxation equation \eqref{eq:beta}, the thermodynamic parameter evolution equations \eqref{eq:Dxi_DE0_HRRF}, and the volume fraction evolution equation \eqref{eq:alpha}, we finally obtain the governing equations for the oscillation-free Homogeneous Relaxation Real Fluid (HRRF) finite volume method:

\begin{subequations}
\label{eq:7eq_system}
\begin{empheq}[left=\empheqlbrace]{align}
&\frac{\partial \rho}{\partial t}+\frac{\partial (\rho u)}{\partial x}=0,\label{eq:7eq_system1} \\ 
 &\frac{\partial (\rho u)}{\partial t}+\frac{\partial (\rho u^2+P)}{\partial x}=0,\label{eq:7eq_system2} \\ 
 &\frac{\partial(\rho e_t)}{\partial t}
 +\frac{\partial\left(\rho u e_t+Pu\right)}{\partial x}=0,\label{eq:7eq_system3} \\
 &\frac{\partial (\rho\beta)}{\partial t}+\frac{\partial (\rho\beta u)}{\partial x}=\rho R,\label{eq:7eq_system4} \\
 &\Dt{\xi}=0,\label{eq:7eq_system5} \\
 &\Dt{E_0}=-\xi R_P,\label{eq:7eq_system6} \\
 &\Dt{\alpha}=K\frac{\partial u}{\partial x}+R_\alpha.\label{eq:7eq_system7} 
\end{empheq}
\end{subequations}
where $e_t= e+u^2 /2$ is the total energy, and $R,R_P,R_{\alpha}$ are the phase-change source terms for the mass fraction, pressure, and volume fraction, respectively. The thermodynamic parameters $\xi, E_0$ are the frozen slope and intercept of the internal energy--pressure relation, satisfying:
\begin{equation}
    \xi =  \frac{h}{c_{W}^2} , \quad E_0 = \rho e - \xi P
    \label{eq:xi_E0_2}
\end{equation}

Now, we specifically discuss the solution procedure for the above equations, the theoretical derivation and modeling of the phase-change source terms, and the energy conservation error introduced by the method.

\subsection{Numerical Solution Procedure}

Defining the primitive variables $\boldsymbol{W}=(\rho_l,\rho_v,u,P)^T$ and thermodynamic parameters $\boldsymbol{\Phi}= (\xi,E_0)^T$, the solution procedure for the system \eqref{eq:7eq_system} advances as follows:

\begin{itemize}
    \item \textbf{Initialization:} At time $t^0$, given primitive variables $\mathbf{W}^0$ and volume fraction $\alpha^0$, compute the mass fraction, internal energy, and frozen speed of sound ($\beta^0 ,c^0_W,e^0$) via the EoS. Then, calculate the conservative variables $\mathbf{U}^0$, and compute the thermodynamic parameters $\boldsymbol{\Phi}^0$ according to Equation \eqref{eq:xi_E0_2}.
    
    \item \textbf{Reconstruction and Evolution:} At time $t^n$, reconstruct the primitive variables $\mathbf{W}^n$ and thermodynamic parameters $\boldsymbol{\Phi}^n$, and compute the fluxes $\mathbf{F}^n$. Evolve the conservation equations \eqref{eq:7eq_system1}--\eqref{eq:7eq_system3} via a first-order Godunov scheme to obtain the temporary conservative variables $\mathbf{U}^{n+1}_\text{temp} = (\rho^{n+1},\rho u^{n+1},\rho e^{n+1}_{t,\text{temp}})^T$ at time $t^{n+1}$, and evolve the advection equations \eqref{eq:7eq_system5} and \eqref{eq:7eq_system6} to obtain the temporary thermodynamic parameters $\boldsymbol{\Phi}^{n+1}_\text{temp} = (\xi^{n+1}_\text{temp}, E^{n+1}_{0,\text{temp}})^T$ for the next time step. Concurrently, evolve the mass fraction equation \eqref{eq:7eq_system4} and volume fraction equation \eqref{eq:7eq_system7} to obtain $\beta^{n+1}$ and $\alpha^{n+1}$ for the next time step.
    
    \item \textbf{Primitive Variable Recovery:} Based on the evolved conservative variables $\mathbf{U}^{n+1}_\text{temp}$ and thermodynamic parameters $\boldsymbol{\Phi}^{n+1}_\text{temp}$, recover the pressure $P^{n+1}$ at time $t^{n+1}$ according to Equation \eqref{eq:P_xi_E0}, and calculate $\rho_l^{n+1},\rho_v^{n+1}$ from $\beta^{n+1}, \alpha^{n+1}$ to form the primitive variables $\mathbf{W}^{n+1}$.
    
    \item \textbf{Thermodynamic Re-projection:} From the primitive variables $\mathbf{W}^{n+1}$, calculate the internal energy and speed of sound ($c_W^{n+1},e^{n+1}$) strictly via the EoS. Re-form the conservative variables $\mathbf{U}^{n+1}= (\rho^{n+1},\rho u^{n+1},\rho e_t^{n+1})^T$ at time $t^{n+1}$, and update the thermodynamic parameters $\boldsymbol{\Phi}^{n+1} = (\xi^{n+1}, E^{n+1}_{0})^T$ for the next time step according to Equation \eqref{eq:xi_E0_2}.
   
    \item \textbf{Time Stepping:} After re-projection, proceed to the reconstruction and evolution for time $t^{n+1}$ until the target calculation time is reached.
\end{itemize}

It should be emphasized that the conservative variables $\mathbf{U}^{n+1}$ strictly calculated from $\mathbf{W}^{n+1}$ and $\alpha^{n+1}$ via the EoS are the true conservative variables required at time $t^{n+1}$. This operation, referred to as thermodynamic re-projection, is a key step of our method and the source of the energy error. To obtain the oscillation-free pressure $P^{n+2}$ at time $t^{n+2}$, as in the preceding step, $\mathbf{W}^{n+1}$, $\mathbf{U}^{n+1}$, and $\boldsymbol{\Phi}^{n+1}$ at time $t^{n+1}$ must satisfy strict thermodynamic relations. Therefore, $\mathbf{U}^{n+1}$ must be obtained by thermodynamic projection from $\mathbf{W}^{n+1}$ and $\beta^{n+1}$, rather than from the natural evolution of the Godunov method.

\subsection{Modeling of Phase-Change Source Terms}

The phase-change source terms comprise the mass fraction $R$, pressure $R_P$, and volume fraction $R_\alpha$. Among them, $R$ is driven by superheat and subcooling, while $R_P$ and $R_\alpha$ are also associated with interphase heat transfer.

Note that the HRRF method is not restricted to any specific source term models; the models adopted in this section represent a generalized extension of existing formulations, intended solely for investigating the relaxation mechanisms of the FeRP rather than for detailed modeling of specific phenomena.

\textbf{1. Mass exchange driven by enthalpy deviation}

Referring to the source term form of superheated liquid evaporation driven by enthalpy difference in \citet{DownarZapolski1996}, we separately define the relaxation times and enthalpy difference driving forces for the evaporation of superheated liquid and the condensation of subcooled vapor to constitute the mass source term:

\begin{equation}
 R=\frac{1-\beta}{\tau_{M,e}}
     \frac{{h_l-h_l^{\text{sat}}}}{h_v^{\text{sat}}-h_l^{\text{sat}}}
   -\frac\beta{\tau_{M,c}}
     \frac{{h_v^{\text{sat}}-h_v}}{h_v^{\text{sat}}-h_l^{\text{sat}}},
 \label{eq:mass_source}
\end{equation}
where $\tau_{M,e},\tau_{M,c}$ are the relaxation times for evaporation and condensation, and the saturated enthalpy is $ h_k^{\text{sat}}=e(\rho_k^{\text{sat}},T^{\text{sat}})+ P/{\rho_k^{\text{sat}}}$. Evaporation is driven by the liquid excess enthalpy relative to saturation at the same pressure, while condensation is driven by the vapor enthalpy deficit, with their contributions weighted by $\beta$.

Referring to the relaxation time formulation of \citet{DownarZapolski1996}, the relaxation times for evaporation and condensation are defined as:

\begin{equation}
 (\tau_{M,e})^{-1}=\frac{\alpha_{e}^{0.54}}{\tau_{M}}
 \left(\frac{P^{\text{sat}}(T_l)-P}{P_c-P^{\text{sat}}(T_l)}\right)^{1.76},
       \quad T^{\text{sat}}<T_l<T_c
 \label{eq:evaporation_time}
\end{equation}
\begin{equation}
 (\tau_{M,c})^{-1}=
 \displaystyle\frac{\alpha_{c}^{0.54}}{\tau_{M}}
 \left(\frac{P-P^{\text{sat}}(T_v)}{P_c-P^{\text{sat}}(T_v)}\right)^{1.76},
 \quad T_v<T^{\text{sat}}<T_c
 \label{eq:condensation_time}
\end{equation}
The original formulation of \citet{DownarZapolski1996} cannot initiate the evaporation of a superheated liquid phase when $\alpha=0$, and the practice in some literature is to always assign a small value to $\alpha$ \citep{Schmidt2010}. To this end, we define the equilibrium reference vapor volume fractions in the above equations as:
\begin{equation}
 \alpha_{e} = \mathrm{max}\left(\alpha, \frac{\rho_l^{\text{sat}}-\rho}{\rho_l^{\text{sat}}-\rho_v^{\text{sat}}}\right), \quad
 \alpha_{c} = \mathrm{min}\left(\alpha, \frac{\rho_l^{\text{sat}}-\rho}{\rho_l^{\text{sat}}-\rho_v^{\text{sat}}}\right).
 \label{eq:alpha_ref}
\end{equation}
In Equations \eqref{eq:evaporation_time} and \eqref{eq:condensation_time}, $\tau_M$ is the mass transfer relaxation time model coefficient, referred to as the characteristic relaxation time. Without specifying its value here, we treat it as an independent parameter in subsequent sections to investigate how relaxation time affects the Riemann solution.

\textbf{2. Closure of interphase heat transfer}

When interphase heat transfer is considered, closing only the mass source term $R$ is insufficient: we only know the destination of the total energy, but not yet the energy allocation between the two individual phases. Since heat transfer represents a relaxation process from frozen equilibrium toward thermodynamic equilibrium, it therefore involves entropy production. As a result, we need to establish the heat transfer timescale and determine the entropy production rate of each phase. For each fluid element, entropy production originates from two independent irreversible thermodynamic dissipations: the heat conduction $\rho Q$ driven by the interphase temperature difference, and the enthalpy carried by mass transfer. Define the exchange enthalpy as $h_\text{tr}$, which depends on the direction of phase transition:
\begin{equation}
 h_\text{tr}=\begin{cases}h_v,&R\geq0,\\ h_l,&R<0.
 \end{cases}
 \label{eq:exchange-enthalpy}
\end{equation}

Under the mechanical equilibrium condition, the entropy evolution equations of the vapor and liquid phases \citep{Lund2012} are written as:
\begin{equation}
 \rho \beta T_v \Dt{s_v} = \rho (h_\text{tr} - h_v)R + \rho Q,
 \label{eq:entropy_v}
\end{equation}
\begin{equation}
 \rho (1-\beta) T_l \Dt{s_l} = -\rho (h_\text{tr} - h_l)R - \rho Q .
 \label{eq:entropy_l}
\end{equation}
\citet{Lund2012} defined the exchange heat as a function related to the chemical potential. For clarity, we simply assume that the heat exchanged during phase transition is carried by one specific phase.

For the temperature-difference heat transfer $Q$, with pressure and phase fraction fixed, it satisfies:
\begin{equation}
 \beta C_{P,v}\totalderiv{T_v}{t}=Q,\qquad
 (1-\beta)C_{P,l}\totalderiv{T_l}{t}=-Q.
\end{equation}
Subtracting the two equations after unifying the coefficients yields the evolution equation of the temperature difference:
\begin{equation}
 \totalderiv{(T_l-T_v)}{t} = -\left(\frac1{(1-\beta)C_{P,l}} + \frac1{\beta C_{P,v}}\right)Q.
\end{equation}
Assuming that the interphase temperature difference follows a linear relaxation law, i.e., $\mathrm{d}(T_l-T_v)/\mathrm{d}t = -(T_l-T_v)/\tau_T$ with $\tau_T$ being the heat transfer relaxation time, we obtain the closed form of the heat transfer source term $Q$:
\begin{equation}
 Q=\frac{1}{\tau_T}
   \frac{\beta(1-\beta)C_{P,l}C_{P,v}}
        {(1-\beta)C_{P,l}+\beta C_{P,v}}(T_l-T_v).
 \label{eq:heat-source}
\end{equation}

\textbf{3. Closure of the pressure source term}

According to the definition \eqref{eq:Rp_def}, expanding the derivatives of the common pressure $P(\rho,\beta,s_l,s_v)$ with respect to $(\beta, s_l, s_v)$ yields:
\begin{equation}
 R_P=
 \rho^2 c_W^2 \left(\frac{1}{\rho_v}-\frac{1}{\rho_l}\right)R+\rho^2 c_W^2 \frac{(1-\beta) \Gamma_l T_l}{\rho_l c_l^2}\Dt{s_l}
 +\rho^2 c_W^2 \frac{\beta \Gamma_v T_v}{\rho_v c_v^2}\Dt{s_v}.
 \label{eq:Rp2}
\end{equation}
where $ \Gamma_k = ({\partial P}/{\partial e})_{\rho}/{\rho}$ is the Gr\"uneisen coefficient of each phase, which can be directly calculated from the EoS. The detailed derivation of the above derivatives is provided in Appendix \ref{app:pressure_derivatives}.

Substituting the phasic entropy evolution equations \eqref{eq:entropy_v} and \eqref{eq:entropy_l} into the pressure source \eqref{eq:Rp2} and collecting terms yields the physical pressure source model:
\begin{equation}
 R_P=\rho^2 c_W^2 \left(\frac{\Gamma_v}{\rho_v c_v^2}-\frac{\Gamma_l}{\rho_l c_l^2}\right)Q
 + \rho^2 c_W^2 \left(\frac{1}{\rho_v}-\frac{1}{\rho_l} + \frac{\Gamma_v(h_\text{tr}-h_v)}{\rho_v c_v^2} - \frac{\Gamma_l(h_\text{tr}-h_l)}{\rho_l c_l^2} \right) R.
 \label{eq:Rp_model}
\end{equation}
The first part represents the thermodynamic pressure response induced by independent heat transfer, while the second part accounts for the combined effect of phase-change mass redistribution and associated enthalpy on pressure.

\textbf{4. Closure of the volume fraction source term}

According to the definition of $R_\alpha$ in \eqref{eq:Ra_def}, we expand the derivatives of the vapor volume fraction $\alpha(\rho,\beta,s_l,s_v)$ with respect to $(\beta, s_l, s_v)$, and obtain:
\begin{equation}
   R_\alpha=
  \frac{\rho(\alpha c_l^2+(1-\alpha)c_v^2)}{\alpha\rho_lc_l^2+(1-\alpha)\rho_vc_v^2}R
  + \frac{-\alpha(1-\alpha)\rho_l\Gamma_lT_l}{\alpha\rho_lc_l^2+(1-\alpha)\rho_vc_v^2}\Dt{s_l}
 +\frac{\alpha(1-\alpha)\rho_v\Gamma_vT_v}{\alpha\rho_lc_l^2+(1-\alpha)\rho_vc_v^2}\Dt{s_v}.
 \label{eq:Ra2}
\end{equation}
The detailed derivation of the above derivatives is provided in Appendix \ref{app:alpha_derivatives}. In addition, in the volume fraction advection equation \eqref{eq:7eq_system7}, the coefficient $K$ has the expression:
\begin{equation}
 K=\frac{\alpha(1-\alpha)(\rho_lc_l^2-\rho_vc_v^2)}
                     {\alpha\rho_lc_l^2+(1-\alpha)\rho_vc_v^2}.
 \label{eq:alpha_K}
\end{equation}

Substituting the phasic entropy evolution equations \eqref{eq:entropy_v} and \eqref{eq:entropy_l} into Equation \eqref{eq:Ra2} yields the volume fraction source term model:
\begin{equation}
 \begin{aligned}
 R_\alpha={}&\frac{\rho[(1-\alpha)\Gamma_v+\alpha\Gamma_l]}
                     {\alpha\rho_lc_l^2+(1-\alpha)\rho_vc_v^2}\,Q\\
 &+\frac{\rho R}{\alpha\rho_lc_l^2+(1-\alpha)\rho_vc_v^2}
 \bigl\{(1-\alpha)[c_v^2+\Gamma_v(h_\text{tr}-h_v)]
 +\alpha[c_l^2+\Gamma_l(h_\text{tr}-h_l)]\bigr\}.
 \end{aligned}
 \label{eq:Ra_model}
\end{equation}

\subsection{Re-projection Energy Error Analysis}
\label{sec:eq_reprojetion}
The HRRF method inherits the re-projection concept of the RFQC method to ensure the correct evolution of mechanical equilibrium, while also converting the deviation of pressure evolution into an energy error through the EoS. As mentioned in the introduction, the energy conservation error is a common challenge faced by all current oscillation-free methods for real fluids, and how to constrain this error through a physically consistent construction lies at the core of the method design. This subsection proves that the energy conservation error introduced by the HRRF method is controllable: it is a second-order small term dominated by the time step in smooth regions, and is proportional to the numerical dissipation in non-smooth regions.

Letting $\boldsymbol{z}=(\beta,s_l,s_v)$ and taking $(P, \boldsymbol{z})$ as independent variables, the volumetric internal energy of the real fluid is written as:
\begin{equation}
    \psi(P,\boldsymbol{z})=\rho e(P,\beta,s_l,s_v).
    \label{eq:psi_HRRF}
\end{equation}
According to the definition of $\xi$ in \eqref{eq:xi_def}, we have $\psi_P=\xi$. Performing a Taylor expansion of $\psi$ at $(P^n,\boldsymbol{z}^n)$ yields:
\begin{equation}
\begin{aligned}
    \psi^{n+1}
    ={}&\psi^n+\xi^n\Delta P
    +\nabla_{\boldsymbol{z}}\psi^n\cdot\Delta\boldsymbol{z}
    +O(\Delta t^2),
\end{aligned}
\label{eq:psi_Taylor_HRRF}
\end{equation}
where $\Delta P=P^{n+1}-P^n$ and $\Delta\boldsymbol{z}=\boldsymbol{z}^{n+1}-\boldsymbol{z}^n$ are the increments within one time step. In smooth regions, both $\Delta P$ and $\Delta\boldsymbol{z}$ are $O(\Delta t)$. Now, we regard $\Delta \boldsymbol{z}$ in the numerical update as consisting of two parts:
\begin{equation}
    \Delta \boldsymbol{z} = \Delta \boldsymbol{z}_R + \Delta \boldsymbol{z}_\sigma,
\end{equation}
where $\Delta \boldsymbol{z}_R$ represents the physical relaxation part in $\Delta \boldsymbol{z}$ that satisfies the EoS constraints, such as the variation of $\beta$ and entropy increase caused by phase transition; $\Delta \boldsymbol{z}_\sigma$ represents the numerical dissipation part in $\Delta \boldsymbol{z}$.
From Equation \eqref{eq:Drhoe_xi_E0} and the governing equation of mechanical equilibrium \eqref{eq:Dxi_DE0_basic}, we obtain:
\begin{equation}
    \Dt{\psi}=\xi \Dt{P}-\xi R_P.
    \label{eq:DpsiDt_HRRF}
\end{equation}
According to the chain rule for the total derivative of $\psi$, we obtain:
\begin{equation}
   -\xi R_P =  \nabla_{\boldsymbol{z}}\psi\cdot\left(\Dt{\boldsymbol{z}}\right)_R.
    \label{eq:psi_relaxation_identity}
\end{equation}
Therefore, the internal energy increment caused by relaxation within one time step satisfies:
\begin{equation}
    \nabla_{\boldsymbol{z}}\psi^n\cdot\Delta\boldsymbol{z}
    =-\xi^n R_P^n\Delta t + \nabla_{\boldsymbol{z}}\psi^n\cdot\Delta\boldsymbol{z}_\sigma.
    \label{eq:psi_relaxation_increment}
\end{equation}
Substituting this into Equation \eqref{eq:psi_Taylor_HRRF} gives the Taylor expansion of the internal energy:
\begin{equation}
    \psi^{n+1}
    =\psi^n+\xi^n\Delta P
    -\xi^nR_P^n\Delta t+\nabla_{\boldsymbol{z}}\psi^n\cdot\Delta\boldsymbol{z}_\sigma+O(\Delta t^2).
    \label{eq:psi_real_HRRF}
\end{equation}

According to the thermodynamic parameter evolution equations \eqref{eq:Dxi_DE0_HRRF}, neglecting spatial discretization errors and performing a first-order time integration along the material pathline yields:
\begin{equation}
    \xi_{\mathrm{temp}}^{n+1}=\xi^n, \quad
    E_{0,\mathrm{temp}}^{n+1}
    =E_0^n-\xi^nR_P^n\Delta t.
\label{eq:Phi_temp_HRRF}
\end{equation}
From the affine relation \eqref{eq:rhoe_xi_E0}, the temporary internal energy before re-projection (i.e., the conservative internal energy) is:
\begin{equation}
    (\rho e)_{\mathrm{temp}}^{n+1}
    =
    P^{n+1}\xi_{\mathrm{temp}}^{n+1}
    +E_{0,\mathrm{temp}}^{n+1}
    =
    \psi^n+\xi^n\Delta P
    -\xi^nR_P^n\Delta t.
\label{eq:psi_temp_HRRF}
\end{equation}
Subtracting the Taylor expansion of internal energy \eqref{eq:psi_real_HRRF} and the conservative internal energy \eqref{eq:psi_temp_HRRF}, we obtain the re-projection error:
\begin{equation}
    \epsilon_p=\psi^{n+1}-(\rho e)_{\mathrm{temp}}^{n+1}=\nabla_{\boldsymbol{z}}\psi^n\cdot\Delta\boldsymbol{z}_\sigma + O(\Delta t^2).
\label{eq:epsilon_p_HRRF}
\end{equation}

It is easy to see that the first-order internal energy variation caused by relaxation is exactly canceled out by the source term $-\xi R_P$ of the thermodynamic parameter equation. Therefore, in smooth regions with finite-rate relaxation, the single-step re-projection error of the HRRF method is a second-order small term dominated by the time step. In non-smooth regions, the re-projection error is controlled by first-order dissipation. Such an error property naturally matches that of Godunov-type shock-capturing schemes.

\section{Method Validation}
In this section, we numerically validate the proposed HRRF method with $n$-dodecane as the working fluid. In advection with phase transition, the HRRF method should eliminate numerical pressure oscillations of conservative methods while preserving physical pressure changes; in supercritical advection without phase transition, the method should maintain complete pressure equilibrium. This section first validates these properties by simulating smooth advection tests in a vapor-liquid mixture near the saturation line and in a supercritical state. Subsequently, a series of Riemann problems are solved to compare the evaporation wave speeds with experimental data, further verifying the accuracy of the method.

The equation of state used throughout this section is the P-R EoS, and the internal energy is calculated via analytical integration of the P-R EoS; specific equations are detailed in Appendix \ref{app:PR_EOS}. The thermodynamic parameters of $n$-dodecane are given in Appendix \ref{app:dodecane}. The calculation of fluid fugacity equilibrium and saturation properties based on the P-R EoS can be found in the appendix of \citet{Bai2026FeRP}. For numerical implementation, all tested schemes employ the HLLC Riemann solver \citep{Toro2009}, with first-order accuracy in both time and space. The detailed numerical discretization of the HRRF method is provided in Appendix \ref{app:Discrete}.

\subsection{Mechanical Equilibrium in Phase-Change Advection}

A density wave advection test is designed near the liquid saturation line of $n$-dodecane. The initial flow field is set with a uniform pressure $P_0 = 1.0 \times 10^6 \mathrm{ Pa}$ and a uniform velocity $u = 200 \mathrm{ m/s}$. At this pressure, the saturated liquid density of $n$-dodecane is $\rho_{\text{sat},l} \approx 389 \mathrm{ kg/m}^3$. The computational domain length is $L=1 \mathrm{ m}$, and the initial density is prescribed as a large-amplitude sinusoidal perturbation in a localized region:

\begin{equation}
\rho(x) = 
\begin{cases} 
\rho_0 + \Delta \rho \sin\left( \frac{2\pi (x - 0.1)}{0.4} \right), & x \in [0.1, 0.5] \\ 
\rho_0, & \text{otherwise}
\end{cases}
\label{eq:density_wave}
\end{equation}
where $\rho_0 = 400\mathrm{ kg/m}^3, \Delta \rho = 100\mathrm{ kg/m}^3$. The fluid state extensively traverses the liquid region and the two-phase mixture; traditional conservative schemes exhibit severe numerical pressure oscillations under similar conditions.

\begin{figure}[htbp]
\centering
\begin{subfigure}{0.46\textwidth}
    \includegraphics[width=\linewidth]{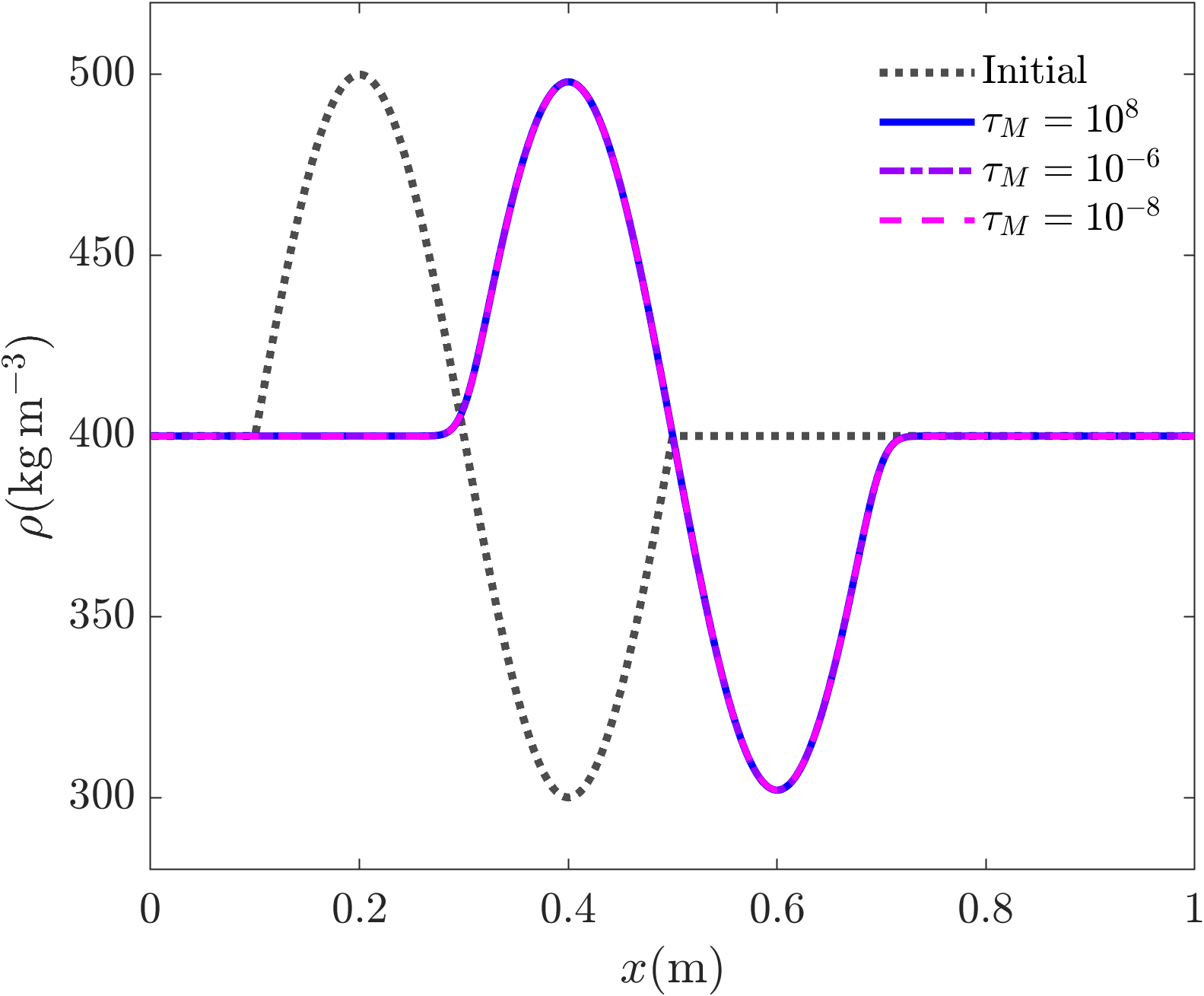}
    \caption{Density}
    \label{fig4a}
\end{subfigure}
\hfill
\begin{subfigure}{0.46\textwidth}
    \includegraphics[width=\linewidth]{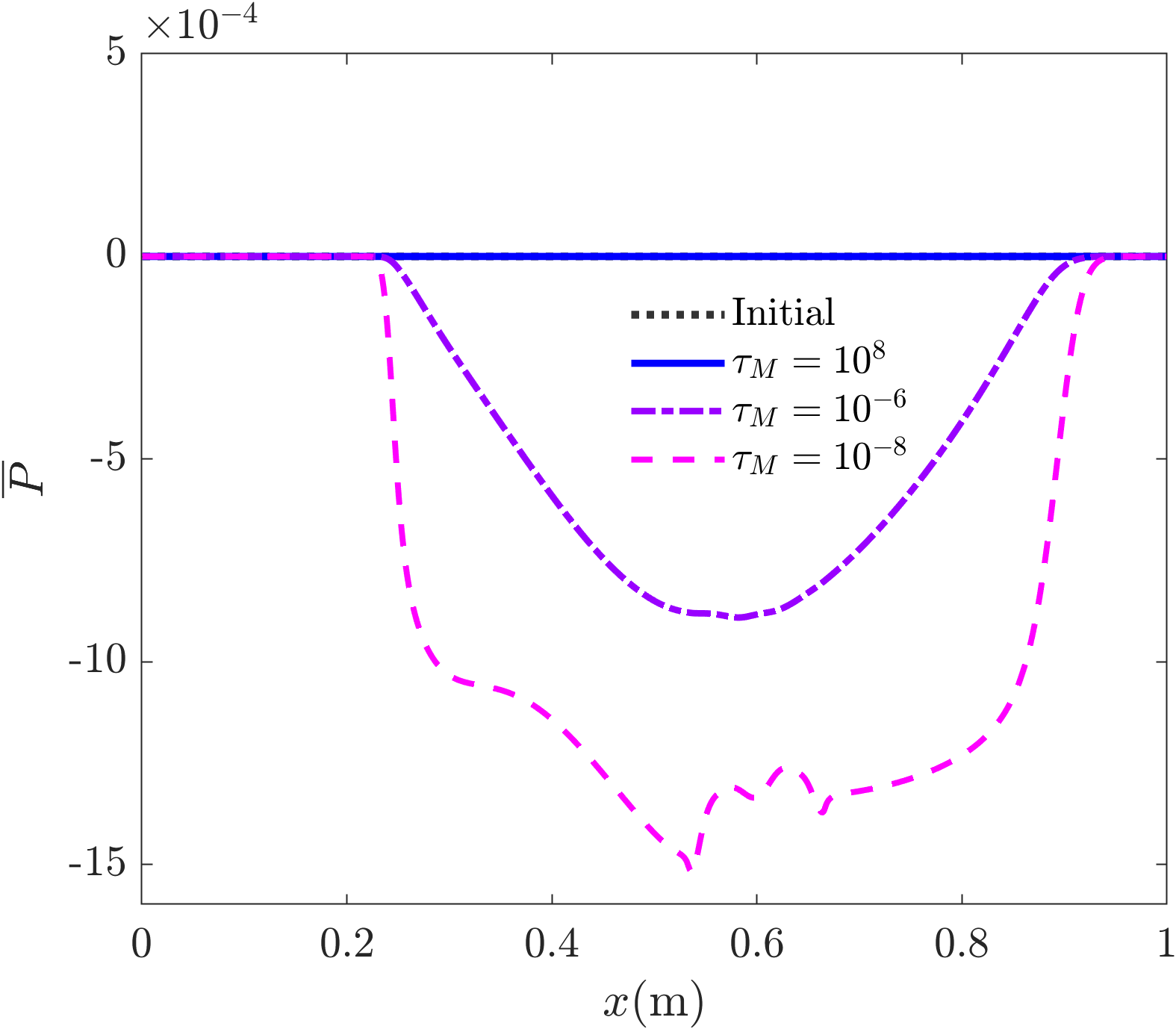}
    \caption{Normalized pressure}
    \label{fig4b}
\end{subfigure}
\caption{Phase change smooth advection test}
\label{fig4}
\end{figure}

Now we set the characteristic mass transfer relaxation times to $\tau_M = 10^8\mathrm{s}, 10^{-6}\mathrm{s}, 10^{-8}\mathrm{s}$, respectively, and employ the HRRF method to compute these cases. Among them, the $\tau_M = 10^8$ condition represents advection under the frozen limit; theoretically, no phase transition occurs between two phases (or it is extremely slow), and the fluid should maintain complete pressure equilibrium, whereas $\tau_M = 10^{-6}, 10^{-8}$ represent finite relaxation at two different rates, where tiny pressure fluctuations arise due to phase transition. In addition, the heat transfer relaxation time is uniformly set to $\tau_H = 10^8$, meaning that its influence is temporarily not considered. The simulation time is set to $t=1 \mathrm{ms}$ and the number of grid cells is $N=1000$. The calculation results are shown in Figure \ref{fig4}.

In Figure \ref{fig4b}, the vertical axis is the normalized pressure $\overline{P}=(P-P_0)/P_0$. After $t=0.001 \mathrm{s}$, the density wave as a whole has moved by $0.2 \mathrm{m}$, and the normalized pressure profile of the $\tau_M = 10^8\mathrm{s}$ configuration adheres strictly to the initial value ($\overline{P}=0$), indicating that the HRRF method completely eliminates numerical pressure oscillations. Meanwhile, the other two finite-rate phase transition configurations exhibit normalized pressure fluctuations on the order of $10^{-4}$, and the amplitude of the fluctuations increases as the relaxation time decreases, which is consistent with basic physical behaviors. This indicates that the HRRF method correctly preserves the physical pressure changes induced by phase transition.

\subsection{Preservation of Mechanical Equilibrium in Supercritical Advection}
\begin{figure}[htbp]
\centering
\begin{subfigure}{0.46\textwidth}
    \includegraphics[width=\linewidth]{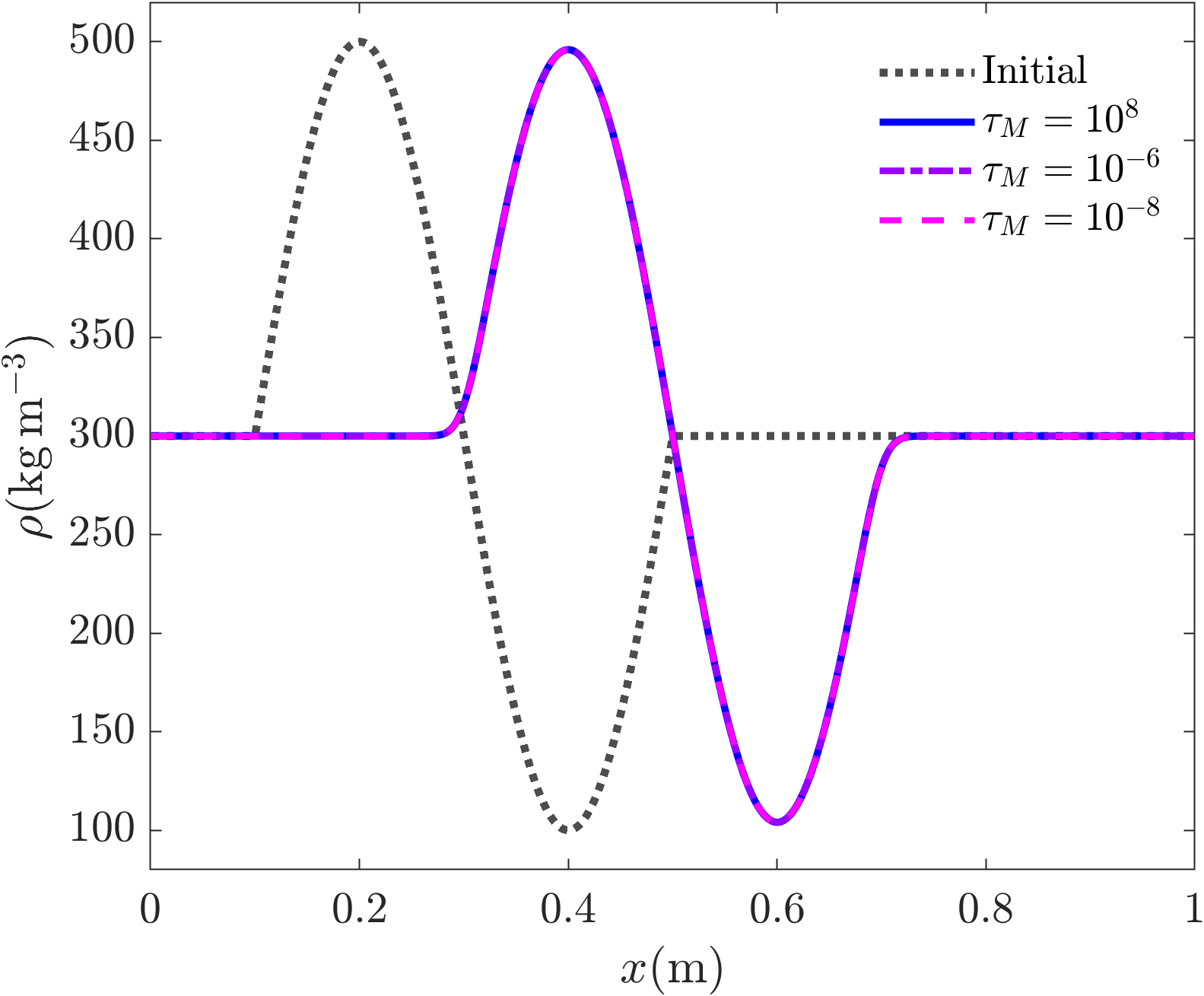}
    \caption{Density}
    \label{fig5a}
\end{subfigure}
\hfill
\begin{subfigure}{0.46\textwidth}
    \includegraphics[width=\linewidth]{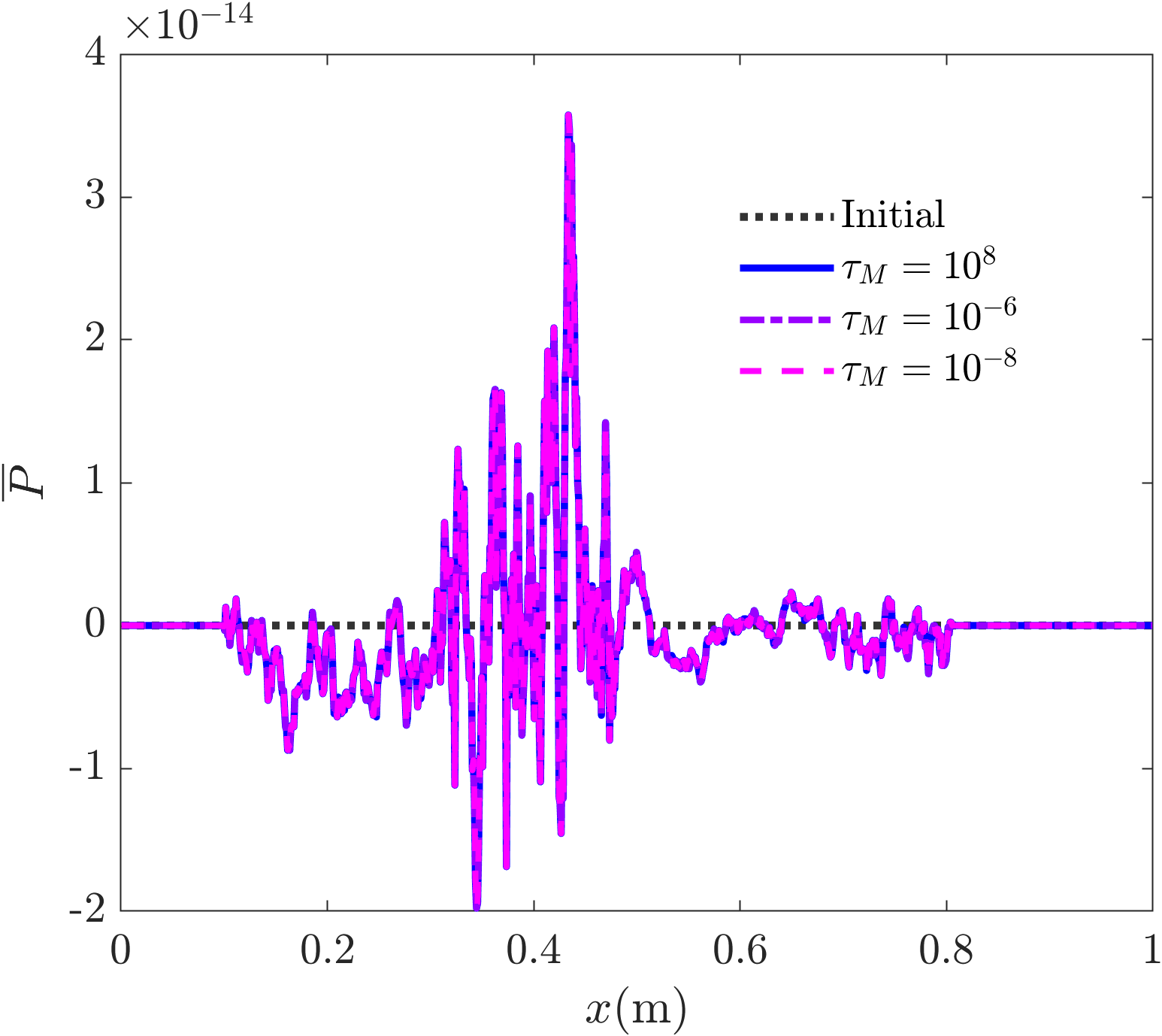}
    \caption{Normalized pressure}
    \label{fig5b}
\end{subfigure}
\caption{Supercritical smooth advection test}
\label{fig5}
\end{figure}

Furthermore, the initial flow field is set with a uniform pressure $P_0 = 2.0 \times 10^6 \mathrm{Pa}$, and the density wave still takes the form of Equation \eqref{eq:density_wave}, but with $\rho_0 = 300\mathrm{ kg/m}^3$ and $\Delta \rho = 200\mathrm{ kg/m}^3$. In this configuration, the density widely crosses the supercritical gas-like and liquid-like regions, with strongly nonlinear variations in thermodynamic properties. The HRRF method is employed to compute this test, and all other settings are identical to those in the phase-change advection configuration. The calculation results are shown in Figure \ref{fig5}.

According to classical thermodynamic theory, supercritical fluids do not undergo phase transition \citep{Landau1980}; therefore, within the framework of the Euler equations, pure supercritical advection should maintain strict pressure equilibrium. The calculation results agree with this theoretical expectation, with the oscillation amplitude of the normalized pressure $\overline{P}$ throughout the entire flow field maintained at the level of machine precision ($10^{-14}$).

\subsection{Validation of Flash Evaporation Wave Speeds Against Experiments}

\citet{SimoesMoreira1999} measured the wave speeds of flash evaporation in superheated $n$-dodecane under a series of conditions, and many multiphase flow methods have adopted these experimental data as a benchmark for numerical testing \citep{Saurel2008,Zein2010,Rodio2015,Pelanti2022}. The computational conditions are listed in Table \ref{tab:test1}.

It should be noted that the left-state conditions of the Riemann problem in this paper are selected directly based on the pre-wave pressures of the evaporation waves measured experimentally by \citet{SimoesMoreira1999}, while the densities are calculated via the P-R EoS through isentropic expansion from the experimental saturation pressures to the pre-wave pressures, which are then used to recalculate the saturation states as the initial conditions. This differs from the selection of initial conditions in the literature mentioned above. The reason for adopting this strategy first stems from the difference in EoS: studies such as \citet{Zein2010} and \citet{Rodio2015} utilized the stiffened gas EoS to compute the reactive Riemann problem and inversely infer the initial pressure, despite a difference from the initial value configured in the experiments. However, for the P-R EoS, we cannot obtain such an initial density by solving the Riemann problem solely from the phase transition onset pressure and initial temperature (and it would also be affected by the relaxation time); for physical clarity, we directly select the pressure upstream of the evaporation wave as the initial pressure. In addition, the experiments of \citet{SimoesMoreira1999} found that when the pressure on the low-pressure side is lower than $0.3 \mathrm{bar}$, the evaporation wave speed varies insignificantly with the outlet pressure; we therefore set the right-side pressure to $0.1 \mathrm{bar}$. The computational domain length is set to $L= 1 \mathrm{m}$, with the initial discontinuity at $x= 0.5\mathrm{m}$, the simulation time is $t=0.5 \mathrm{ms}$, and the number of grid cells is $N=1000$.

\begin{table}[htbp]
    \centering
    \caption{Initial Values for the Flash Evaporation Test}
    \label{tab:test1}
    \begin{tabular}{cccc} 
         \toprule
        State & $P$ (bar) & $\rho$ ($\mathrm{kg \,m^{-3}}$) & $u$ ($\mathrm{m \, s^{-1}}$) \\
        \midrule
        Left & \makecell{0.24, 0.33, 0.44, 0.59, \\ 0.83, 1.19, 1.91, 2.12} 
            & \makecell{582.11, 566.72, 553.15, 540.21,\\ 519.63, 496.03, 468.64, 452.76} 
             & 0 \\
        Right & $0.1$ & \makecell{0.4553, 0.4357, 0.4212, 0.4093, \\ 0.3934, 0.3787,    0.3651, 0.3587}  & 0 \\
        \bottomrule
    \end{tabular}\par
\end{table}

\begin{figure}[htbp]
\centering
\begin{subfigure}{0.48\textwidth}
    \includegraphics[width=\linewidth]{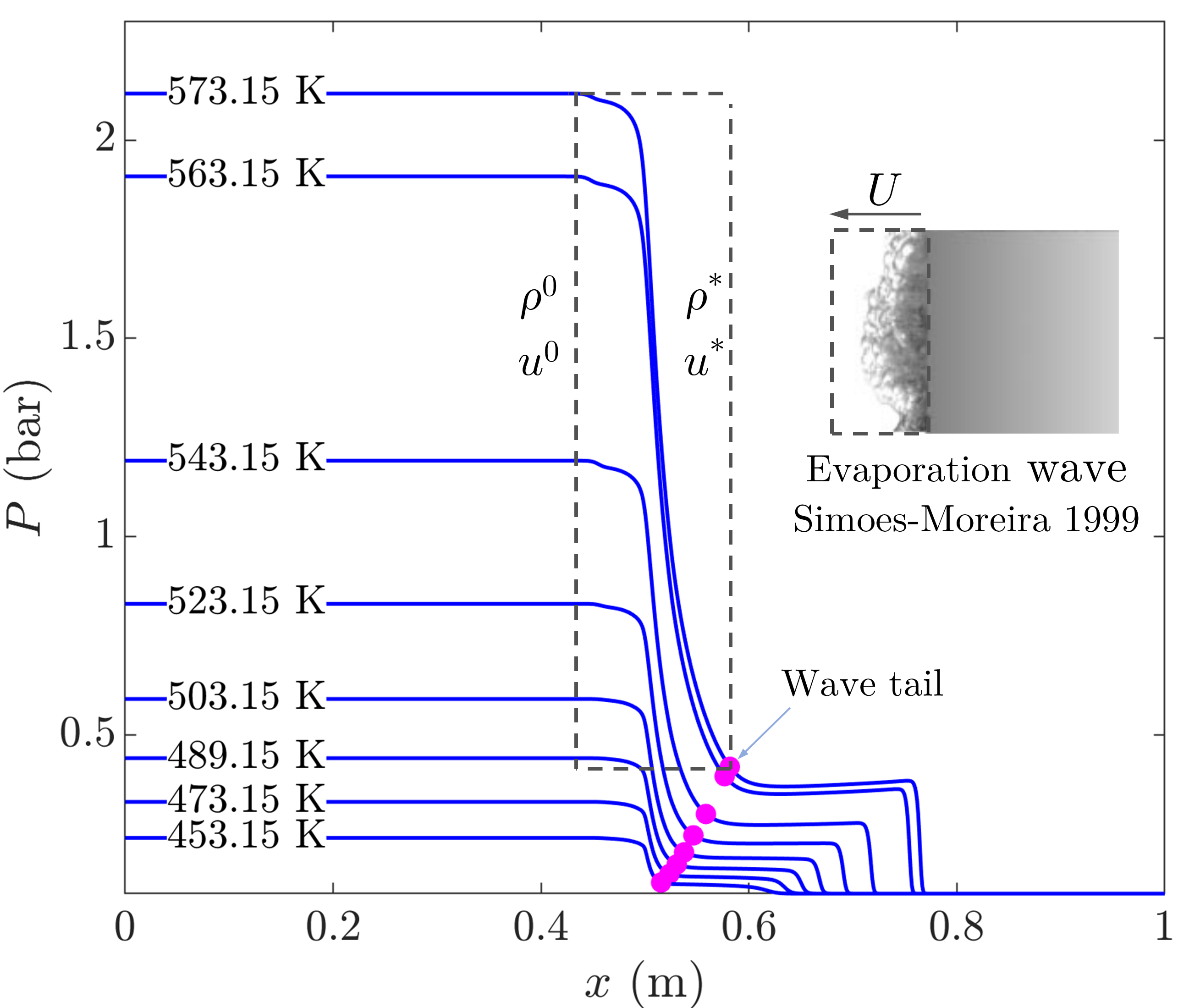}
    \caption{Pressure ($\tau_M = 10^{-8} \mathrm{s}$)}
    \label{fig6a}
\end{subfigure}
\hfill
\begin{subfigure}{0.48\textwidth}
    \includegraphics[width=\linewidth]{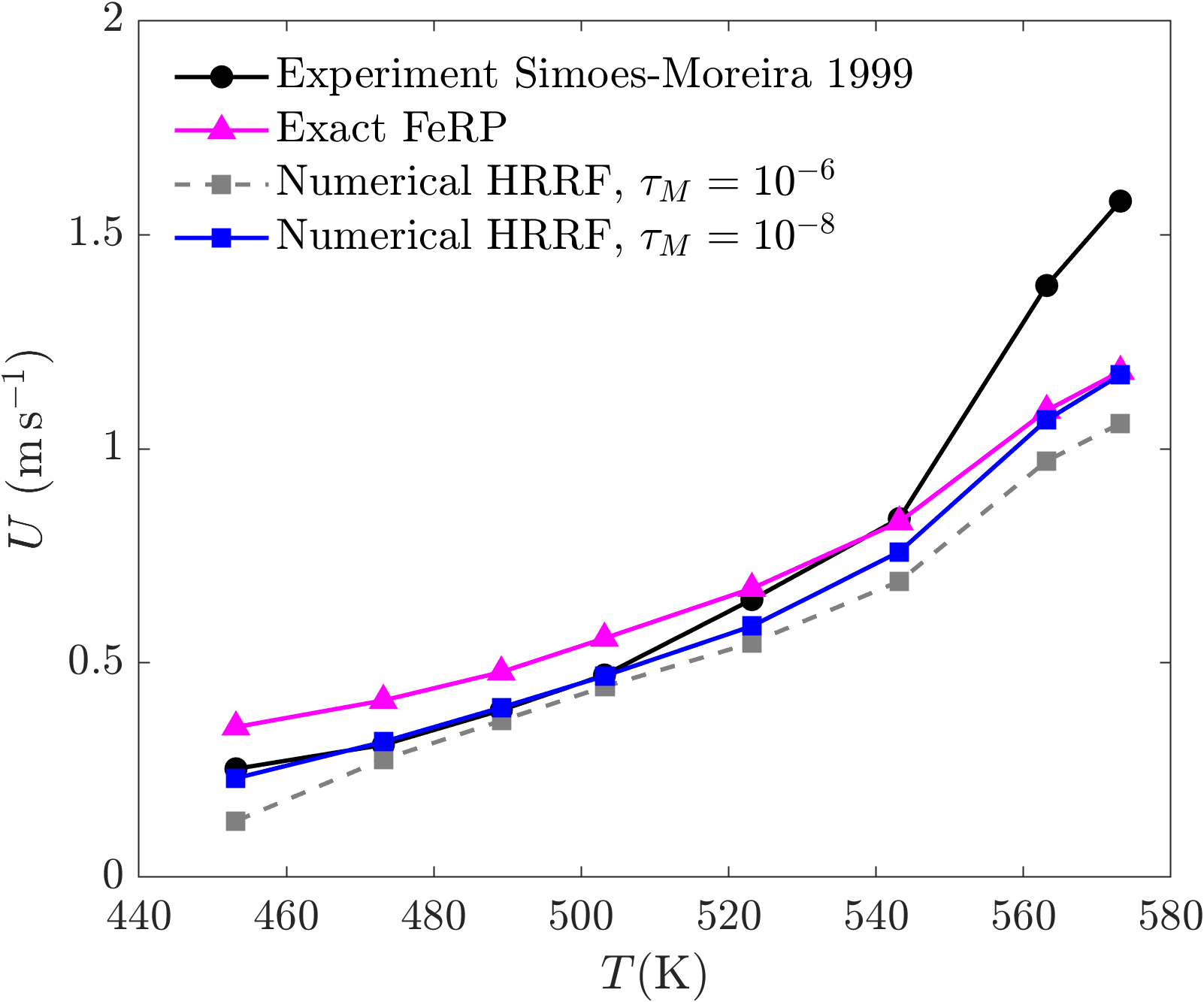}
    \caption{Wave speed}
    \label{fig6b}
\end{subfigure}
\caption{Flash evaporation wave test}
\label{fig6}
\end{figure}

Figure \ref{fig6} shows the calculation results of the HRRF method with mass transfer relaxation times $\tau_M = 10^{-8}\mathrm{s}, 10^{-6}\mathrm{s}$ and heat transfer relaxation time $\tau_T = 10\tau_M$, along with the wave speed calculation results of the exact FeRP solution. Here, the wave speed is determined according to the following relation:
\begin{equation}
    U = \frac{\rho^* u^* - \rho^0 u^0}{\rho^* - \rho^0}
\end{equation}
where $\rho^0, u^0$ denote the pre-wave state. For $\rho^*, u^*$, we take the state at the tail of the rarefaction wave: for the exact solution, it corresponds to the left intermediate state of the Riemann solution; for the HRRF results, we select the first grid point that simultaneously reaches $90\%$ of both the intermediate pressure and the maximum velocity.

The calculation results show that as the relaxation time decreases (the phase transition rate accelerates), the evaporation wave speed gradually increases and progressively approaches the exact solution. The overall variation trend of the wave speed is consistent with the experimental results, although the wave speed predicted by the current method tends to be lower as the degree of superheat increases. Preliminary analysis indicates that this deviation is attributed to the setting of initial conditions: although we make the initial pressure and density theoretically consistent with the experimental pre-wave state, in order to compromise with the assumption in our model that metastable states cannot be maintained for a long time and the assumption in the exact FeRP solution that no metastable state exists, we set the initial state to be saturated. This prevents the numerical wave speed from accelerating as the initial superheat increases. Nevertheless, the current method agrees well with the experimental wave speeds under most configurations. The above numerical validations demonstrate that the HRRF method reliably maintains pressure equilibrium and effectively captures flash evaporation phenomena.

\section{Mechanisms of Mass and Heat Transfer Relaxation on the FeRP Solution}

In the FeRP framework \citep{Bai2026FeRP}, three wave structures of flash evaporation were formulated: the (split) rarefaction wave (R), the rarefaction--expansion shock composite wave (RS), and the rarefaction--expansion shock--rarefaction composite wave (RSR). Among them, the expansion shock is a non-classical wave induced by a $\delta$-type singularity of the Landau fundamental derivative $\Gamma$. Meanwhile, the Landau fundamental derivative also represents the curvature of the isentrope; dynamically, the expansion shock results from the positive jump in the speed of sound as the isentrope enters the vapor region from the two-phase region.

However, thus far, experiments have observed various non-classical waves in the $\Gamma < 0$ region of BZT fluids and during phase transitions in molecularly complex fluids \citep{Thompson1972, Thompson1986, Borisov1983}, whereas expansion shocks have never been observed during the phase transition of dodecane, despite being theoretically predicted. This prompts us to ask: in physical reality, do theoretically predicted non-classical expansion shocks caused by phase transition always exist? As the relaxation time increases, does this expansion shock tend to weaken, shift, or disappear? What influence does finite thermodynamic relaxation have on the characteristic quantities of the Flash evaporation Riemann solution, such as pressure and velocity? 

Previously, we attempted to answer these questions by introducing a two-phase non-equilibrium speed of sound into the exact solution of the FeRP (Section 4 of \citealt{Bai2026FeRP}), and drew a series of qualitative conclusions: as thermodynamic non-equilibrium increases, the speed of sound jump decreases, the expansion shock weakens, and the intermediate pressure and velocity of the Riemann solution decrease. However, we also pointed out that this artificial substitution is thermodynamically inconsistent and can only serve as a preliminary understanding; to formally answer these questions, the complete real-fluid thermodynamic relaxation system must be solved. Now, equipped with the HRRF method, we solve a series of FeRPs to comprehensively elucidate the mechanisms by which finite thermodynamic relaxation affects the Riemann solution.

\subsection{Degeneration of the Non-classical Expansion Shock}

\label{subsec:expansion_shock}
To investigate the non-classical expansion shock in relaxation systems, we select a condition where $n$-dodecane undergoes complete flash evaporation; the computational settings are listed in Table \ref{tab:search1}. The computational domain is $L= 0.1 \mathrm{m}$ with the initial discontinuity located at $x_0= 0.04\mathrm{m}$, discretized by $N=1000$ grid cells with a total physical time of $t=2 \mathrm{ms}$.

\begin{table}
    \centering
    \caption{Initial Values For the Expansion Shock Test of FeRP}
    \label{tab:search1}
    \begin{tabular}{ccccc} 
         \toprule
        State & $P$ ($\mathrm{MPa}$) & $\rho$ ($\mathrm{kg \,m^{-3}}$) & $T$ ($\mathrm{K}$) & $u$ ($\mathrm{m \, s^{-1}}$) \\
        \midrule
        Left & 2.0  & 300 & 652.4 & 0 \\
        Right & 0.1 & 2 & 1026.8& 0\\
        \bottomrule
    \end{tabular}\par
\end{table}

In our model, multiple time parameters $\tau_{M,e}, \tau_{M,c}, \tau_H$ are involved, where the mass-transfer parameters $\tau_{M,e}$ and $\tau_{M,c}$ are determined locally by Equations \eqref{eq:evaporation_time} and \eqref{eq:condensation_time}, thus varying from cell to cell. To avoid introducing excessive timescales that would break self-similarity, in this subsection we uniformly set $\tau = \tau_{M,e}=\tau_{M,c}=0.1\tau_H$. Four configurations with $\tau = 10^{-10}, 10^{-9}, 10^{-8}, 10^{-7}\mathrm{s}$ are simulated, while the equilibrium FeRP is also computed with the RFQC method under identical discretization and computational settings for comparison. 

\begin{figure}[ht]
\centering
\begin{subfigure}[b]{0.48\textwidth}
    \includegraphics[width=\textwidth]{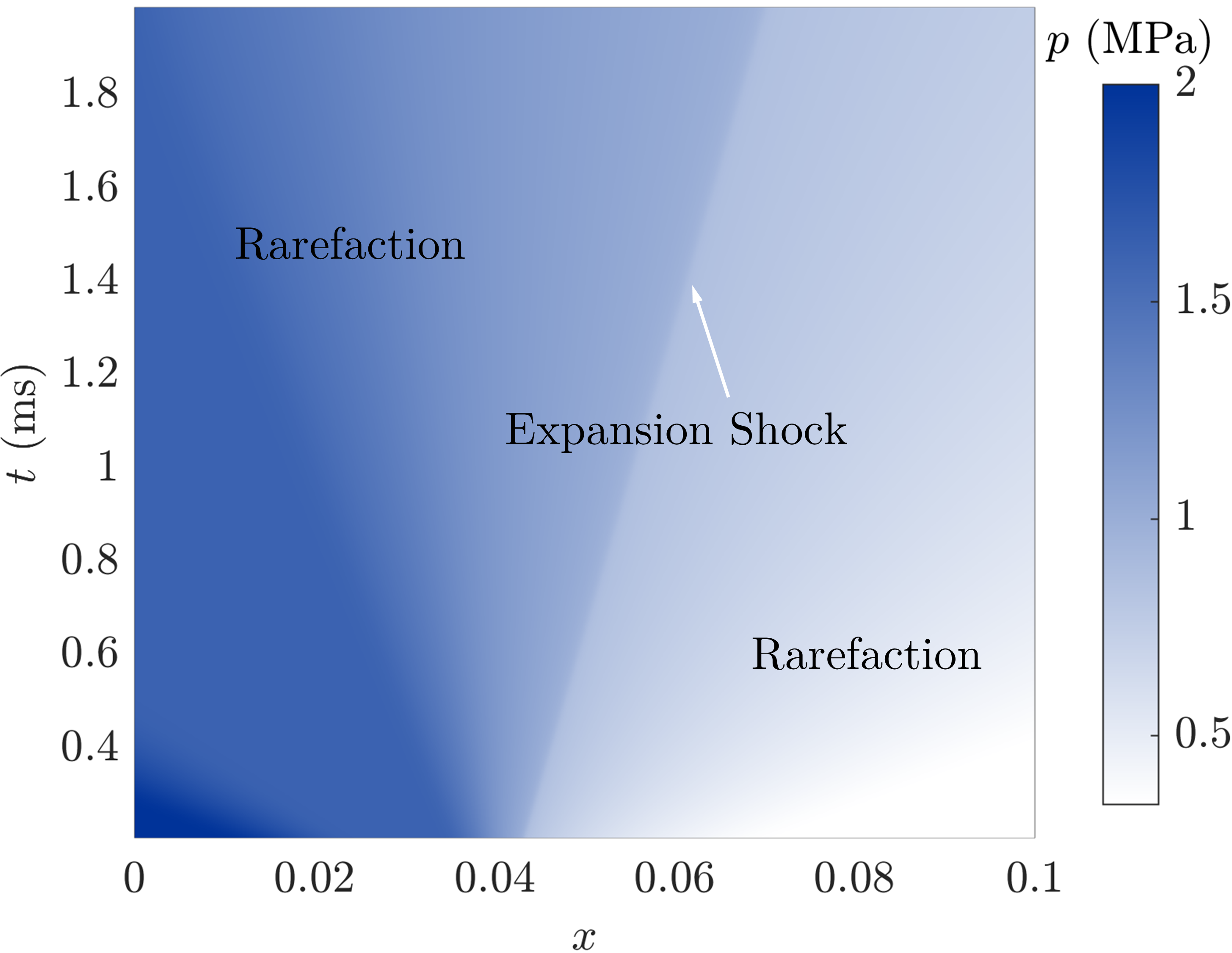}
    \caption{$\tau =10^{-10}\mathrm{s} $}
    \label{fig7a}
\end{subfigure}
\hfill 
\begin{subfigure}[b]{0.48\textwidth}
    \includegraphics[width=\textwidth]{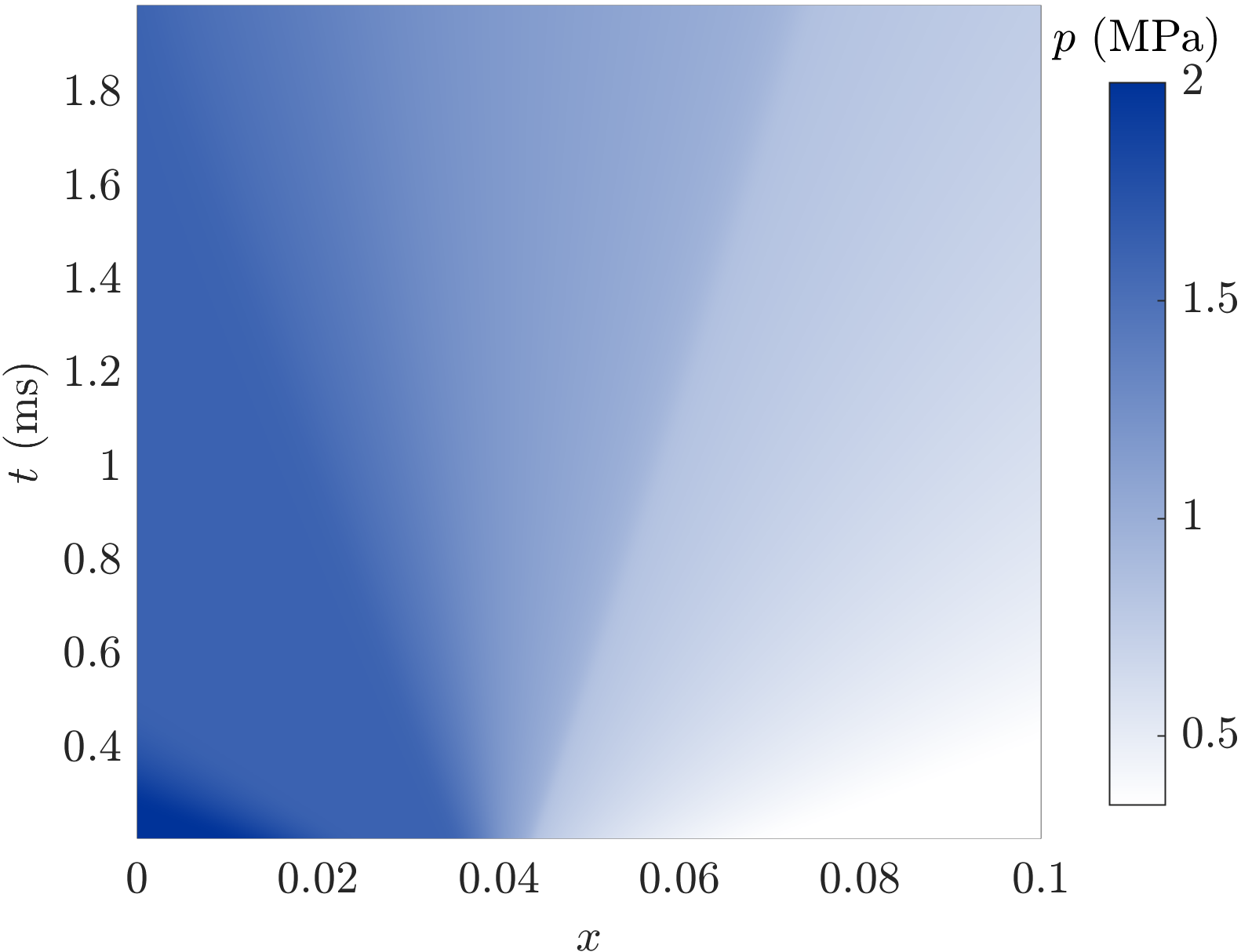}
    \caption{$\tau =10^{-9}\mathrm{s}$}
    \label{fig7b}
\end{subfigure}
\begin{subfigure}[b]{0.48\textwidth}
    \includegraphics[width=\textwidth]{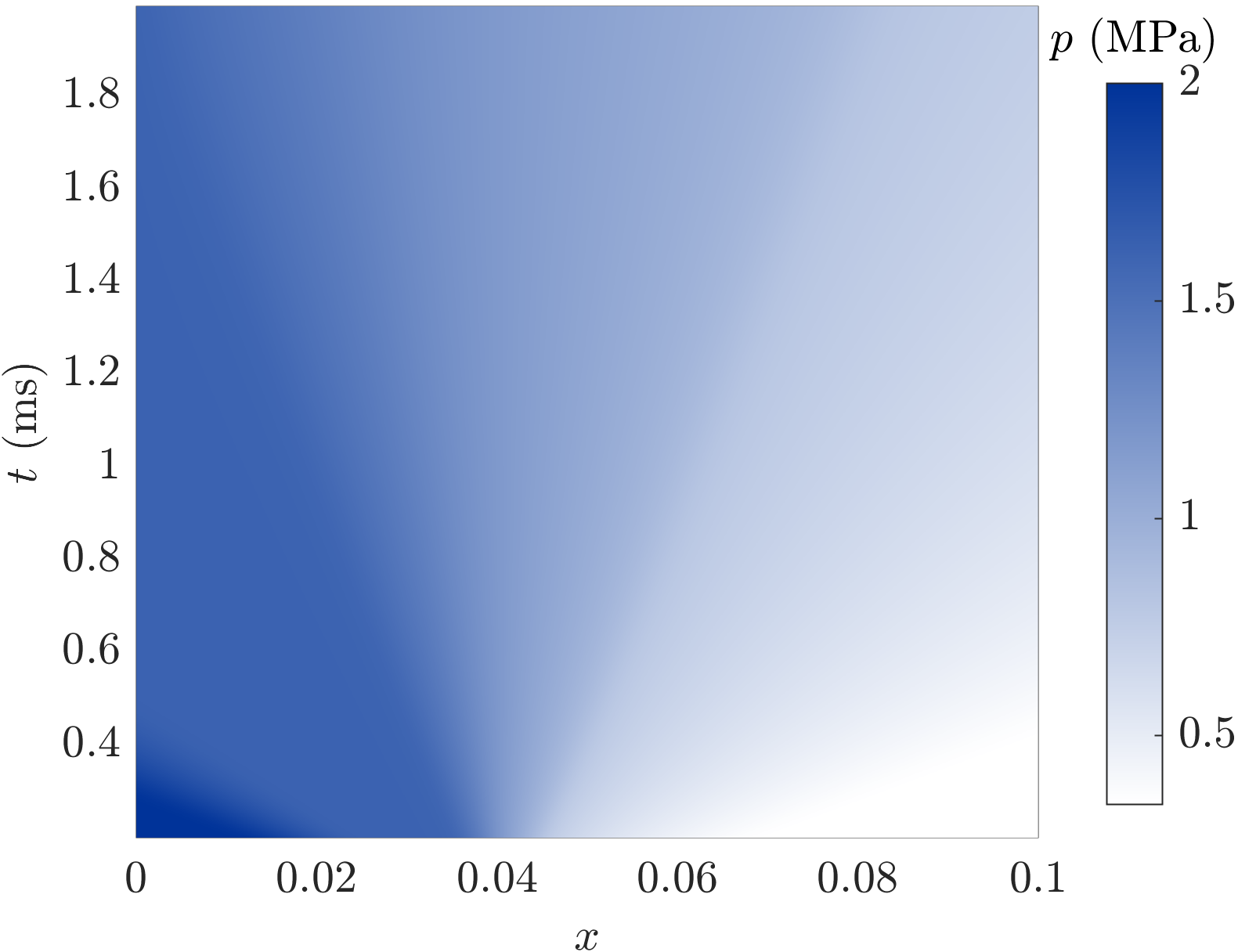}
    \caption{$\tau =10^{-8}\mathrm{s}$}
    \label{fig7c}
\end{subfigure}
\hfill 
\begin{subfigure}[b]{0.48\textwidth}
    \includegraphics[width=\textwidth]{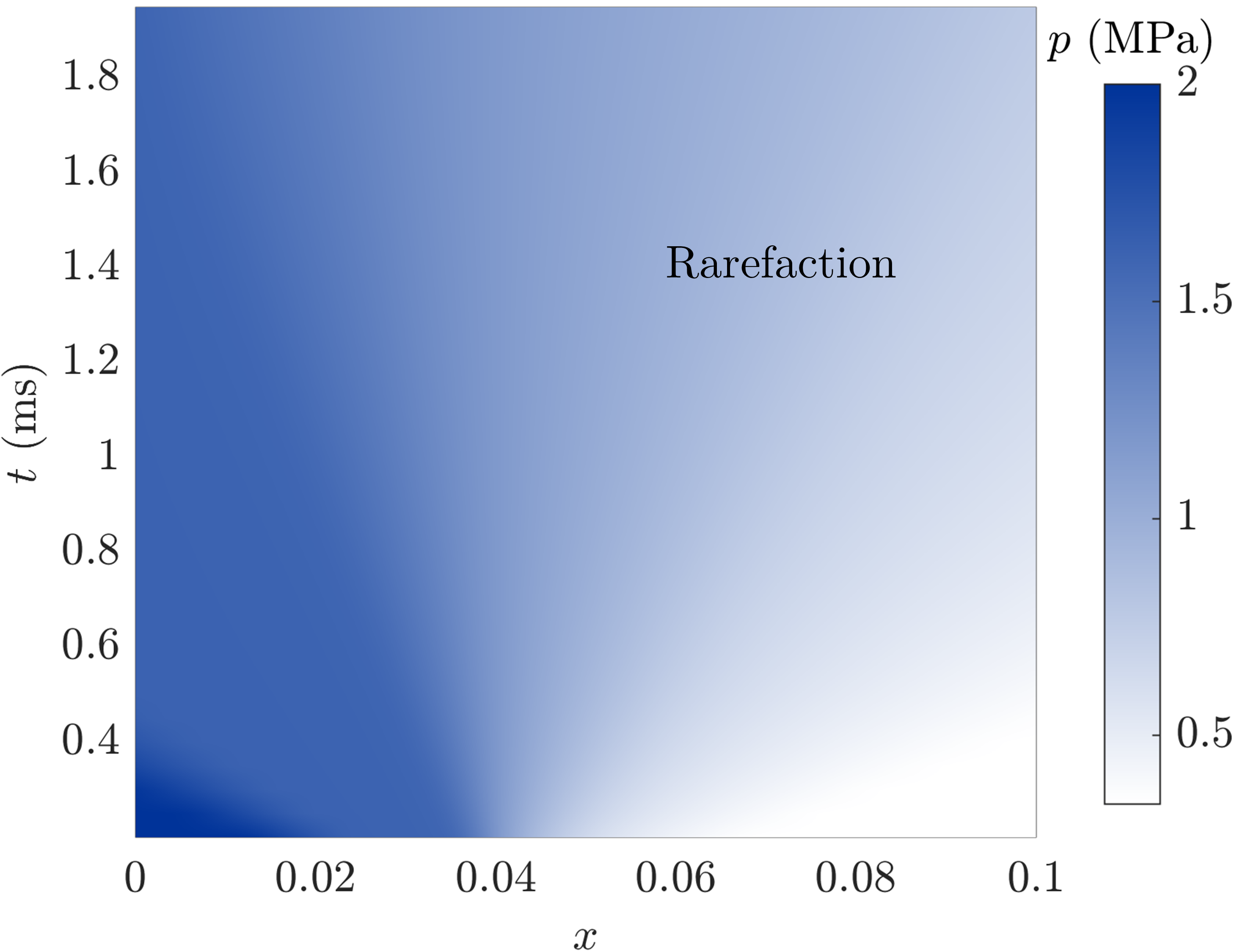}
    \caption{$\tau =10^{-7}\mathrm{s}$}
    \label{fig7d}
\end{subfigure}
\caption{Pressure contours of FeRP on the $x-t$ plane.}
\label{fig7}
\end{figure}

Figure \ref{fig7} shows the pressure contours on the $x-t$ plane, where an RSR composite wave is expected in the equilibrium limit. In the configuration with the shortest relaxation time ($\tau =10^{-10}\mathrm{s}$, Figure \ref{fig7a}), a distinct discontinuity exists between the two rarefaction waves. As the relaxation time increases ($\tau =10^{-9},10^{-8}\mathrm{s}$, Figures \ref{fig7b} and \ref{fig7c}), this discontinuity becomes progressively blurred. When the relaxation time reaches $\tau =10^{-7}\mathrm{s}$ (Figure \ref{fig7d}), hardly any distinct discontinuity can be discerned in the flow field.

\begin{figure}[ht]
\centering
\begin{subfigure}[b]{0.48\textwidth}
    \includegraphics[width=\textwidth]{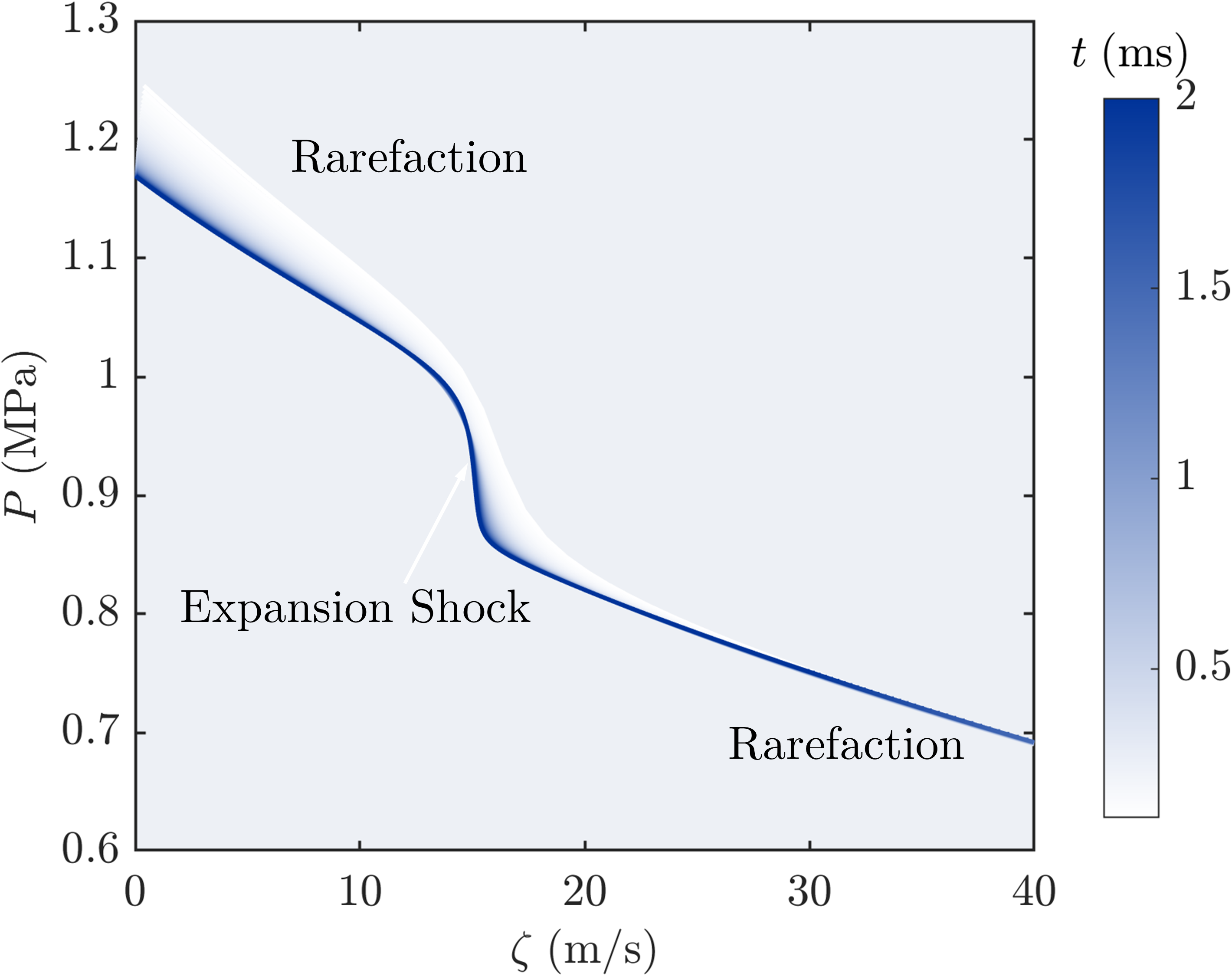}
    \caption{$\tau =10^{-10}\mathrm{s} $}
    \label{fig8a}
\end{subfigure}
\hfill 
\begin{subfigure}[b]{0.48\textwidth}
    \includegraphics[width=\textwidth]{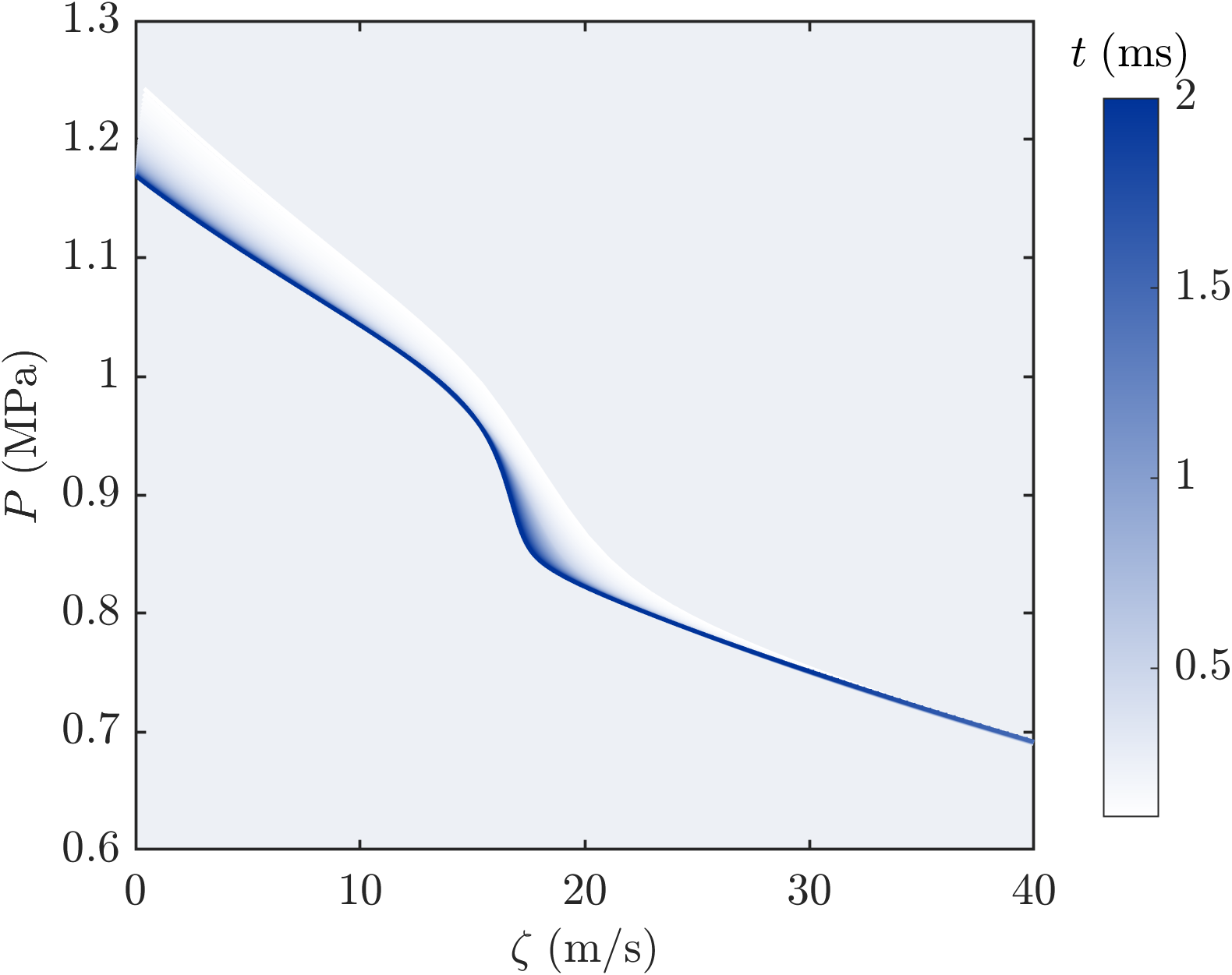}
    \caption{$\tau =10^{-9}\mathrm{s}$}
    \label{fig8b}
\end{subfigure}

\begin{subfigure}[b]{0.48\textwidth}
    \includegraphics[width=\textwidth]{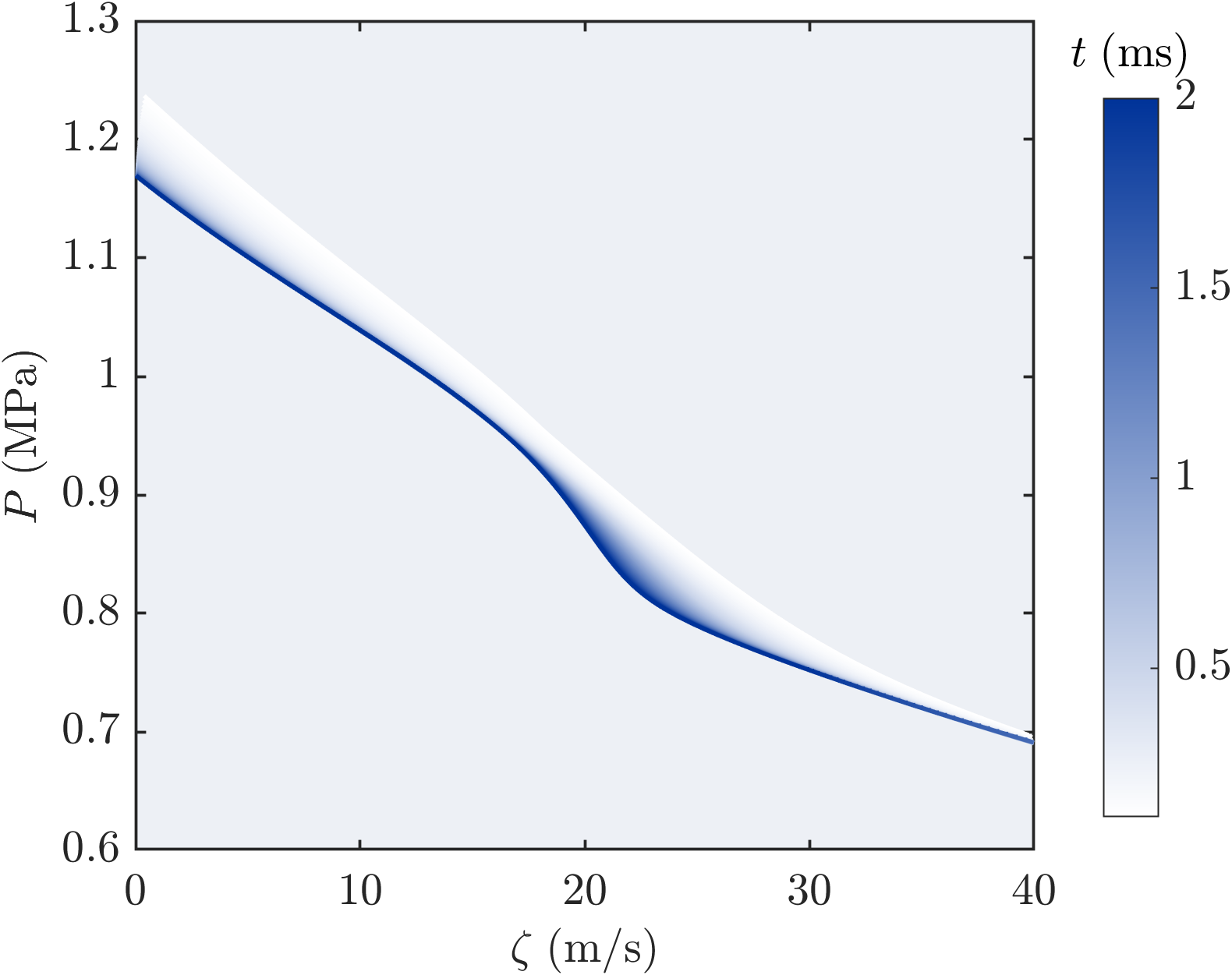}
    \caption{$\tau =10^{-8}\mathrm{s}$}
    \label{fig8c}
\end{subfigure}
\hfill
\begin{subfigure}[b]{0.48\textwidth}
    \includegraphics[width=\textwidth]{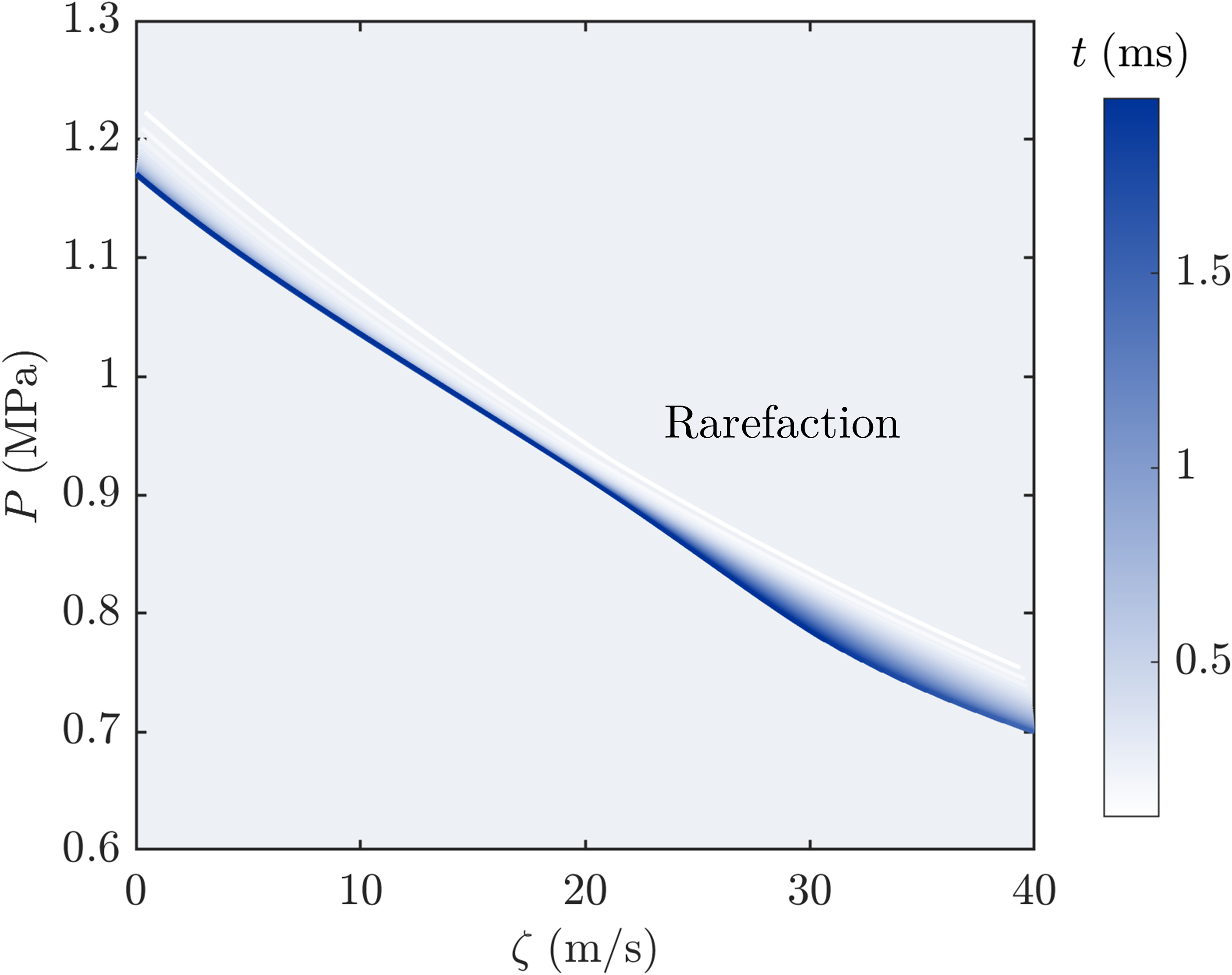}
    \caption{$\tau =10^{-7}\mathrm{s}$}
    \label{fig8d}
\end{subfigure}
\caption{Pressure curves in self-similar coordinates.}
\label{fig8}
\end{figure}

To further distinguish such ``blurred'' discontinuities in Figures \ref{fig7b} and \ref{fig7c}, especially to distinguish whether it is a traveling wave or a rarefaction wave with a relatively large gradient, we plot the pressure distributions at different times against the self-similar parameter $\zeta$, as shown in Figure \ref{fig8}, where $\zeta$ is defined as:
\begin{equation}
    \zeta = \frac{x-x_0}{t}
\end{equation}

In physical coordinates, a rarefaction wave manifests as an expansion fan that continuously spreads over time, whereas in self-similar coordinates, the shape of a self-similar rarefaction wave remains unchanged. In contrast, a traveling wave with a fixed thickness in physical coordinates converges toward a discontinuity in self-similar coordinates.

In Figure \ref{fig8}, under the condition of $\tau = 10^{-10}\mathrm{s}$, we can clearly see that the pressure curve in the self-similar coordinate system continuously approaches a discontinuity as time increases. This indicates that there exists a traveling wave in physical space with nearly constant width over time; therefore, we can conclude that under the extremely small relaxation time of $\tau = 10^{-10}\mathrm{s}$, the expansion shock generated by phase transition still exists. However, in the configurations with $\tau = 10^{-9},10^{-8}\mathrm{s}$, the pressure curves also exhibit a steepening trend, but even though the simulation time is already much longer than the relaxation time, the pressure curves neither converge to a discontinuity nor flatten out. This indicates that the expansion shock is degenerating, but has not yet completely become a self-similar rarefaction wave; its wave width grows over time, but the growth rate is smaller than the spreading rate of a self-similar wave. In the configuration with $\tau = 10^{-7}\mathrm{s}$, the shape of the pressure curve hardly changes with time, indicating that the expansion branch in this regime is already very close to a self-similar rarefaction wave, and the expansion shock has almost completely degenerated.

To quantitatively verify this conclusion, we introduce a traveling-wave characteristic function $G$. For the pressure $P(\zeta, t)$ in self-similar coordinates, $G$ is defined as its maximum gradient:
\begin{equation}
    G(t) = \text{max}\left(-\pderiv{P}{\zeta}{t} \right) = \text{max}\left(-t\pderiv{P}{x}{t} \right) 
\end{equation}
For a stable wave, it can be approximated as:
\begin{equation}
     \pderiv{P}{x}{t} \approx \frac{\Delta P}{\delta_x} 
\end{equation}
where $\delta_x$ is the physical thickness of the wave and $\Delta P$ is the pressure increment across the wave. Therefore,
\begin{equation}
    G(t) = \text{max}\left(-t\pderiv{P}{x}{t} \right) \approx  \frac{t\Delta P}{\delta_x} 
\end{equation}
For a self-similar rarefaction wave, its physical width grows approximately linearly with time ($\delta_x \propto t$), so $G$ should satisfy:
\begin{equation}
    G(t) \propto t^0 
\end{equation}
For a traveling wave with a time-invariant physical width, $G$ should satisfy:
\begin{equation}
    G(t) \propto t 
\end{equation}
And if the width of the wave grows with time, but at a rate lower than the linear growth rate of a rarefaction wave, $G$ satisfies sublinear growth with a characteristic exponent $C$:
\begin{equation}
    G(t) \propto t^C,\quad 0<C<1 
\end{equation}

\begin{figure}[htbp]
\centering
    \includegraphics[width=0.5\textwidth]{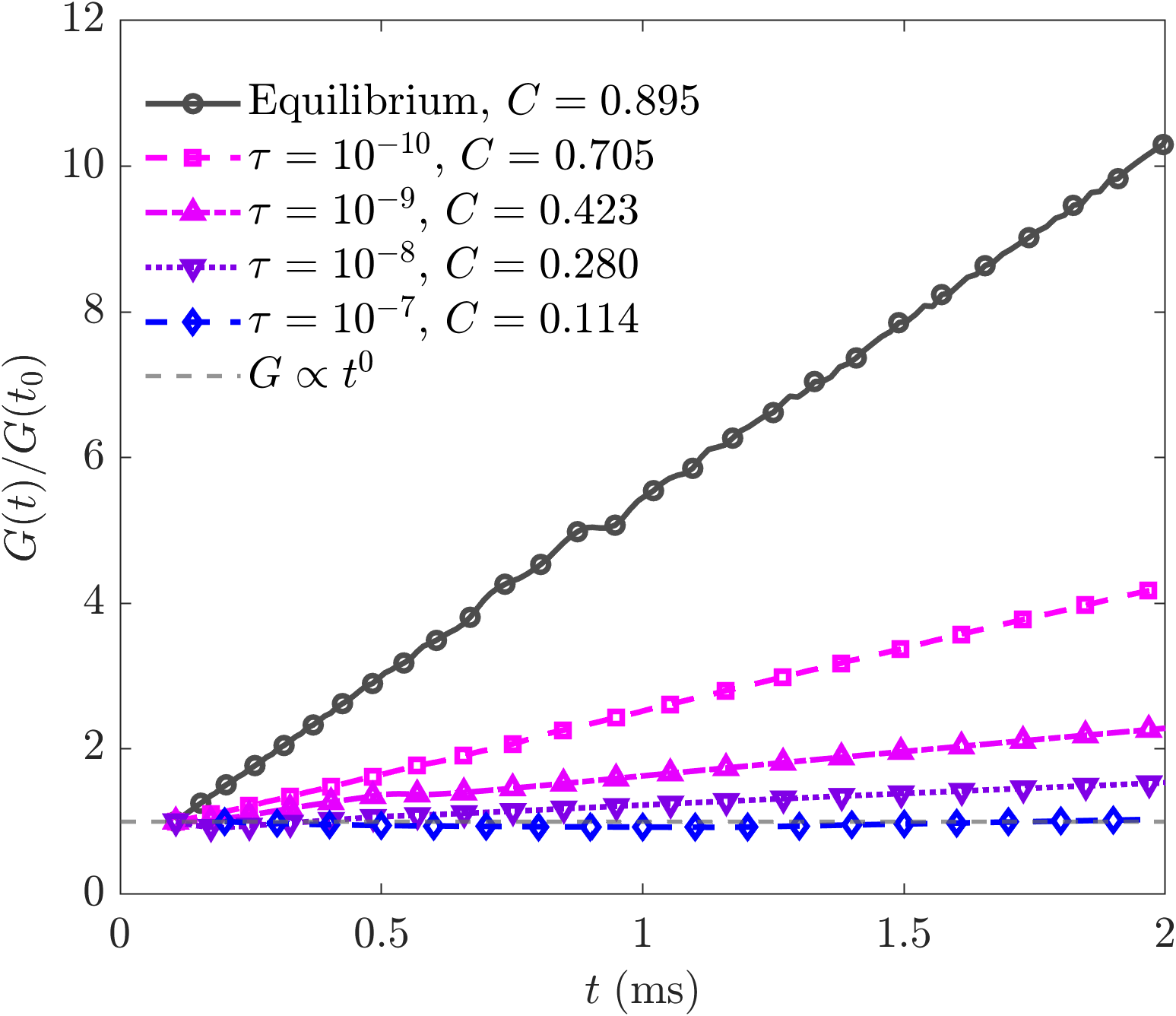}
    \caption{Wave characteristic function  curve}
\label{fig9}
\end{figure}

Following this scaling analysis, we plot the variation curves of the traveling wave characteristic function $G$ with time for the above cases, and determine the characteristic exponent $C$ via logarithmic fitting, as shown in Figure \ref{fig9}. For comparison, the equilibrium result computed by the RFQC method serves as the standard reference for the expansion shock. Note that we do not directly adopt the exact solution as the reference here, because the calculation results of the RFQC method have numerical dissipation similar to that of the HRRF method under identical numerical discretizations; adopting it as the reference can largely avoid attributing dissipation-induced broadening to relaxation effects. The RFQC method agrees well with the exact solution in the current test case; see \citet{Bai2026RFQC1} for details.

\begin{figure}[htbp]
\centering
\begin{subfigure}{0.48\textwidth}
    \includegraphics[width=\linewidth]{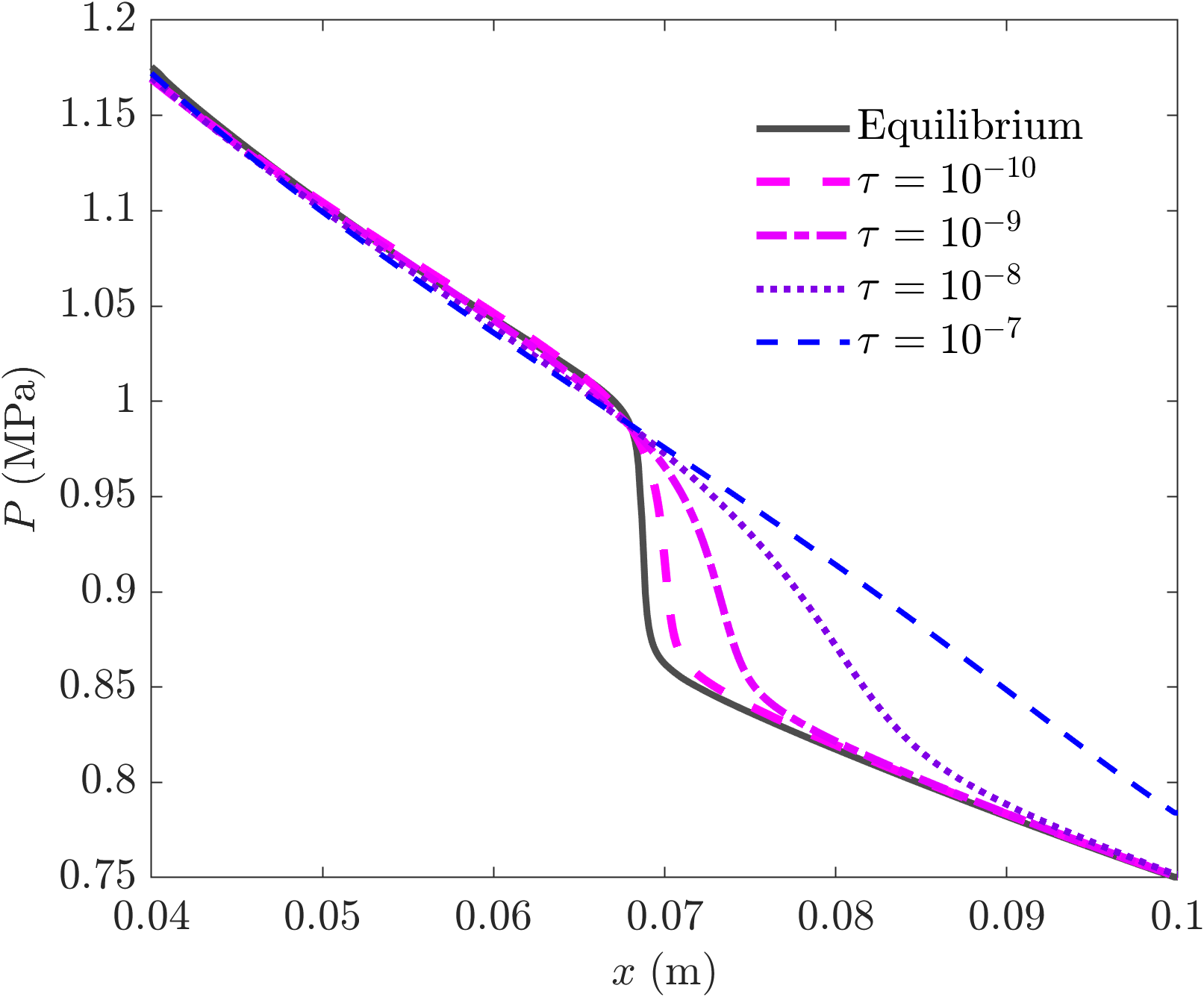}
    \caption{Pressure curve}
    \label{fig10a}
\end{subfigure}
\hfill
\begin{subfigure}{0.48\textwidth}
    \includegraphics[width=\linewidth]{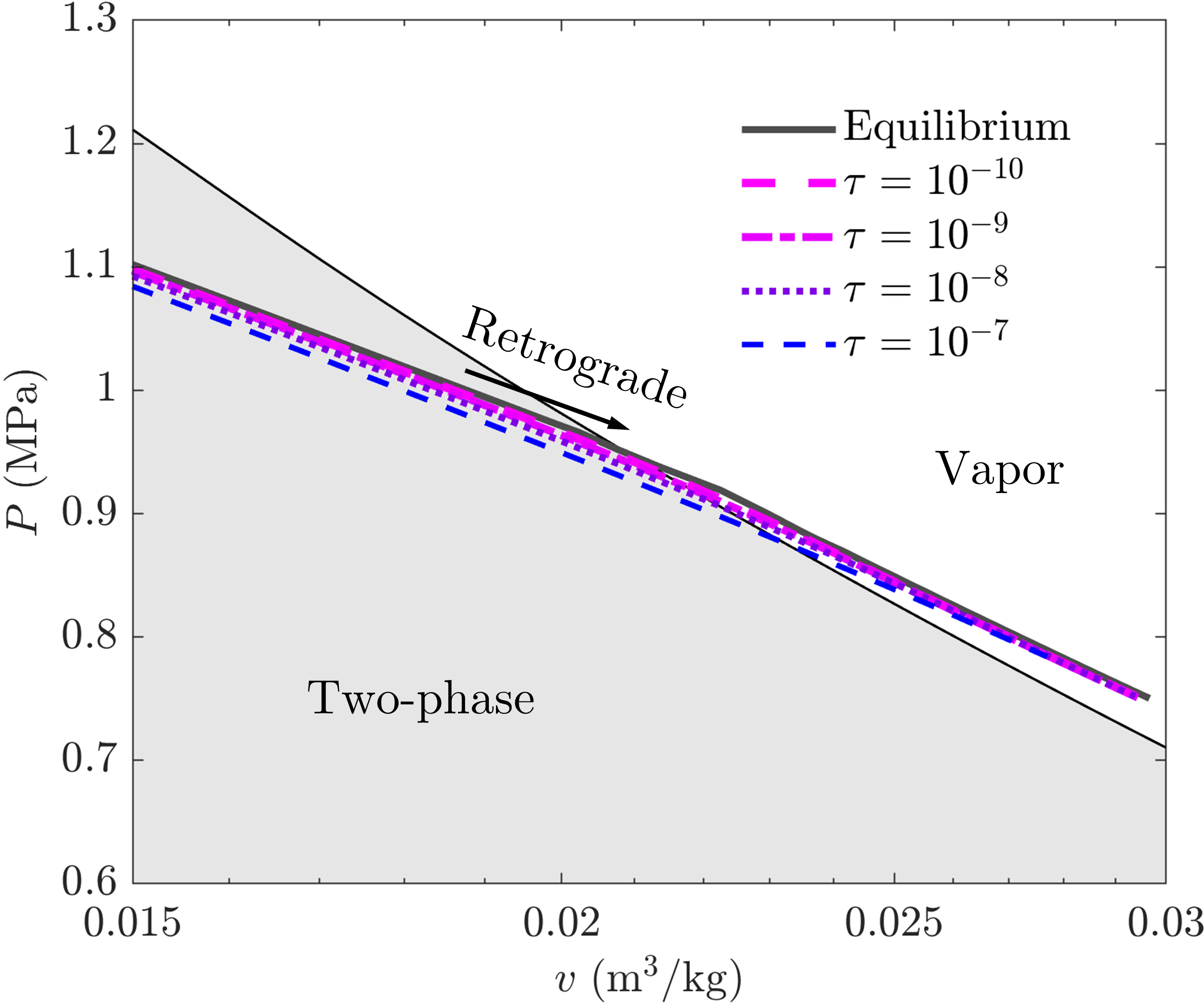}
    \caption{State trajectory on the $P-v$ plane}
    \label{fig10b}
\end{subfigure}
\caption{Calculation results for different $\tau$ at $t=0.2\mathrm{ms}$}
\label{fig10}
\end{figure}

Figure \ref{fig9} shows that in the complete equilibrium state, the expansion shock has a characteristic exponent of $C=0.895$, which is slightly lower than the theoretical value of 1 due to numerical dissipation. The wave characteristic function curves of the HRM solutions show that as the order of magnitude of the relaxation time increases, the characteristic exponent decreases from near 1 toward 0; at $\tau = 10^{-10}\mathrm{s}$, the characteristic exponent is $C=0.705$, moderately lower than $0.895$ of the equilibrium state. Therefore, dynamically speaking, the expansion shock already exhibits certain degeneration characteristics in this configuration, and the shock width slowly increases over time. However, in terms of the wave structure in physical space, the HRM solution under this configuration is very close to the HEM solution (Figure \ref{fig10a}); hence, we consider that it remains an expansion shock. At $\tau = 10^{-7}\mathrm{s}$, the characteristic exponent is only 0.114, which is already close to zero, indicating that the expansion shock has almost completely degenerated. Meanwhile, as can be seen from Figure \ref{fig10a} and Figure \ref{fig7}, the increase in width of the expansion shock during the degeneration process is directional: its wave head remains relatively fixed, while its tail continuously shifts downstream. This is because the differences induced by phase transition relaxation manifest only after retrogradely crossing the saturation line (reaching the end of phase transition); prior to this, the trajectories on the phase diagram for different relaxation times (Figure \ref{fig10b}) almost overlap, and the phase transition end points are very close, all crossing the saturation line from the two-phase region into the vapor phase.

From the above findings, we summarize the behavior of the expansion shock with varying relaxation time: as the relaxation time increases, the expansion shock degenerates toward a self-similar rarefaction wave. Under an extremely small relaxation time ($O(10^{-10})$), a very strong expansion shock feature is still retained within the expansion branch of the Riemann solution; as the relaxation time increases, the expansion shock degenerates, and its width grows continuously downstream over time, but at a growth rate lower than that of a self-similar rarefaction wave---this stage can be regarded as a non-self-similarly developing expansion wave. When the relaxation time is relatively long ($O(10^{-7}\mathrm{s})$), the expansion shock can barely be discerned and degenerates into a self-similar rarefaction wave.

In addition, for real $n$-dodecane, a characteristic mass transfer time on the order of $O(10^{-7})\mathrm{s}$ is already extremely short. The model of \citet{DownarZapolski1996} suggests that the relaxation time approaches zero as the expansion path nears the critical point. However, real fluids also exhibit statistical-mechanical effects such as critical slowing down. Therefore, we consider that the expansion shock phenomenon during $n$-dodecane phase transition is extremely difficult to observe in experiments.

\subsection{Mass and Heat Transfer Relaxation Times}

This subsection sets a series of different mass and heat transfer relaxation times to examine the overall influence of relaxation on the flash evaporation Riemann solution. It is worth noting that, unlike some previous studies on this type of problem \citep{Saurel2008,DeLorenzo2018}, the mass transfer relaxation time here cannot be arbitrarily large. We cannot solve a completely frozen (infinite mass transfer time) flash evaporation Riemann problem by directly ``turning off'' relaxation. Because our method is coupled with a real-fluid EoS, the thermodynamic path of the fluid must be strictly constrained outside the spinodal line, which cannot be guaranteed under an excessively long relaxation time. Similarly, an overly short heat transfer time cannot be specified, as it would cause part of the liquid to be excessively heated and enter inside the spinodal line. These characteristics will be further clarified in our subsequent analysis. We select the following configuration for investigation:

\begin{table}
    \centering
    \caption{Initial values for the relaxation time effects test}
    \label{tab:search2}
    \begin{tabular}{ccccc} 
        \toprule
        State & $P$ ($\mathrm{MPa}$) & $\rho$ ($\mathrm{kg \,m^{-3}}$) & $T$ ($\mathrm{K}$) & $u$ ($\mathrm{m \, s^{-1}}$) \\
        \midrule
        Left & 2.0  & 400 & 617.9 & 0 \\
        Right & 0.1 & 2 & 1026.8& 0 \\
        \bottomrule
    \end{tabular}\par
\end{table}
Meanwhile, the computational domain is $L = 1\mathrm{m}$ with the initial discontinuity located at $x_0 = 0.5\mathrm{m}$, discretized by $N = 1000$ grid cells with a total physical time of $t = 0.8\mathrm{ms}$.

\textbf{1. Influence of characteristic mass transfer time}

\begin{figure}[htbp]
    \centering

 \begin{subfigure}[b]{0.32\textwidth}
        \centering
        \includegraphics[width=\linewidth]{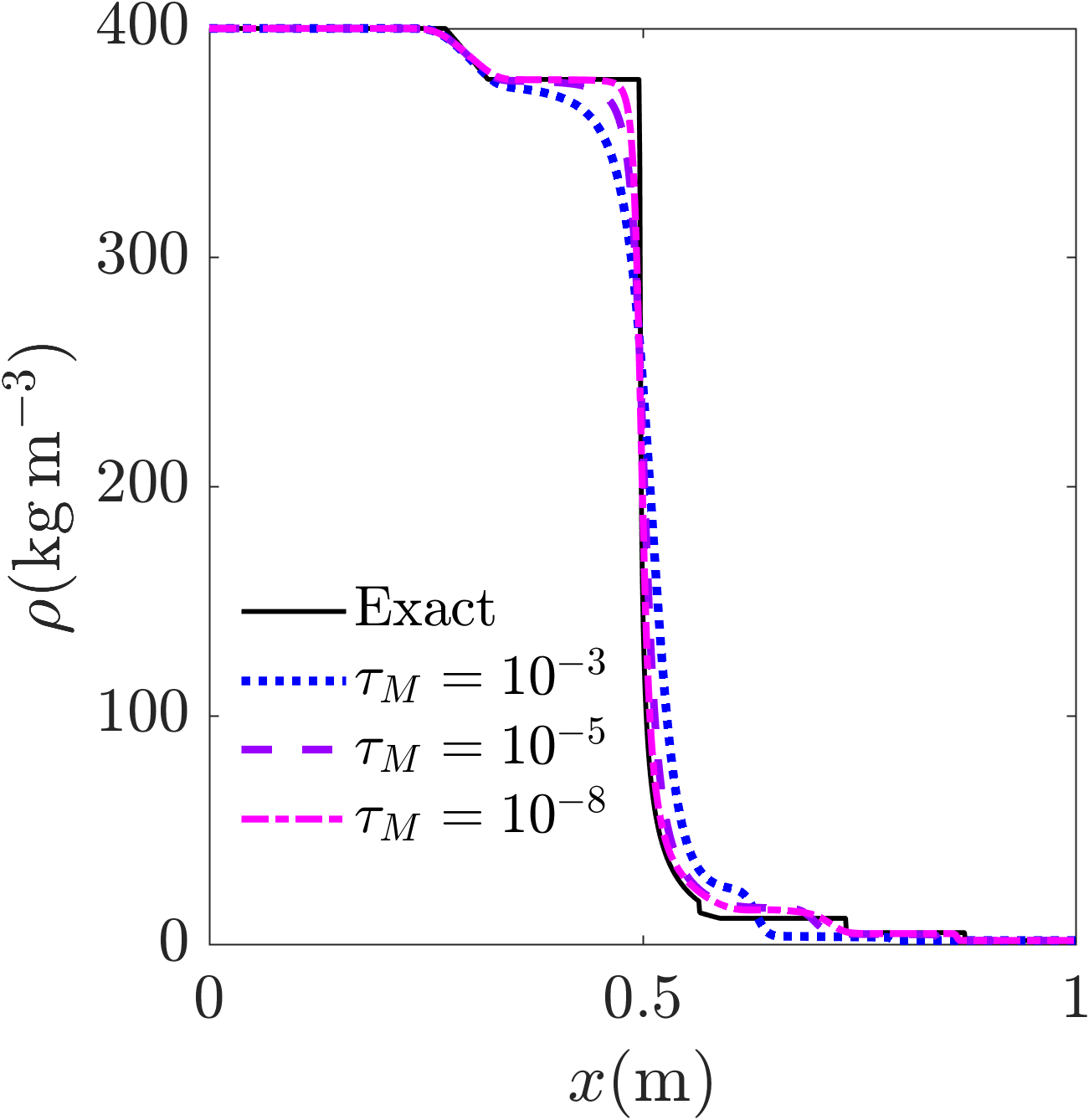}
        \caption{Density} 
        \label{fig11a}
    \end{subfigure}
    \hfill 
    \begin{subfigure}[b]{0.32\textwidth}
        \centering
        \includegraphics[width=\linewidth]{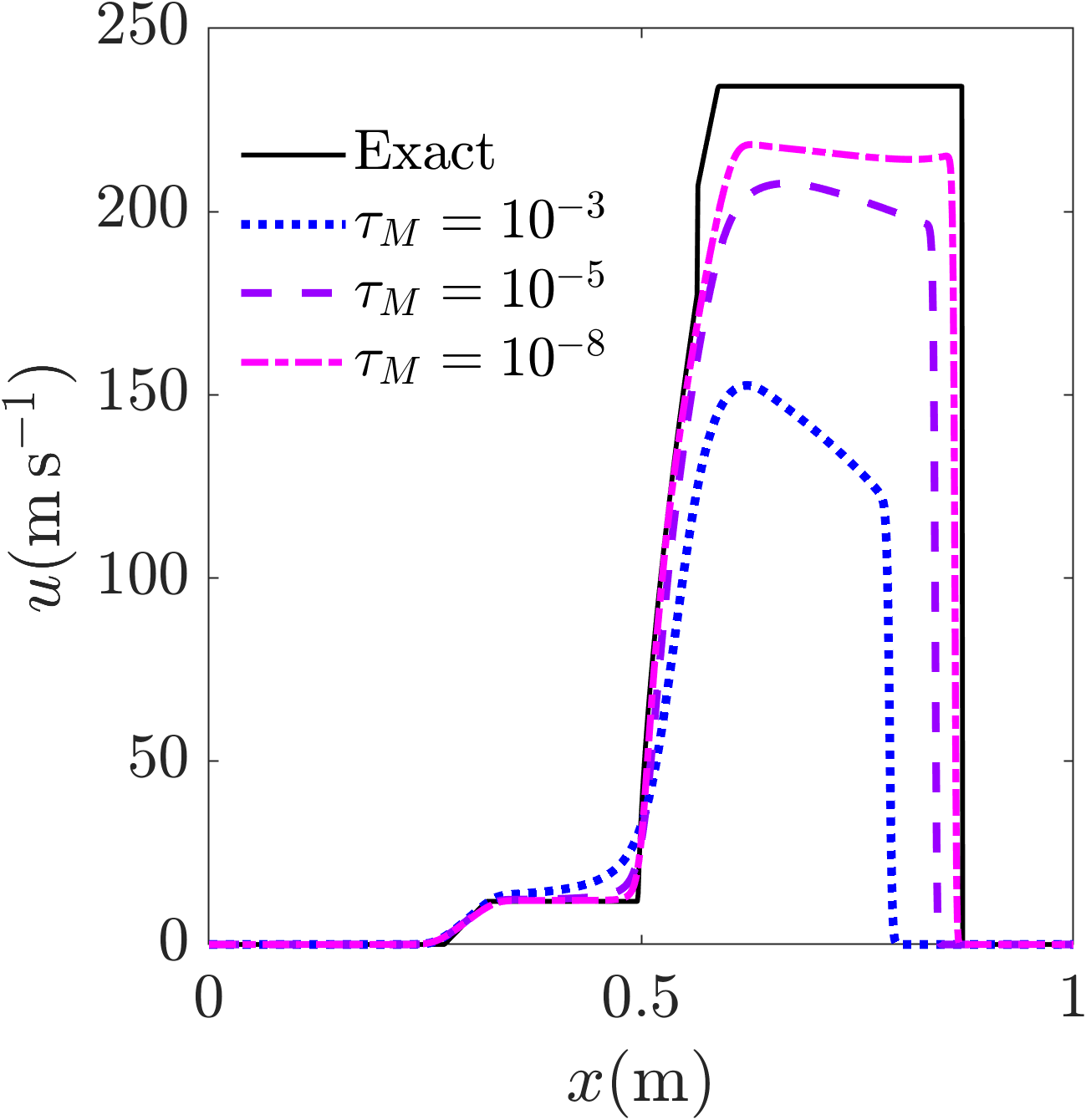}
        \caption{Velocity}
        \label{fig11b}
    \end{subfigure}
    \hfill
    \begin{subfigure}[b]{0.32\textwidth}
        \centering
        \includegraphics[width=\linewidth]{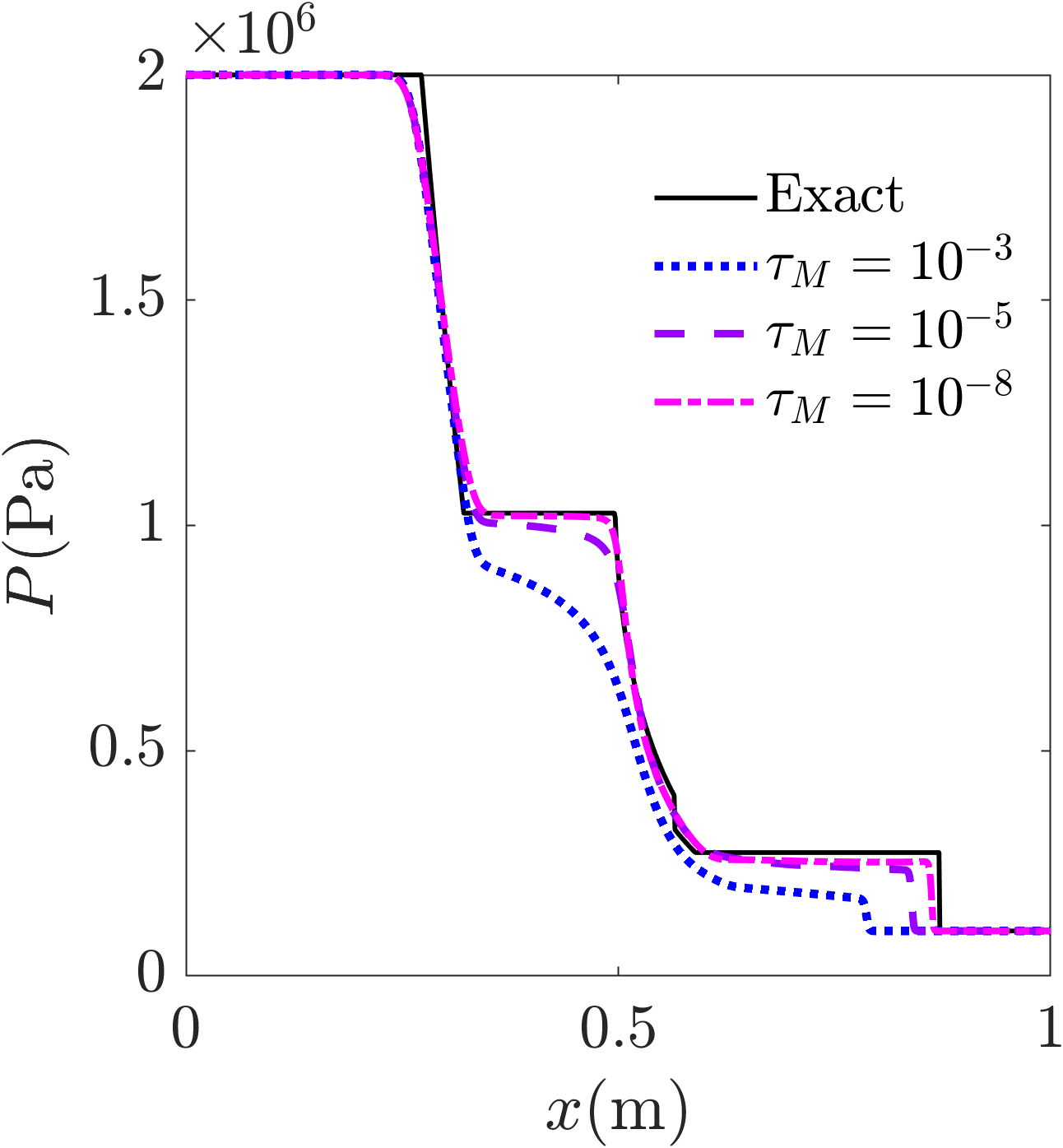}
        \caption{Pressure}
        \label{fig11c}
    \end{subfigure}
    \vspace{0.5em} 
    
    \begin{subfigure}[b]{0.32\textwidth}
        \centering
        \includegraphics[width=\linewidth]{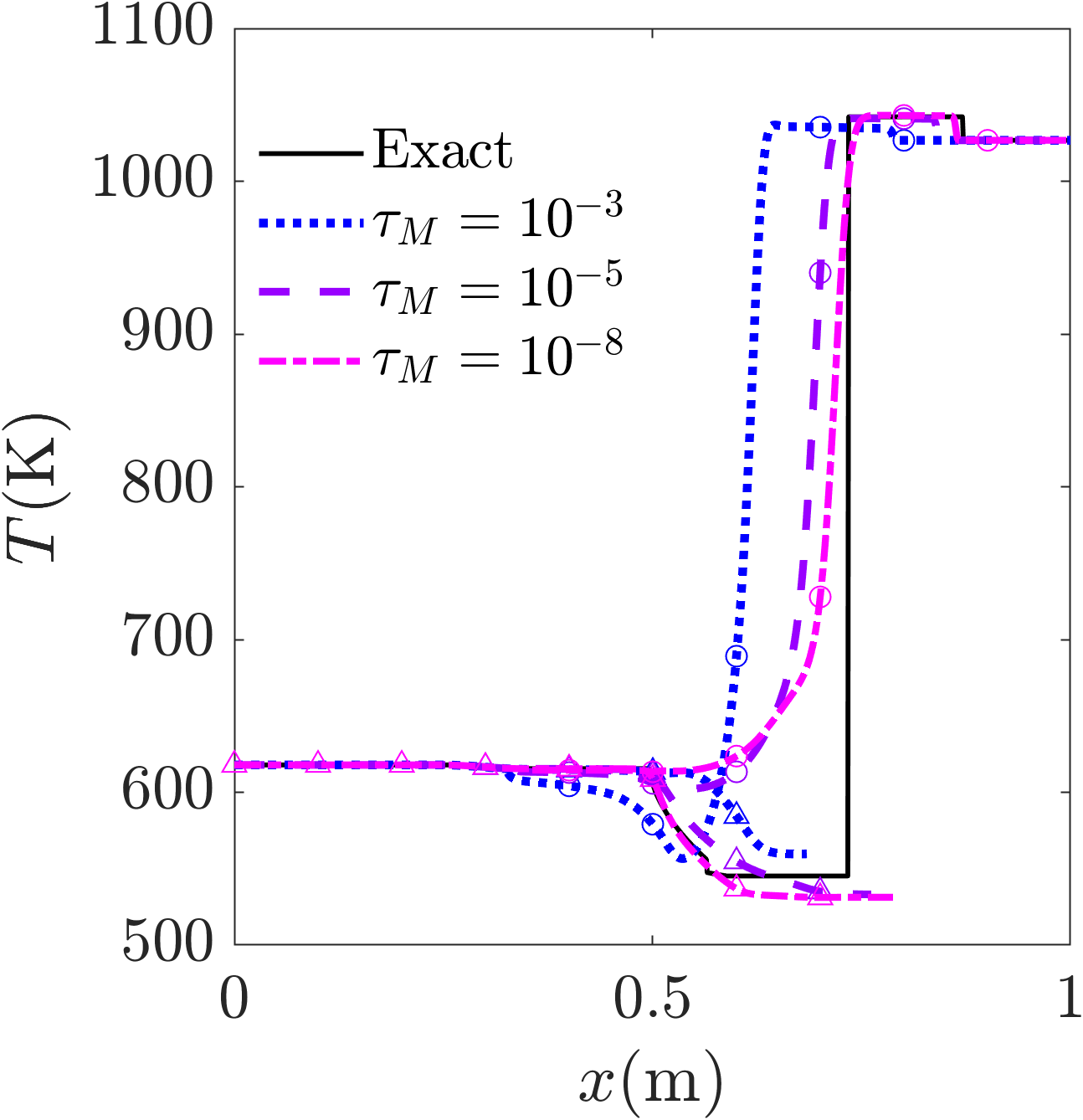}
        \caption{Temperature}
        \label{fig11d}
    \end{subfigure}
    \hfill
    \begin{subfigure}[b]{0.32\textwidth}
        \centering
        \includegraphics[width=\linewidth]{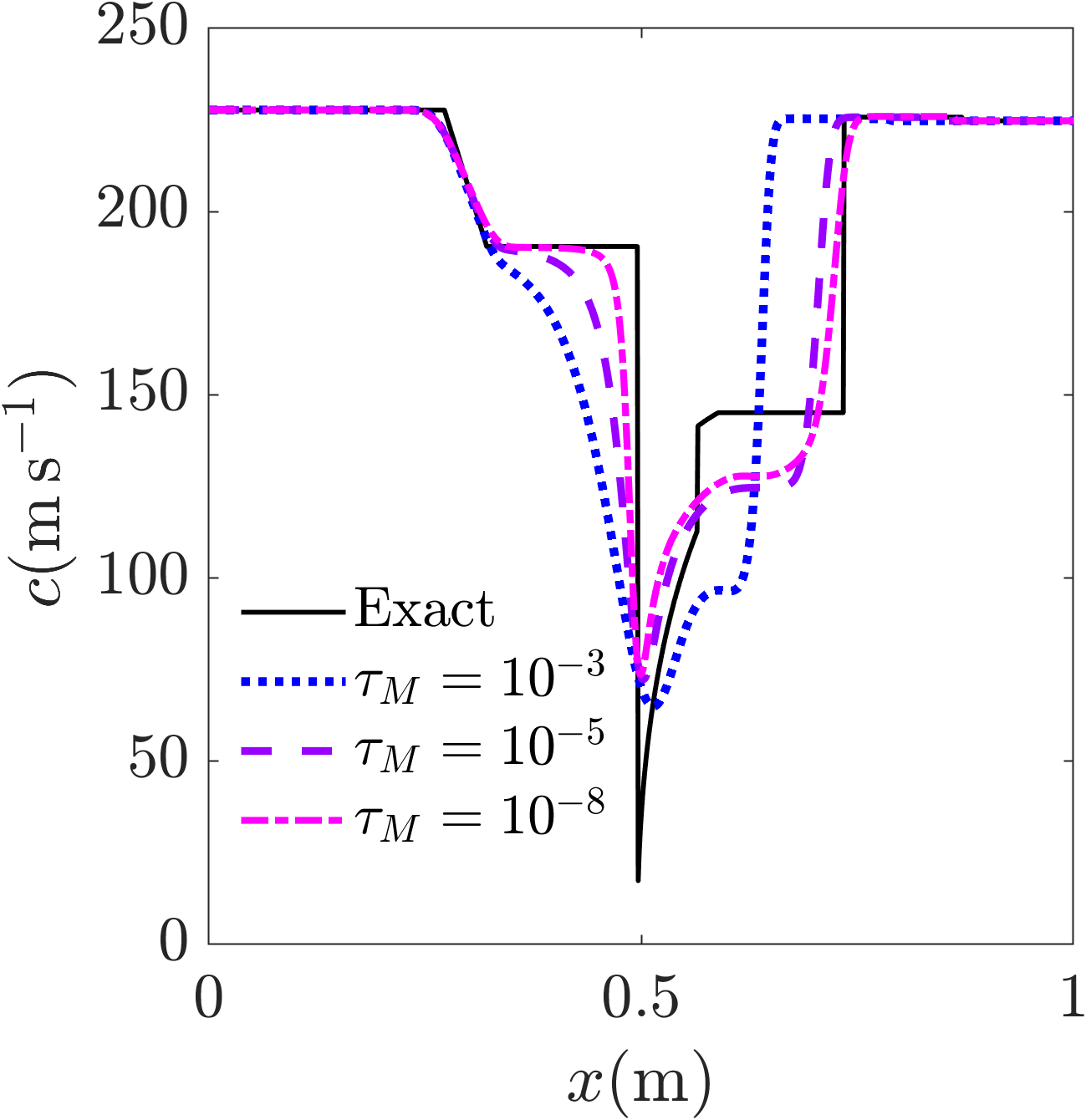}
        \caption{Speed of sound}
        \label{fig11e}
    \end{subfigure}
    \hfill
    \begin{subfigure}[b]{0.32\textwidth}
        \centering
        \includegraphics[width=\linewidth]{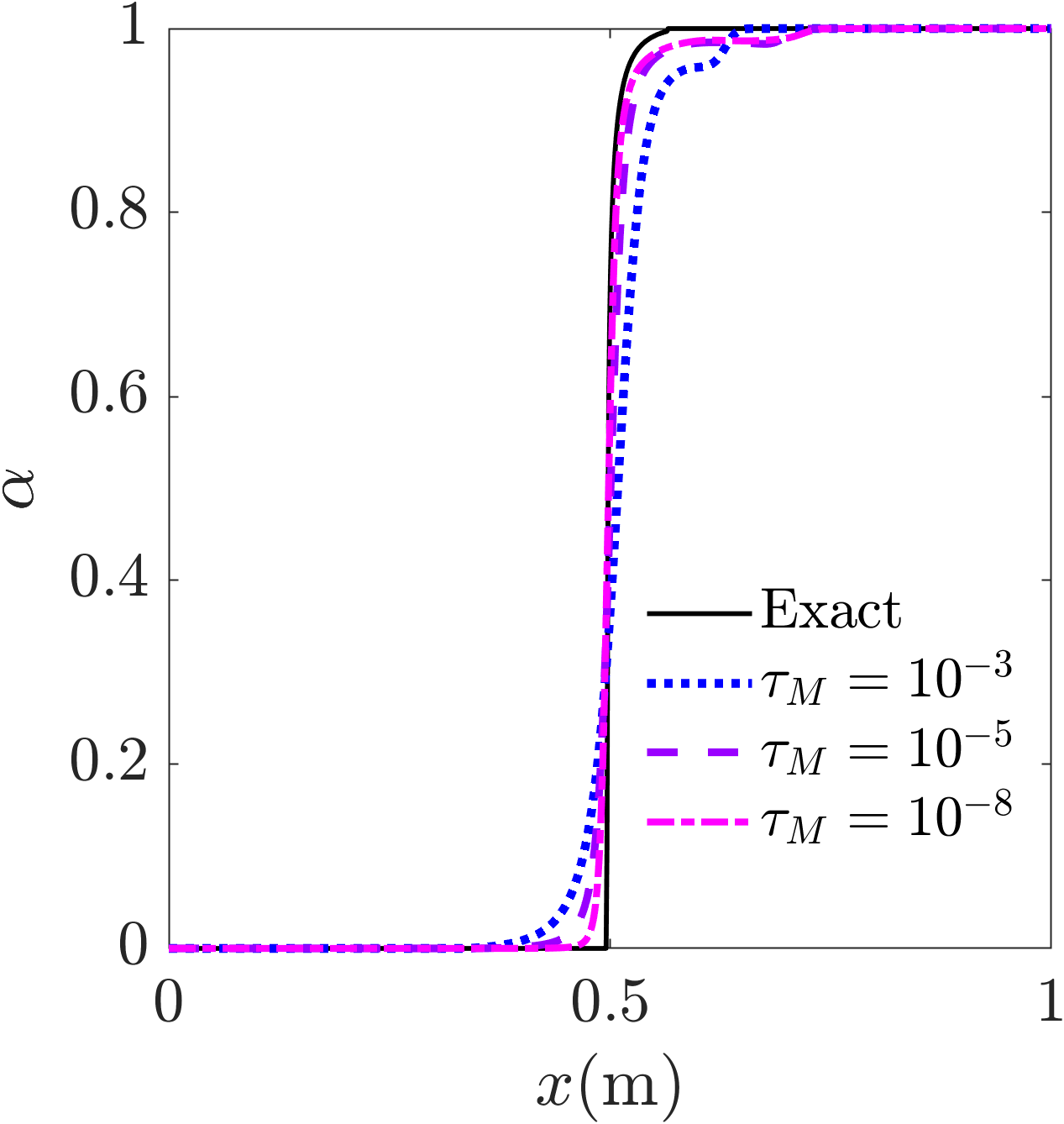}
        \caption{Vapor volume fraction}
        \label{fig11f}
    \end{subfigure}
    \vspace{0.5em}     
    
    \begin{subfigure}[b]{0.58\textwidth}
        \centering
        \includegraphics[width=\linewidth]{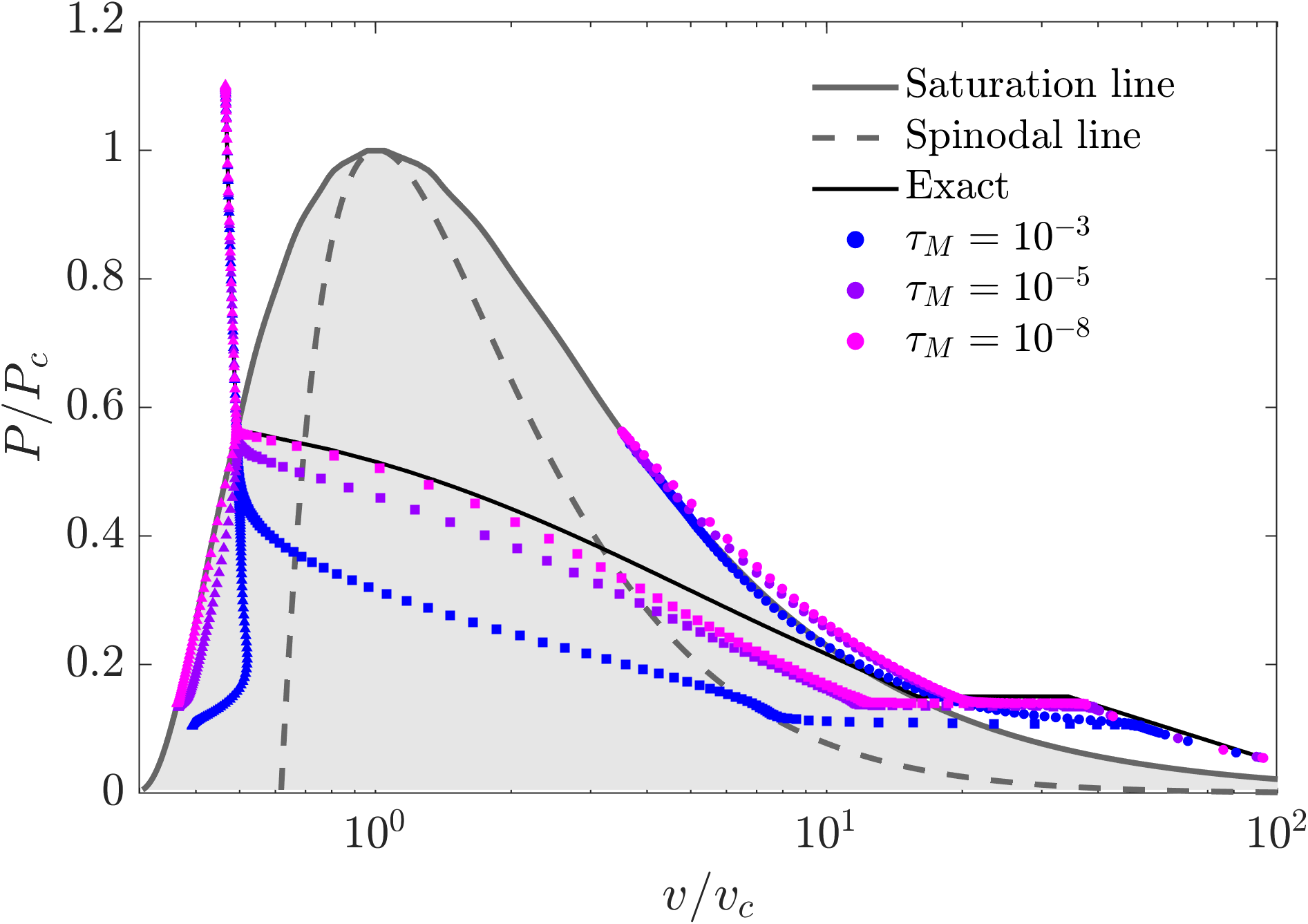}
        \caption{Solution on $P-v$ coordinates}
        \label{fig11g}
    \end{subfigure}
    
    \caption{Comparison between the exact equilibrium solution and numerical solutions with different mass transfer relaxation times. Triangles: liquid phase; circles: vapor phase; squares: mixture phase.}
    \label{fig11}
\end{figure}

First, we fix a sufficiently long heat transfer time and analyze the influence of different characteristic mass transfer times on the solution. We set $\tau_T = 10^{-1}\mathrm{s}$ and $\tau_M = 10^{-3},10^{-5},10^{-8}\mathrm{s}$. Note that, unlike the subsection \ref{subsec:expansion_shock}, the characteristic mass transfer time here is a model parameter, and the actual $\tau_{M,e},\tau_{M,c}$ are calculated from Equations \eqref{eq:evaporation_time} and \eqref{eq:condensation_time}. The calculation results are shown in Figure \ref{fig11}.

From the phase diagram in Figure \ref{fig11g}, the mass transfer time directly changes the expansion trajectory, thereby affecting the profile of the Riemann solution curves. As the relaxation time increases, the onset of liquid phase transition is delayed and the initial phase transition rate decreases, which manifests on the phase diagram as the liquid extending deeper into the metastable region; this is most evident at $\tau_M = 10^{-3}\mathrm{~s}$. Due to the lower phase transition rate, less vapor is newly generated in the mixture, resulting in a weaker work capacity. This leads to lower intermediate equilibrium pressure and velocity in the Riemann solution (Figures \ref{fig11b} and \ref{fig11c}), which further reduces the shock strength and shock speed on the compression branch. Meanwhile, since the configuration with a larger $\tau_M$ must reach a lower equilibrium pressure, the vapor phase undergoes a larger share of the expansion within the mixture. The vapor temperature drops due to over-expansion, even falling below the local liquid temperature (Figure \ref{fig11d}). At the same time, Wood's speed of sound can be expressed in terms of $\alpha$ as:

\begin{equation}
    \frac{1}{\rho c_W^2} = \frac{\alpha}{\rho_v c_v^2} + \frac{1-\alpha}{\rho_l c_l^2}
\end{equation}
which is the harmonic mean weighted by the volume fraction $\alpha$. Consequently, during the initial stage of phase transition, the larger $\alpha$ induced by slow relaxation results in a lower local speed of sound (Figure \ref{fig11e} and \ref{fig11f}).

We point out that under finite relaxation, the expansion branch exhibits a ``pseudo-splitting'' behavior. In the equilibrium limit, the speed of sound discontinuously drops as the expansion trajectory enters the two-phase region, splitting the rarefaction wave into two parts. The propagation speed of the upstream rarefaction wave tail is lower than that of the downstream wave head, thereby generating a plateau state in physical space that stretches over time. In the $P-v$ state space, this splitting plateau in the $x-t$ physical space is compressed into a single point. Under finite relaxation, the foregoing analysis regarding Wood's speed of sound leads to another conclusion: evaporation triggered by small pressure and density variations within the two-phase region causes a rapid drop in the speed of sound. This rapidly decreasing speed of sound causes the fluid to traverse a wide range of characteristic wave speeds over a tiny interval of thermodynamic state variations. Consequently, the minute variations of the $P-v$ state space are ``stretched'' into ``pseudo-splitting'' plateau in physical space.

Notably, in Figure \ref{fig11f}, $\alpha$ in the $x<0.5\mathrm{m}$ region is higher under relaxation and increases with $\tau$, whereas the phase transition onset in the equilibrium state lies near $x=0.5\mathrm{m}$, making it appear as if phase transition occurred earlier under relaxation. Actually, this is an artifact, which we refer to as the ``pseudo-advance'' of the phase transition onset. As just mentioned, the rapid drop in the speed of sound stretches the variations of state variables in physical space; the closer the state approaches equilibrium, the more the spatial stretching intensifies, until a single point in state space is stretched into a line in physical space at complete equilibrium. Therefore, the phase transition onset observed in physical space is already a stretched result. In complete equilibrium, the phase transition onset actually lies at the splitting point of the expansion wave, namely near $x=0.3\mathrm{m}$, but is stretched to $x=0.5\mathrm{m}$ due to wave speed splitting. As the relaxation time increases, this spatial stretching weakens, causing the phase transition onset under slow relaxation to shift back toward $x=0.3\mathrm{m}$, ultimately producing the observed ``pseudo-advance''.

\textbf{2. Influence of characteristic heat transfer time}

\begin{figure}[htbp]
    \centering

 \begin{subfigure}[b]{0.32\textwidth}
        \centering
        \includegraphics[width=\linewidth]{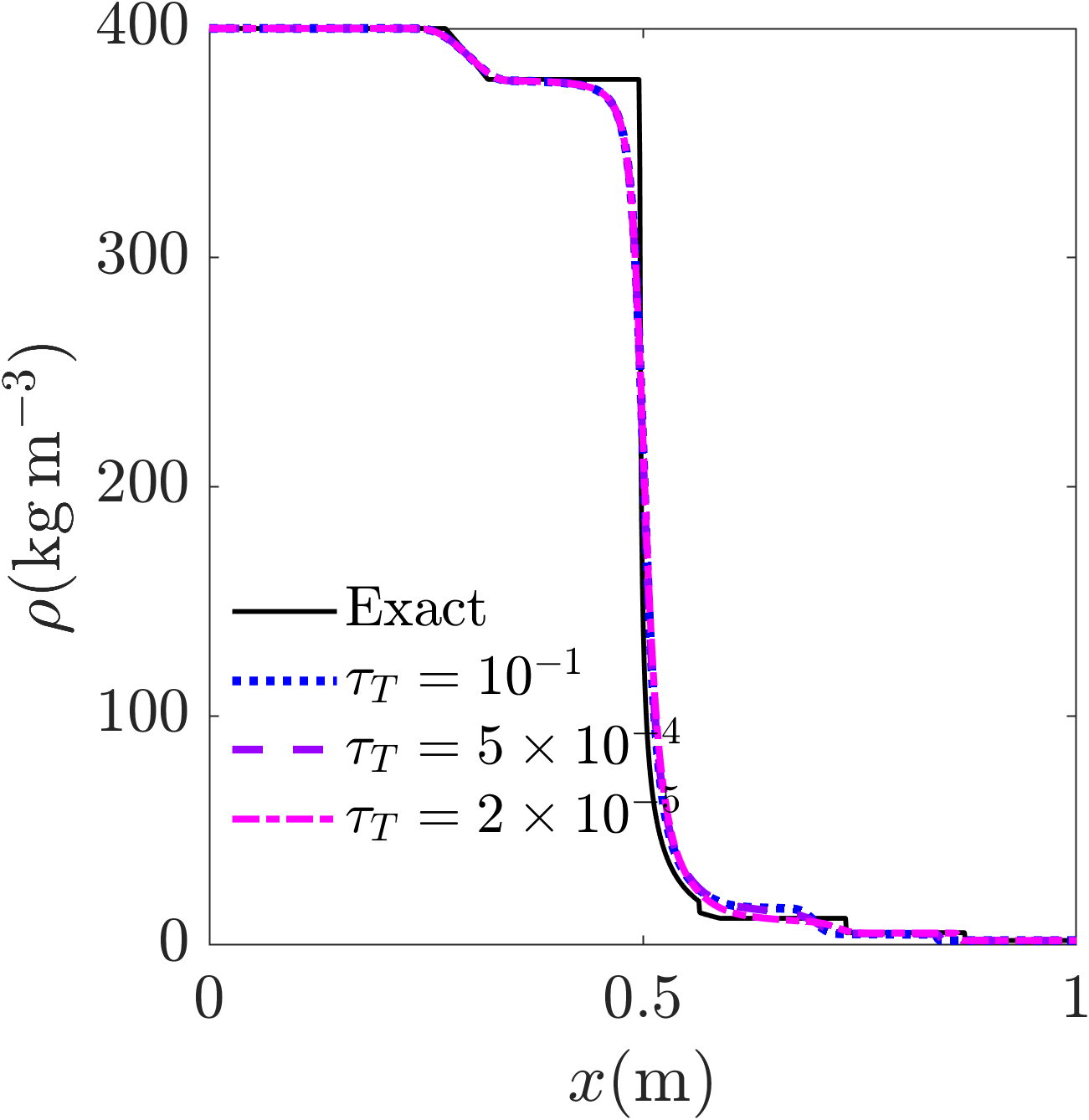}
        \caption{Density} 
        \label{fig12a}
    \end{subfigure}
    \hfill 
    \begin{subfigure}[b]{0.32\textwidth}
        \centering
        \includegraphics[width=\linewidth]{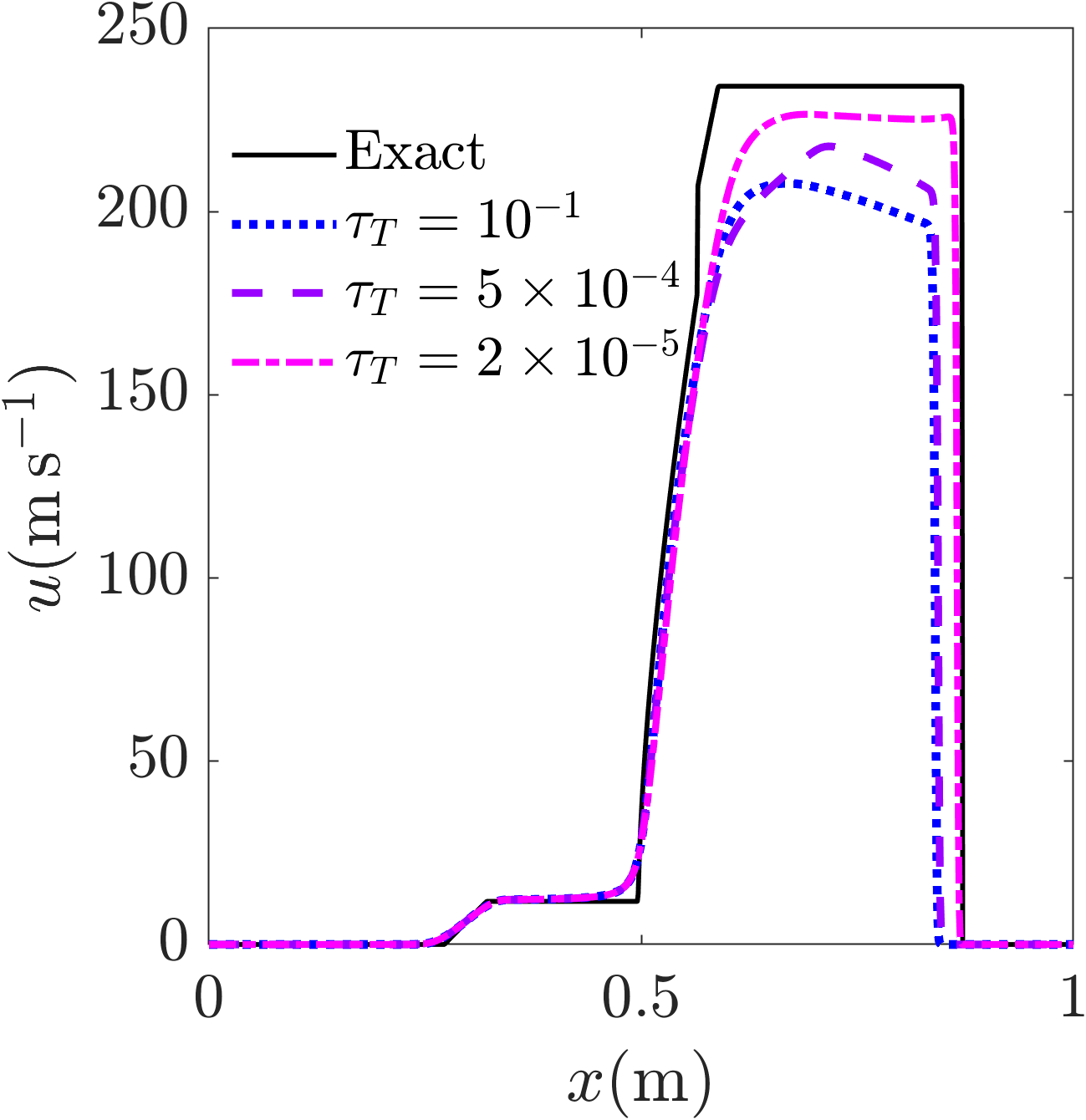}
        \caption{Velocity}
        \label{fig12b}
    \end{subfigure}
    \hfill
    \begin{subfigure}[b]{0.32\textwidth}
        \centering
        \includegraphics[width=\linewidth]{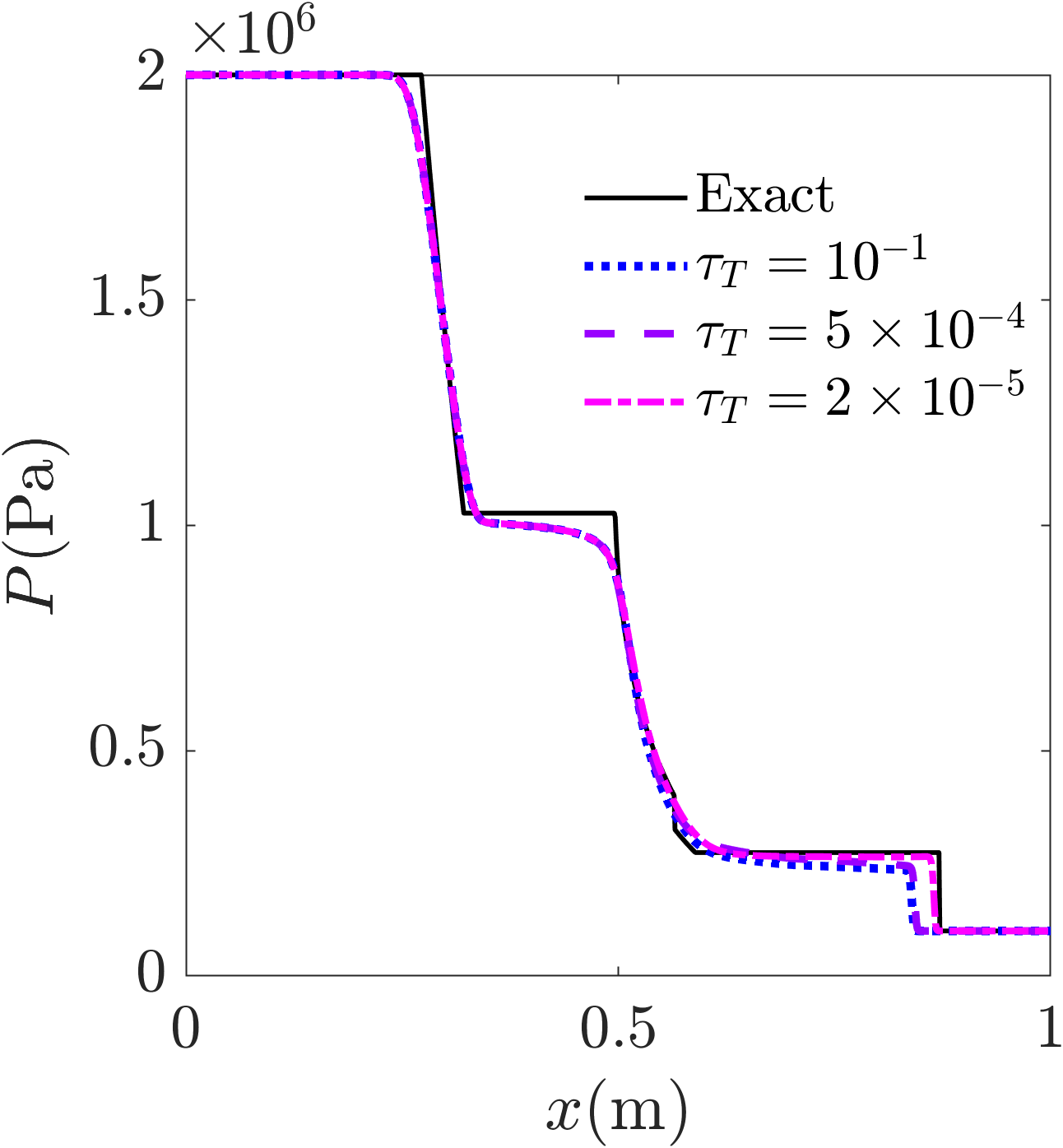}
        \caption{Pressure}
        \label{fig12c}
    \end{subfigure}
    \vspace{0.5em} 
    
    \begin{subfigure}[b]{0.32\textwidth}
        \centering
        \includegraphics[width=\linewidth]{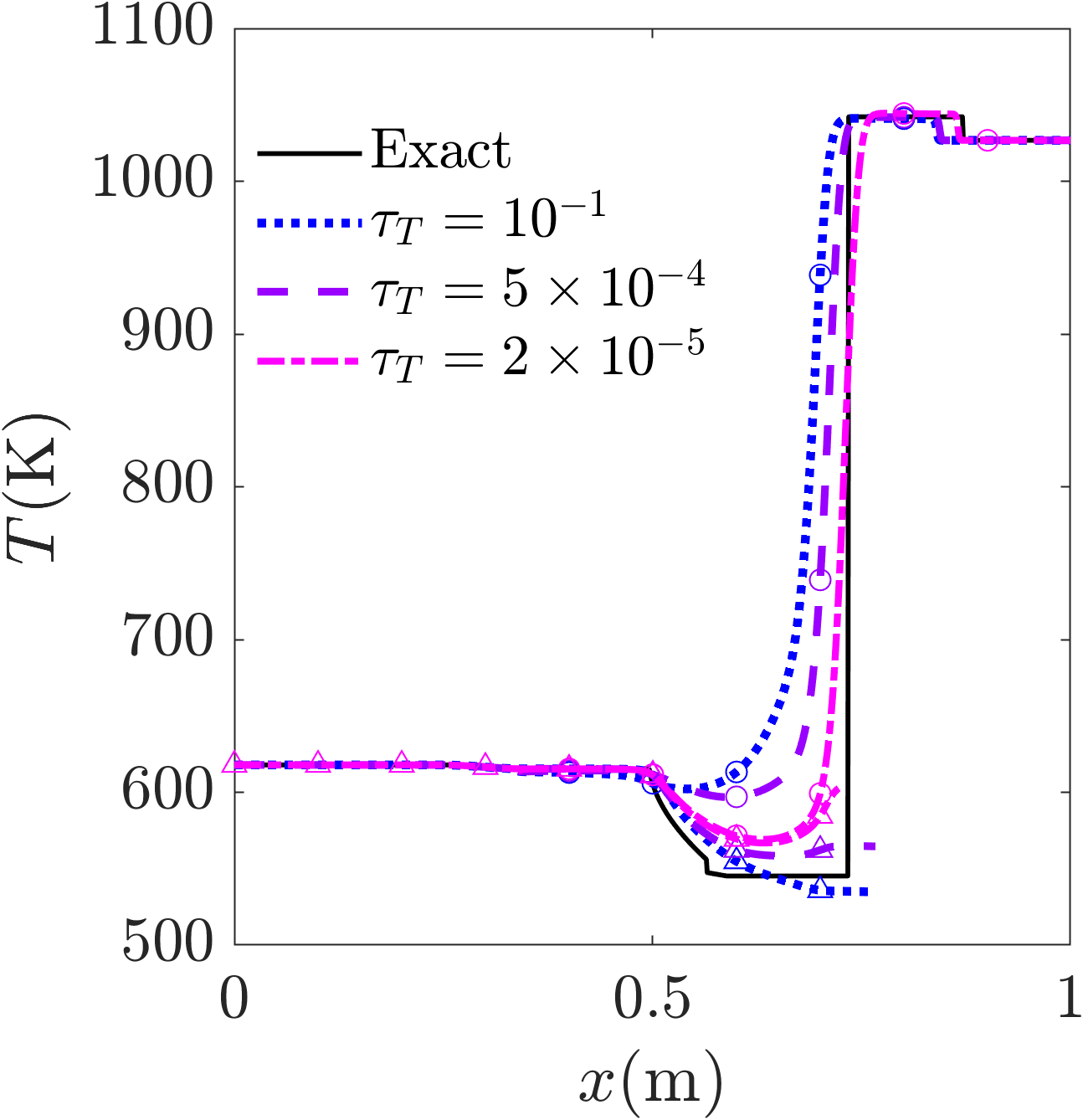}
        \caption{Temperature}
        \label{fig12d}
    \end{subfigure}
    \hfill
    \begin{subfigure}[b]{0.32\textwidth}
        \centering
        \includegraphics[width=\linewidth]{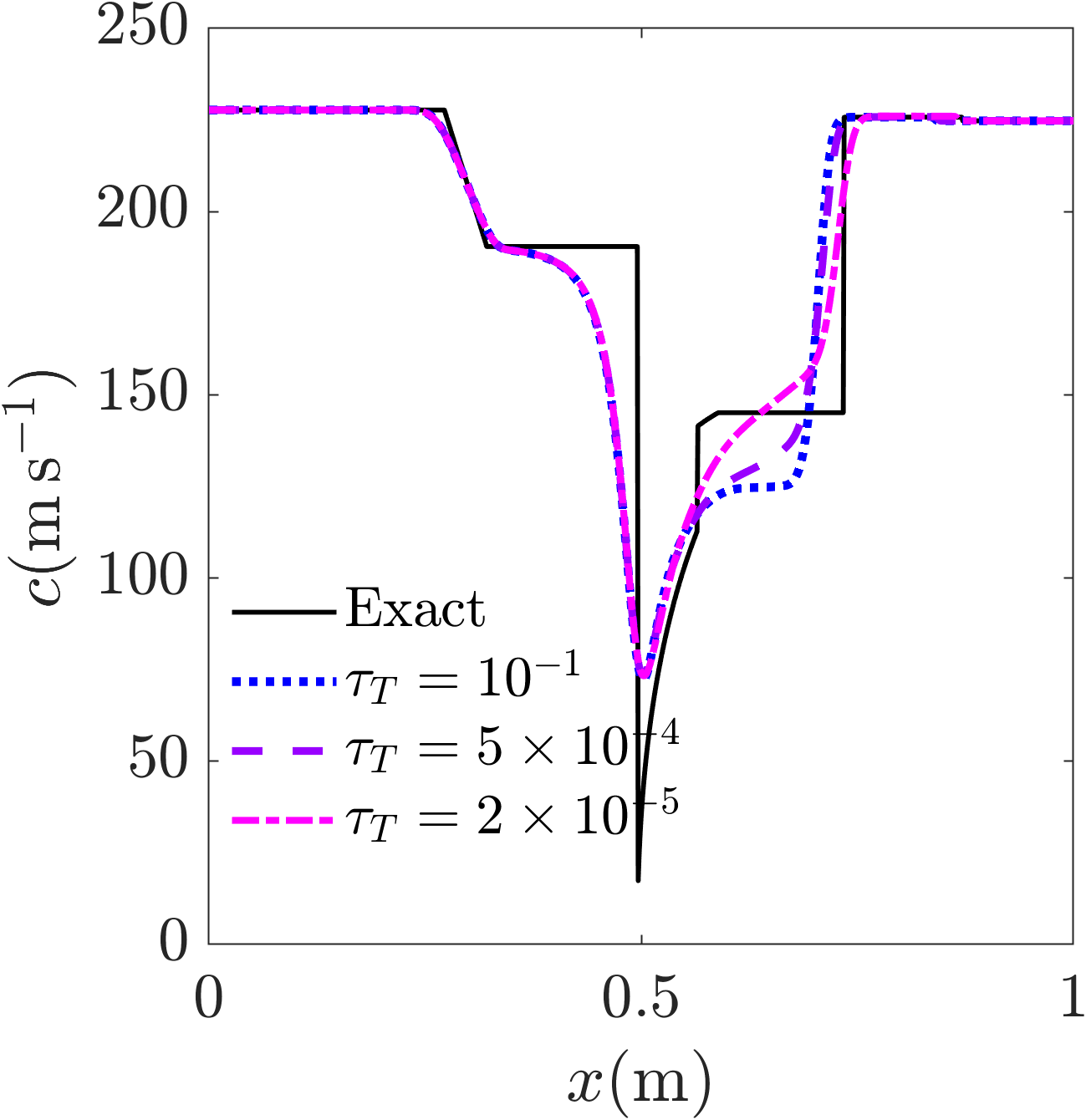}
        \caption{Speed of sound}
        \label{fig12e}
    \end{subfigure}
    \hfill
    \begin{subfigure}[b]{0.32\textwidth}
        \centering
        \includegraphics[width=\linewidth]{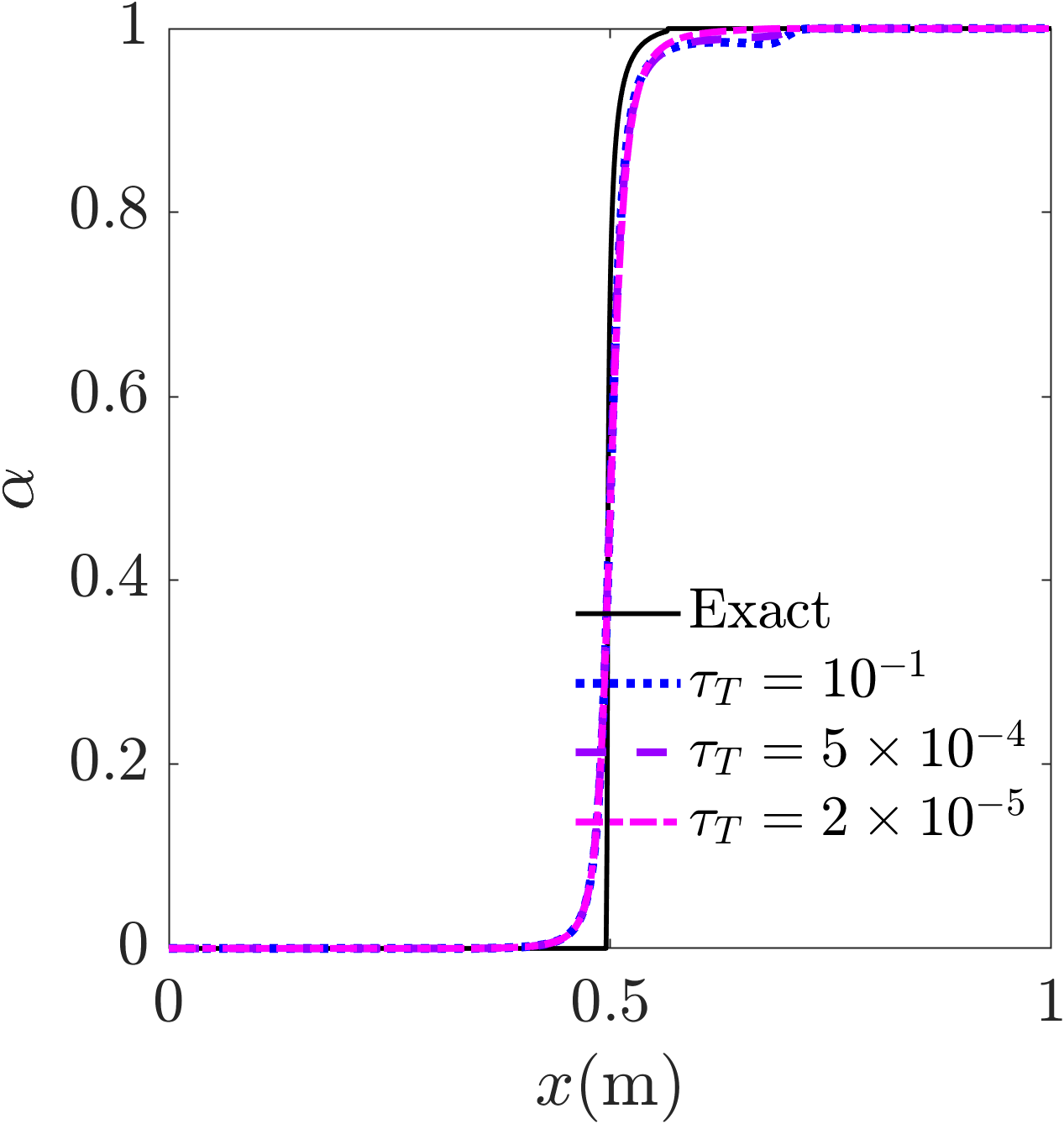}
        \caption{Vapor volume fraction}
        \label{fig12f}
    \end{subfigure}
    \vspace{0.5em}     
    
    \begin{subfigure}[b]{0.58\textwidth}
        \centering
        \includegraphics[width=\linewidth]{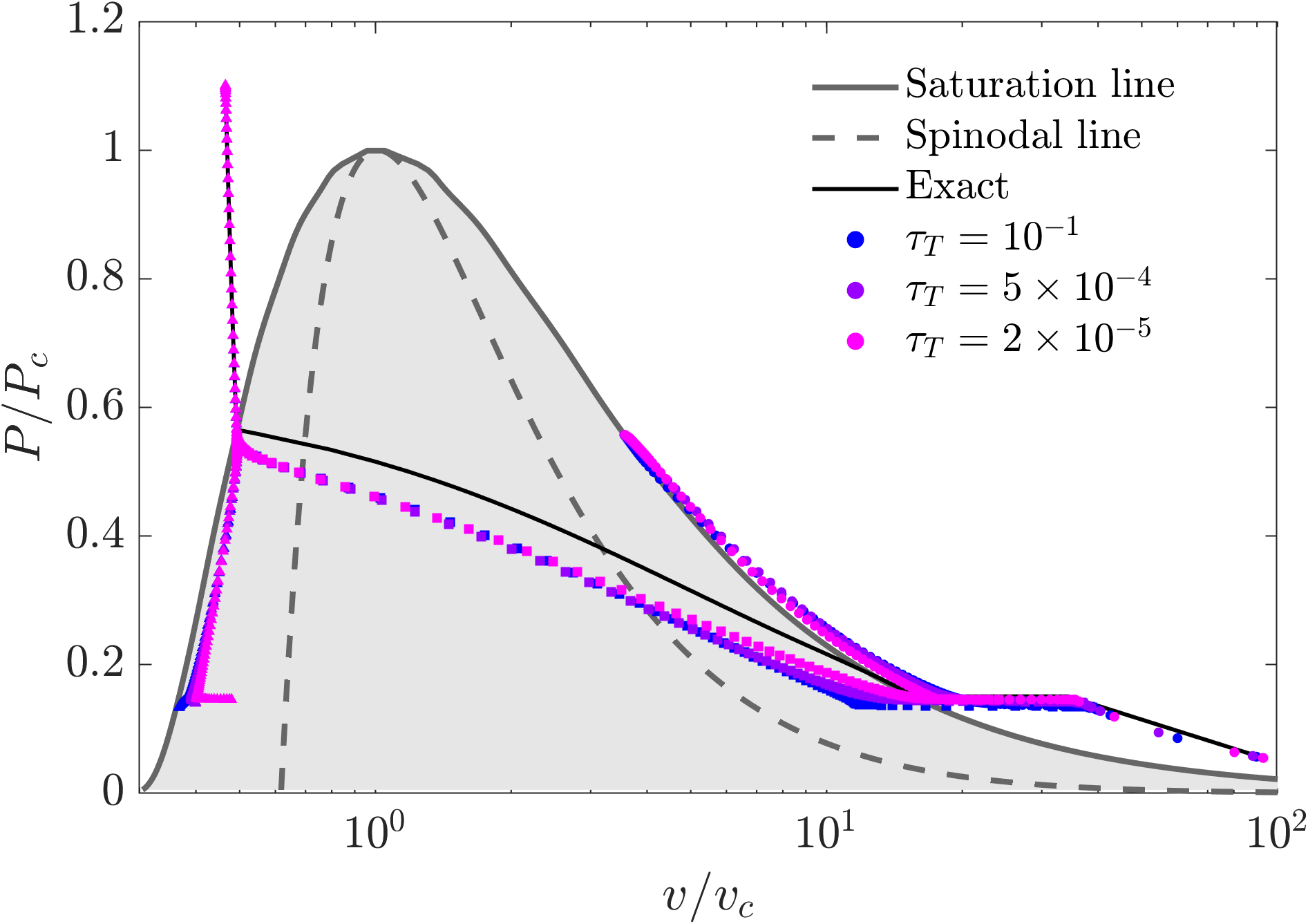}
        \caption{Solution on $P-v$ coordinates}
        \label{fig12g}
    \end{subfigure}
    
    \caption{Comparison between the exact equilibrium solution and numerical solutions with different heat transfer relaxation times. Triangles: liquid phase; circles: vapor phase; squares: mixture phase.}
    \label{fig12}
\end{figure}

Fixing the characteristic mass transfer time at $\tau_M = 10^{-5}\mathrm{s}$ and taking the heat transfer relaxation times as $\tau_H = 10^{-1},5\times10^{-4}, 2\times 10^{-5}\mathrm{s}$, the calculation results are shown in Figure \ref{fig12}.

It can be seen that even though $\tau_H$ crosses multiple orders of magnitude, its influence on the FeRP solution is relatively weak compared to that of $\tau_M$. In fact, since our governing equations \eqref{eq:7eq_system} do not explicitly contain a heat transfer source term, the influence of $\tau_H$ is indirect. A smaller $\tau_H$ accelerates the heat exchange between the two phases in the mixture, allowing them to reach thermal equilibrium faster. As shown in Figure \ref{fig12d}, as $\tau_H$ decreases, the temperature curves of the two phases gradually approach each other, and at $\tau_H= 2\times 10^{-5}\mathrm{s}$, the two temperature curves have almost merged. It should be noted that the mixture temperature exceeds its saturation temperature, leaving the liquid phase locally superheated. On the phase diagram in Figure \ref{fig12g}, as $\tau_H$ decreases, the liquid phase instead shifts slightly toward the metastable region during the expansion process, especially at the phase transition end point near $P=0.15\mathrm{MPa}$: affected by the heat transfer from the high-temperature vapor on the right side of the contact discontinuity, the liquid phase at the phase transition end point in the $\tau_H= 2\times 10^{-5}\mathrm{s}$ configuration is severely superheated and has extended deep into the metastable region. Therefore, under finite mass transfer relaxation, reducing $\tau_H$ accelerates the heat transfer between the superheated vapor phase and the liquid phase, thereby increasing $h_l$ of the metastable liquid phase and enhancing the mass transfer rate through Equation \eqref{eq:mass_source}. This in turn increases the intermediate pressure and velocity in the Riemann solution and strengthens the shock wave on the compression side (Figures \ref{fig12b} and \ref{fig12c}).

Having clarified the mechanisms of how the mass and heat transfer relaxation times affect the FeRP, we now answer the question raised at the beginning of this subsection regarding the specification of relaxation times. There is an upper limit on specifying the mass transfer relaxation time, because an excessively long mass transfer time causes the liquid to extend continuously into the metastable region and touch the spinodal line; under a finite mass transfer time, there is a lower limit on specifying the heat transfer time, because overly rapid heat transfer causes the liquid phase in the mixture to absorb excessive heat, extending deep into the metastable region and touching the spinodal line as well. This problem particularly occurs during the early evolution of the Riemann solution, when the wave structure is concentrated within a few grid cells, the temperature difference within the mixture in the grid cells is extremely large, and heat transfer proceeds violently. For the model in this paper, achieving stable computation generally requires $\tau_M \le 1.5 \tau_T$. The above properties also provide insights for the design of heat and mass transfer models: the heat and mass transfer models must be coordinated with each other, and in particular, a sufficiently strong phase-change driving force needs to be provided for fluids located deep within the metastable region to drive them toward the saturation state.

\subsection{Discussion on the Non-equilibrium Speed of Sound}

Finally, we revisit the discussion on the non-equilibrium speed of sound in Section 4 of \citet{Bai2026FeRP}. Motivated by the formulation of the dynamic speed of sound \eqref{eq:c_dynamic} in relaxation systems, we introduced the non-equilibrium speed of sound $c_\eta$ directly into the exact solution of the Riemann problem:
\begin{equation}
   c_{\eta}^{-2} = (1 - \eta) c_W^{-2} + \eta c_{eq}^{-2}
\end{equation}
where $0\le \eta\le 1$ is a coefficient representing the degree of equilibrium: when $\eta=0$, $c_\eta = c_W$, and when $\eta=1$, $c_\eta = c_{eq}$. Specifying $c_\eta$ in the two-phase region to find the exact solution actually extends the self-similar solution of the Riemann problem by introducing a non-equilibrium degree of freedom. Now, through complete HRM modeling, we can re-examine the effects brought by this practice.

\begin{figure}[htbp]
\centering 
\includegraphics[width=0.58\textwidth]{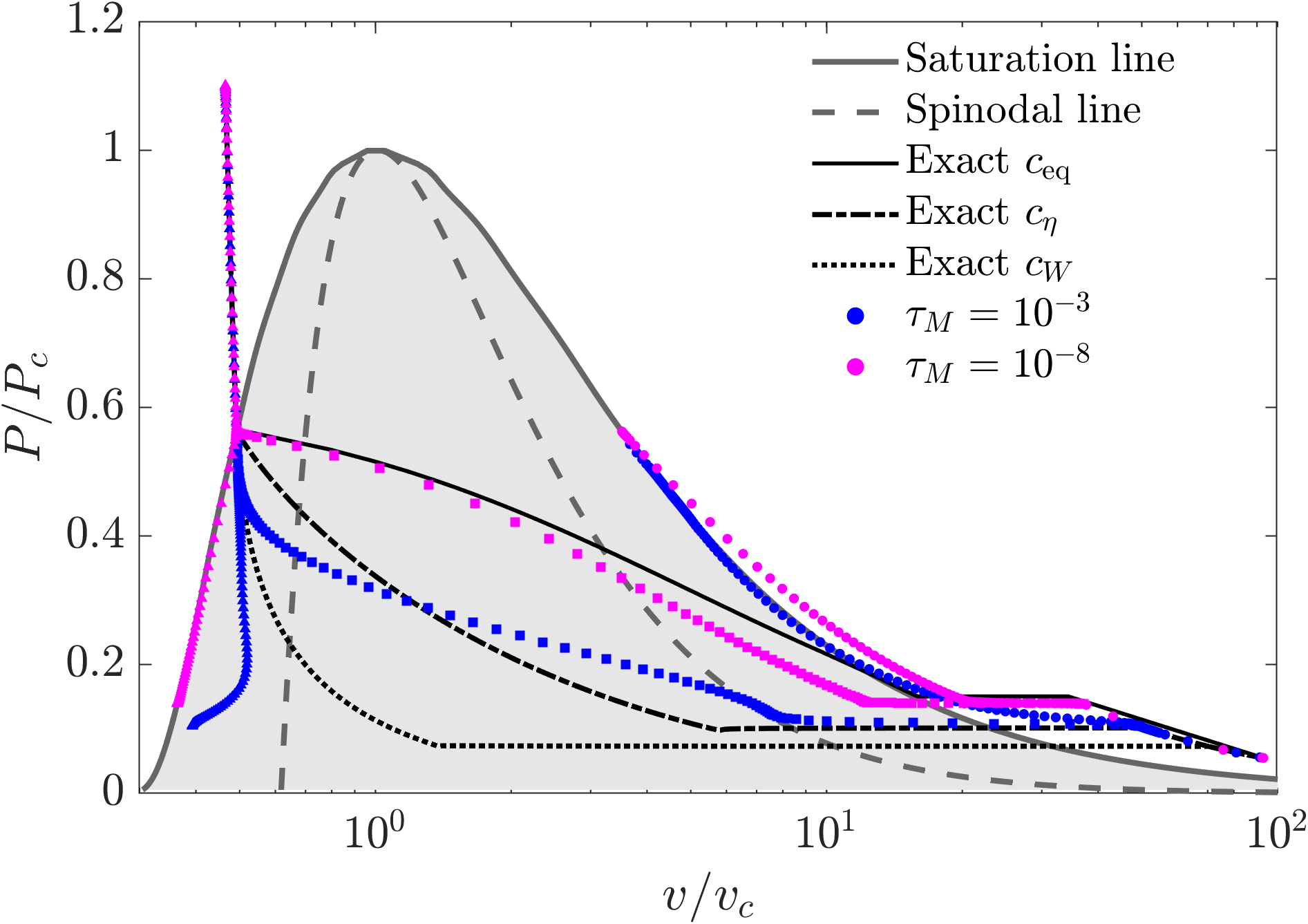} 
\caption{Comparison of self-similar exact solutions with different speeds of sound and numerical solutions with various relaxation times in the $P-v$ plane.}
\label{fig13}
\end{figure}

Figure \ref{fig13} shows the self-similar exact solutions with different two-phase speeds of sound (lines) and the numerical solutions with different mass transfer relaxation times (points) for the Riemann problem in Table \ref{tab:search2}, where the non-equilibrium speed of sound is computed with $\eta = 0.1$. As can be seen from the figure, as the degree of non-equilibrium increases, the intermediate pressure of the Riemann problem decreases continuously, which is consistent with the results solved by the complete homogeneous relaxation model. For the self-similar solution with $c_{\eta = 0.1}$, its intermediate pressure is very close to the result of $\tau_M=10^{-3}\mathrm{s}$. Correspondingly, the self-similar Riemann solution with the non-equilibrium speed of sound also exhibits properties such as a relatively lower intermediate velocity and a weaker compression shock strength on the right side (Figures 10--12 in \citealt{Bai2026FeRP}).

However, specifying a non-equilibrium speed of sound in the exact solution also has many limitations. In addition to the ``entropy decrease'' effect analyzed previously, constrained by its physical assumptions, it cannot genuinely approximate the solution curves of the actual relaxation system. Since phase equilibrium in the exact solution is completed instantaneously (i.e., the VLE assumption), as long as $\eta \neq 0$, $c_\eta$ will always exhibit a jump at the phase boundary, and non-classical expansion shocks and expansion wave splitting will appear in the exact solution; there will be no continuous degeneration of the expansion shock toward a rarefaction wave, nor will there be ``pseudo-splitting'' of the expansion wave or ``pseudo-advance'' of the phase transition point, because these properties are all caused by finite-rate phase transition delay within the state space.

\clearpage

\clearpage

\section{Conclusion}

To elucidate the mechanisms of real-fluid thermodynamic relaxation effects on the flash evaporation Riemann problem solution, this paper first proposes an oscillation-free Homogeneous Relaxation Real Fluid finite volume method, and systematically explores the evolution characteristics of the FeRP wave structures under finite heat and mass transfer relaxation. The main conclusions are as follows:

The pressure oscillation problem of real fluids in conservative finite volume methods originates from the assumption of instantaneous thermodynamic equilibrium. Based on this physical understanding, the HRRF method supplements mechanical equilibrium information by additionally evolving the internal energy--pressure affine parameters in the relaxation system, thereby eliminating numerical pressure oscillations. Meanwhile, by evolving the vapor volume fraction equation combined with thermodynamic re-projection, the HRRF method achieves natural degeneration into a single-phase real-fluid oscillation-free method. When solving the FeRP with finite relaxation, HRRF demonstrates favorable physical consistency and equilibrium convergence properties. As an initial development, the HRRF framework in this work leaves several algorithmic aspects for future research, particularly high-order spatial reconstruction, stiff source term integration and multidimensional extensions.

The non-classical expansion shock in the FeRP continuously degenerates as the thermodynamic relaxation time increases. For $n$-dodecane, when the mass transfer relaxation time is on the order of $O(10^{-10}\mathrm{s})$, the expansion shock still retains strong traveling wave characteristics. On the order of $O(10^{-9}\mathrm{s})-O(10^{-8}\mathrm{s})$, the expansion shock undergoes degeneration, and its width increases continuously over time, though still significantly slower than the spreading rate of a self-similar expansion wave. On the order of $O(10^{-7}\mathrm{s})$, the expansion shock is close to complete degeneration, and the expansion wave pattern as a whole recovers the characteristics of a self-similar rarefaction wave. Such degeneration behavior leads us to predict that, for certain fluids, the existence of expansion shocks during phase transition is related to the rate of phase transition, and the theoretically predicted expansion shocks may be difficult to observe in actual experiments.

The mass transfer relaxation time has the most significant influence on the FeRP solution. Increasing the mass transfer time reduces the liquid evaporation rate, causing the liquid to extend deep into the metastable region during expansion, and further leads to lower intermediate equilibrium pressure and velocity in the Riemann solution. When the vapor phase becomes superheated, increasing the heat transfer rate causes the liquid phase in the mixture to shift deeper into the metastable region, thereby indirectly enhancing the mass transfer rate.

The HRRF method does not rely on specific source term models, and the relaxation closure adopted in this work serves as a representative generalization. Nevertheless, the physical insights gained from the HRRF results provide practical considerations for designing such models: the relaxation time models for heat and mass transfer should be coordinated with each other, and the heat transfer rate should be constrained by the mass transfer rate to prevent the liquid from absorbing excessive heat and entering the spinodal state. Meanwhile, for fluids located deep within the metastable region, the mass transfer model needs to provide a sufficiently strong driving force to shift them toward the saturation state.

\begin{appendix}

\section{Partial Derivatives in the Pressure Evolution Equation}
\label{app:pressure_derivatives}

Consider a single-phase fluid $k$ ($k=l,v$). Selecting $(\rho_k, s_k)$ as independent state parameters, the total differential of its pressure $P = P(\rho_k, s_k)$ is:
\begin{equation}
 \mathrm{d}P = \pderiv{P}{\rho}{s} \mathrm{d}\rho + \pderiv{P}{s}{\rho}\mathrm{d}s.
 \label{eq:app_dP_k}
\end{equation}
Introducing the Gr\"uneisen coefficient:
\begin{equation}
 \Gamma = \frac{1}{\rho} \pderiv{P}{e}{\rho}= \frac{1}{\rho C_v} \pderiv{P}{T}{\rho}.
 \label{eq:app_Gamma}
\end{equation}
Combining with the fundamental thermodynamic relation $\mathrm{d}e = T \mathrm{d}s + (P / \rho^2)\mathrm{d}\rho$, under the isochoric ($\mathrm{d}\rho=0$) condition we have $\mathrm{d}e = T \mathrm{d}s$. According to the chain rule:
\begin{equation}
 \pderiv{P}{s}{\rho}= \pderiv{P}{e}{\rho} \pderiv{e}{s}{\rho} = \rho \Gamma T.
 \label{eq:app_dP_ds}
\end{equation}
Substituting the above relation and the definition of the isentropic speed of sound into Equation \eqref{eq:app_dP_k}, the total differential of the single-phase pressure is rewritten as:
\begin{equation}
 \mathrm{d}P = c^2 \mathrm{d}\rho + \rho \Gamma T \mathrm{d}s.
 \label{eq:app_dP_k_final}
\end{equation}
Under the mechanical equilibrium assumption, the vapor and liquid phases share a common pressure increment at any instant, namely $\mathrm{d}P_l = \mathrm{d}P_v = \mathrm{d}P$. Rearranging Equation \eqref{eq:app_dP_k_final}, the density differentials of the two phases are expressed in terms of the pressure differential and entropy differentials:
\begin{equation}
 \mathrm{d}\rho_v = \frac{1}{c_v^2}\mathrm{d}P - \frac{\rho_v \Gamma_v T_v}{c_v^2}\mathrm{d}s_v,
 \label{eq:app_drho_v}
\end{equation}
\begin{equation}
 \mathrm{d}\rho_l = \frac{1}{c_l^2}\mathrm{d}P - \frac{\rho_l \Gamma_l T_l}{c_l^2}\mathrm{d}s_l.
 \label{eq:app_drho_l}
\end{equation}

The density $\rho$ of the fluid mixture satisfies the specific volume rule:
\begin{equation}
 \frac{1}{\rho} = \frac{\beta}{\rho_v} + \frac{1-\beta}{\rho_l}.
 \label{eq:app_rho_mix}
\end{equation}
Taking the total differential on both sides of Equation \eqref{eq:app_rho_mix} yields:
\begin{equation}
 -\frac{1}{\rho^2}\mathrm{d}\rho = \left( \frac{1}{\rho_v} - \frac{1}{\rho_l} \right)\mathrm{d}\beta - \frac{\beta}{\rho_v^2}\mathrm{d}\rho_v - \frac{1-\beta}{\rho_l^2}\mathrm{d}\rho_l.
 \label{eq:app_drho_mix}
\end{equation}
Substituting Equations \eqref{eq:app_drho_v} and \eqref{eq:app_drho_l} into Equation \eqref{eq:app_drho_mix} eliminates $\mathrm{d}\rho_v$ and $\mathrm{d}\rho_l$:
\begin{equation}
 \begin{aligned}
 -\frac{1}{\rho^2}\mathrm{d}\rho ={}& \left( \frac{1}{\rho_v} - \frac{1}{\rho_l} \right)\mathrm{d}\beta 
 - \frac{\beta}{\rho_v^2} \left[ \frac{1}{c_v^2}\mathrm{d}P - \frac{\rho_v \Gamma_v T_v}{c_v^2}\mathrm{d}s_v \right] \\
 &- \frac{1-\beta}{\rho_l^2} \left[ \frac{1}{c_l^2}\mathrm{d}P - \frac{\rho_l \Gamma_l T_l}{c_l^2}\mathrm{d}s_l \right].
 \end{aligned}
 \label{eq:app_substitution}
\end{equation}
Moving the terms containing $\mathrm{d}P$ to the left-hand side of the equation and combining them gives:
\begin{equation}
 \left( \frac{\beta}{\rho_v^2 c_v^2} + \frac{1-\beta}{\rho_l^2 c_l^2} \right) \mathrm{d}P = \frac{1}{\rho^2}\mathrm{d}\rho + \left( \frac{1}{\rho_v} - \frac{1}{\rho_l} \right)\mathrm{d}\beta + \frac{\beta \Gamma_v T_v}{\rho_v c_v^2}\mathrm{d}s_v + \frac{(1-\beta) \Gamma_l T_l}{\rho_l c_l^2}\mathrm{d}s_l.
 \label{eq:app_grouped}
\end{equation}
Substituting the definition of $c_W$ \eqref{eq:c_W} into Equation \eqref{eq:app_grouped} and multiplying both sides by $\rho^2 c_W^2$, we obtain the total differential of the two-phase common pressure $P(\beta, \rho, s_l, s_v)$:
\begin{equation}
 \mathrm{d}P = c_W^2 \mathrm{d}\rho + \rho^2 c_W^2 \left( \frac{1}{\rho_v} - \frac{1}{\rho_l} \right) \mathrm{d}\beta + \rho^2 c_W^2 \frac{(1-\beta) \Gamma_l T_l}{\rho_l c_l^2} \mathrm{d}s_l + \rho^2 c_W^2 \frac{\beta \Gamma_v T_v}{\rho_v c_v^2} \mathrm{d}s_v.
 \label{eq:app_dP_final}
\end{equation}
According to the definition of the total differential of a multivariable function:
\begin{equation}
 \mathrm{d}P = \pderiv{P}{\rho}{\beta,s_l,s_v}\mathrm{d}\rho + \pderiv{P}{\beta}{\rho,s_l,s_v}\mathrm{d}\beta + \pderiv{P}{s_l}{\rho,\beta,s_v}\mathrm{d}s_l + \pderiv{P}{s_v}{\rho,\beta,s_l}\mathrm{d}s_v.
 \label{eq:app_dP_def}
\end{equation}
Comparing the coefficients of each term in Equation \eqref{eq:app_dP_final} and Equation \eqref{eq:app_dP_def} yields the expressions for the partial derivatives of pressure with respect to each variable:
\begin{equation}
\begin{aligned}
\pderiv{P}{\rho}{\beta,s_l,s_v} &= c_W^2, \\
\pderiv{P}{\beta}{\rho,s_l,s_v} &= \rho^2 c_W^2 \left(\frac{1}{\rho_v}-\frac{1}{\rho_l}\right), \\
 \pderiv{P}{s_l}{\rho,\beta,s_v} &= \rho^2 c_W^2 \frac{(1-\beta) \Gamma_l T_l}{\rho_l c_l^2}, \\
 \pderiv{P}{s_v}{\rho,\beta,s_l} &= \rho^2 c_W^2 \frac{\beta \Gamma_v T_v}{\rho_v c_v^2}.  
\end{aligned}
\end{equation}

\section{Partial Derivatives in the Vapor Volume Fraction Evolution Equation}
\label{app:alpha_derivatives}

In the homogeneous model with mechanical equilibrium, the volume fraction is determined by the independent thermodynamic state variables $(\rho, \beta, s_l, s_v)$. From the definition of phasic densities:
\begin{equation}
    \rho_v=\frac{\rho\beta}{\alpha},
    \qquad
    \rho_l=\frac{\rho(1-\beta)}{1-\alpha}.
\end{equation}
Taking the logarithmic differential of the above equations yields the relative variations of the phase densities:
\begin{equation}
    \begin{aligned}
        \frac{\mathrm{d}\rho_v}{\rho_v}
        &=
        \frac{\mathrm{d}\rho}{\rho}
        +\frac{\mathrm{d}\beta}{\beta}
        -\frac{\mathrm{d}\alpha}{\alpha},\\
        \frac{\mathrm{d}\rho_l}{\rho_l}
        &=
        \frac{\mathrm{d}\rho}{\rho}
        -\frac{\mathrm{d}\beta}{1-\beta}
        +\frac{\mathrm{d}\alpha}{1-\alpha}.
    \end{aligned}
\end{equation}
For each phase, the pressure differential is given by Equation \eqref{eq:app_dP_k_final}:
\begin{equation}
    \mathrm{d}P_k
    =
    \rho_kc_k^2\frac{\mathrm{d}\rho_k}{\rho_k}
    +\rho_k\Gamma_kT_k\mathrm{d}s_k.
\end{equation}
Substituting the relative differential variations of the phasic densities and imposing the mechanical equilibrium condition $\mathrm{d}P_v = \mathrm{d}P_l = \mathrm{d}P$ gives:
\begin{equation}
    \begin{aligned}
        &\rho_vc_v^2\left(
        \frac{\mathrm{d}\rho}{\rho}
        +\frac{\mathrm{d}\beta}{\beta}
        -\frac{\mathrm{d}\alpha}{\alpha}
        \right)
        +\rho_v\Gamma_vT_v\mathrm{d}s_v\\
        ={}&
        \rho_lc_l^2\left(
        \frac{\mathrm{d}\rho}{\rho}
        -\frac{\mathrm{d}\beta}{1-\beta}
        +\frac{\mathrm{d}\alpha}{1-\alpha}
        \right)
        +\rho_l\Gamma_lT_l\mathrm{d}s_l.
    \end{aligned}
\end{equation}
Moving the terms containing $\mathrm{d}\alpha$ to the left-hand side of the equation and the remaining terms to the right-hand side for rearrangement yields:
\begin{equation}
    \begin{aligned}
        \left(
        \frac{\rho_vc_v^2}{\alpha}
        +\frac{\rho_lc_l^2}{1-\alpha}
        \right)\mathrm{d}\alpha
        ={}
        (\rho_vc_v^2-\rho_lc_l^2)\frac{\mathrm{d}\rho}{\rho}
        &+
        \left(
        \frac{\rho_vc_v^2}{\beta}
        +\frac{\rho_lc_l^2}{1-\beta}
        \right)\mathrm{d}\beta\\
        &+
        \left(
        \rho_v\Gamma_vT_v\mathrm{d}s_v
        -\rho_l\Gamma_lT_l\mathrm{d}s_l
        \right).
    \end{aligned}
\end{equation}
To simplify the expression, both sides of the equation are multiplied by $\alpha(1-\alpha)$. Using the substitution relations $\rho_v/\beta=\rho/\alpha$ and $\rho_l/(1-\beta)=\rho/(1-\alpha)$, the coefficient in front of $\mathrm{d}\beta$ can be simplified to $\rho(\alpha c_l^2+(1-\alpha)c_v^2)$. This yields the total differential expression for $\alpha$:
\begin{equation}
\begin{aligned}
  \mathrm{d}\alpha
  ={}&
  \frac{\alpha(1-\alpha)(\rho_vc_v^2-\rho_lc_l^2)}
  {\rho( \alpha\rho_lc_l^2+(1-\alpha)\rho_vc_v^2)}\mathrm{d}\rho
  +\frac{\rho(\alpha c_l^2+(1-\alpha)c_v^2)}{\alpha\rho_lc_l^2+(1-\alpha)\rho_vc_v^2}\mathrm{d}\beta\\
  &-\frac{\alpha(1-\alpha)\rho_l\Gamma_lT_l}
  {\alpha\rho_lc_l^2+(1-\alpha)\rho_vc_v^2}\mathrm{d}s_l+\frac{\alpha(1-\alpha)\rho_v\Gamma_vT_v}
  {\alpha\rho_lc_l^2+(1-\alpha)\rho_vc_v^2}\mathrm{d}s_v.
\end{aligned}
\label{eq:alpha_total_differential}
\end{equation}
Comparing with the total differential definition of a multivariable function:
\begin{equation}
    \mathrm{d}\alpha = 
    \pderiv{\alpha}{\rho}{\beta,s_l,s_v}\mathrm{d}\rho
    +\pderiv{\alpha}{\beta}{\rho,s_l,s_v}\mathrm{d}\beta
    +\pderiv{\alpha}{s_l}{\rho,\beta,s_v}\mathrm{d}s_l
    +\pderiv{\alpha}{s_v}{\rho,\beta,s_l}\mathrm{d}s_v,
\end{equation}
the partial derivatives of the volume fraction with respect to each state variable can be extracted:
\begin{equation}
\begin{aligned}
    \pderiv{\alpha}{\rho}{\beta,s_l,s_v} &= \frac{\alpha(1-\alpha)(\rho_vc_v^2-\rho_lc_l^2)}{\rho\bigl[\alpha\rho_lc_l^2+(1-\alpha)\rho_vc_v^2\bigr]}, \\
    \pderiv{\alpha}{\beta}{\rho,s_l,s_v} &= \frac{\rho(\alpha c_l^2+(1-\alpha)c_v^2)}{\alpha\rho_lc_l^2+(1-\alpha)\rho_vc_v^2}, \\
    \pderiv{\alpha}{s_l}{\rho,\beta,s_v} &= \frac{-\alpha(1-\alpha)\rho_l\Gamma_lT_l}{\alpha\rho_lc_l^2+(1-\alpha)\rho_vc_v^2},\\
    \pderiv{\alpha}{s_v}{\rho,\beta,s_l} &= \frac{\alpha(1-\alpha)\rho_v\Gamma_vT_v}{\alpha\rho_lc_l^2+(1-\alpha)\rho_vc_v^2}. \\
\end{aligned}
\end{equation}

\section{Appendix: P-R Equation of State and Thermodynamic Quantities}
\label{app:PR_EOS}
All thermodynamic calculations in this paper employ the single-component P-R equation of state. The P-R equation of state is defined as:
\begin{equation}
P = \frac{RT}{V - b} - \frac{a}{V(V + b) + b(V - b)}
\end{equation}
where $R=8.31446[ \text{J}/( \text{mol}\cdot \text{K})] $ is the universal gas constant, $V$ is the molar volume (the relation between molar volume $V$ and specific volume $v$ is $v = V/M$), $a$ is the energy parameter, and $b$ is the covolume parameter. The coefficients of the P-R equation of state are defined as:
\begin{align}
    &a(T) = a_c \cdot \alpha(T) = a_c \left[ 1 + c\left( 1 - \sqrt{T/T_c} \right) \right]^2 \\
    &a_c = 0.457236\frac{R^2 T_c^2}{P_c}\\
    &b = 0.077796 \frac{R T_c}{P_c} \\
    &c = 0.37464 + 1.54226\omega - 0.26992\omega^2 
\end{align}
where $\omega$ is the acentric factor. Calculating the internal energy and specific heat capacities requires the following derivatives:
\begin{align}
    \pderiv{p}{T}{V}&=\frac{R}{V-b}-\frac{{\mathrm{d}a}/{\mathrm{d}T}}{V^2+2Vb-b^2}\\
    \pderiv{p}{V}{T}&=-\frac{RT}{(V-b)^2} +\frac{2a(V+b)}{(V^2+2Vb-b^2)^2}\\
    \frac{\mathrm{d}a}{\mathrm{d}T} &= - \frac{c \sqrt{a \cdot a_c}}{\sqrt{T_c T}}\\
    \frac{\mathrm{d}^2 a}{\mathrm{d} T^2} &= \frac{a_c c (1+c)}{2 T \sqrt{T T_c}}
\end{align}

For general real fluids, thermodynamic parameters are usually calculated from ideal gas values and departure functions. Hereinafter, the superscript zero (0) denotes the ideal gas value of a thermodynamic quantity. For the P-R EoS, the analytical integral expression for internal energy is:

\begin{equation}
e = e^0(T) + \frac{K_0}{M} \left( a - T \frac{\mathrm{d}a}{\mathrm{d}T} \right).
\end{equation}
where $K_0$ is a function related to the specific volume:
\begin{equation}
K_0 = \int_{+\infty}^{V} \frac{1}{V^2 + 2bV - b^2} dV = \frac{1}{2\sqrt{2}b} \ln \left( \frac{V + (1 - \sqrt{2})b}{V + (1 + \sqrt{2})b} \right).
\end{equation}

Similarly, the specific entropy can be expressed as a combination of the ideal gas value and a departure function. For the P-R EoS, its integrated value is as follows:

\begin{equation}
s = s^0(T) + \frac{R}{M} \ln \frac{V - b}{V} - \frac{K_0}{M} \left( \frac{\mathrm{d}a}{\mathrm{d}T} \right).
\end{equation}

The specific heat capacity at constant volume is calculated as follows:
\begin{equation}
C_v = \left( \frac{\partial e}{\partial T} \right)_{v} = C_v^0 - \frac{K_0}{M} T \left( \frac{\mathrm{d}^2a}{\mathrm{d}T^2} \right),
\end{equation}
and the specific heat capacity at constant pressure is calculated as follows:
\begin{equation}
C_p = \left( \frac{\partial h}{\partial T} \right)_{p} = C_p^0 - R - \frac{K_0}{M} T \left( \frac{\mathrm{d}^2a}{\mathrm{d}T^2}  \right)- \frac{T}{M} \frac{(\partial p / \partial T)_{V}^2}{(\partial p / \partial V)_{T}},
\end{equation}
where $C_v^0$ is the ideal gas specific heat capacity at constant volume, and $C_p^0$ is the ideal gas specific heat capacity at constant pressure.

\section{Thermophysical Parameters of Dodecane}
\label{app:dodecane}

The parameters required to calculate the saturation properties of dodecane using the P-R equation of state are listed in tables~\ref{tab:dodecane1} and~\ref{tab:dodecane2}.

\begin{table}
    \centering
    \caption{Thermodynamic parameters of Dodecane}
    \label{tab:dodecane1}
    \begin{tabular}{ccccc}
        \toprule
        $M$ ($\mathrm{kg\,kmol^{-1}}$) & $\omega$ & $T_c$ (K) & $\rho_c$ ($\mathrm{kg \,m^{-3}}$) & $P_c$ (MPa) \\
        \midrule
        170.33& 0.574 & 658.1 & 186 & 1.82\\
        \bottomrule
    \end{tabular}\par
\end{table}

\begin{table}
    \centering
    \caption{The parameters of the energy and entropy model}
    \label{tab:dodecane2}
    \begin{tabular}{cccc}
           \toprule
          $C_{v,\infty}^c$ ($\mathrm{J \, kg^{-1} \,K^{-1}}$) & $n $ & $e_\text{ref}$ ($\mathrm{J \, kg^{-1}}$) & $s_\text{ref}$ ($\mathrm{J \, kg^{-1}\, K^{-1}}$) \\
          \midrule
        $2.970 \times 10^3$& 0.613 & $1 \times 10^6$ & $1.4 \times 10^3$ \\
        \bottomrule
    \end{tabular}\par
\end{table}

Detailed definitions of the parameters above are available in \citet{GuardoneArgrow2005}. Among them, $\rho_c$ is derived from the PR equation of state based on $T_C$ and $P_c$, which is smaller than the NIST data \citep{NIST2023} ($\rho_c = 226\mathrm{kg \,m^{-3}}$). The calculated results for key thermodynamic variables, such as saturation line properties and saturated speed of sound, exhibit good agreement with NIST data.

\section{Discrete Scheme of the HRRF Method}
\label{app:Discrete}

The discretization of the HRRF method combined with the HLLC solver is generally consistent with that of the RFQC method \citep{Bai2026RFQC1} and Shyue's original scheme \citep{Shyue1998}, mainly following the work of Johnsen and Colonius \citep{Johnsen2006}. Specifically, the discretization of the Euler equations is identical to the standard Godunov finite volume method. At cell $i$, the semi-discrete update equation is given by:
\begin{equation}
\frac{\mathrm{d} \mathbf{U}_i}{\mathrm{d} t} = 
- \frac{1}{\Delta x} \left( \mathbf{F}^{\mathrm{HLLC}}_{i+\frac12} - \mathbf{F}^{\mathrm{HLLC}}_{i-\frac12} \right),
\label{eq:Euler_discrete}
\end{equation}
where $\mathbf{F}^{\mathrm{HLLC}}_{i+\frac12}$ is the flux obtained by the HLLC solver \citep{Toro2009}. Denoting the mass flux computed by the HLLC solver as $(\rho u)^{\mathrm{HLLC}}$, the vapor mass equation is discretized in conservative form with a source term:
\begin{equation}
\frac{\mathrm{d}(\rho\beta)_i}{\mathrm{d}t}
=
-\frac{1}{\Delta x}
\left(
(\rho u)^{\mathrm{HLLC}}_{i+\frac12}
\beta^{\mathrm{up}}_{i+\frac12}
-
(\rho u)^{\mathrm{HLLC}}_{i-\frac12}
\beta^{\mathrm{up}}_{i-\frac12}
\right)
+\rho_i R_i,
\label{eq:beta_semidiscrete}
\end{equation}
where the upwind value of $\beta$ is determined by the sign of the mass flux:
\begin{equation}
\beta^{\mathrm{up}}_{i+\frac12}
=
\begin{cases}
\beta_i,
& (\rho u)^{\mathrm{HLLC}}_{i+\frac12}\geq 0,\\
\beta_{i+1},
& (\rho u)^{\mathrm{HLLC}}_{i+\frac12}<0.
\end{cases}
\label{eq:beta_upwind}
\end{equation}

Particular attention should be paid to the advection equations, for which the basic principle of discretization is that the thermodynamic parameters $\xi, E_0$ and the vapor volume fraction $\alpha$ need to remain consistent with the mass flux of the HLLC solver. Specifically, following the formulation of conservative variables in the finite volume method, we can equivalently rewrite the advection equations \eqref{eq:7eq_system5}--\eqref{eq:7eq_system7} in the ``flux + source term'' form:
\begin{equation}
\frac{\partial \boldsymbol{\phi}}{\partial t} + \frac{\partial (u\boldsymbol{\phi})}{\partial x}
=
(\boldsymbol{\phi}+\boldsymbol{k})\frac{\partial u}{\partial x}+S_\phi,
\label{eq:advFluxSource}
\end{equation}
where
\begin{equation}
\boldsymbol{\phi}
=
\begin{pmatrix}
\xi\\
E_0\\
\alpha
\end{pmatrix},
\qquad
\boldsymbol{k}
=
\begin{pmatrix}
0\\
0\\
K
\end{pmatrix},
\qquad
\boldsymbol{S}_{\phi}
=
\begin{pmatrix}
0\\
-\xi R_P\\
R_\alpha
\end{pmatrix}.
\end{equation}
At cell $i$, the semi-discrete update equation is:
\begin{equation}
\frac{\mathrm{d} \boldsymbol{\phi}_i}{\mathrm{d} t}
=
-\frac{1}{\Delta x}
\left(
u^{\mathrm{HLLC}}_{i+\frac12}
\boldsymbol{\phi}^{\mathrm{up}}_{i+\frac12}
-
u^{\mathrm{HLLC}}_{i-\frac12}
\boldsymbol{\phi}^{\mathrm{up}}_{i-\frac12}
\right)
+
\frac{\boldsymbol{\phi}_i+\boldsymbol{k}_i}{\Delta x}
\left(
u^{\mathrm{HLLC}}_{i+\frac12}
-
u^{\mathrm{HLLC}}_{i-\frac12}
\right)
+
\boldsymbol{S}_{\phi,i}.
\label{eq:Phi_split}
\end{equation}
$\boldsymbol{\phi}^{\mathrm{up}}$ is also determined by the sign of the HLLC mass flux:
\begin{equation}
\boldsymbol{\phi}^{\mathrm{up}}_{i+\frac12}
=
\begin{cases}
\boldsymbol{\phi}_{i}, &
(\rho u)^{\mathrm{HLLC}}_{i+\frac12}\ge 0,\\
\boldsymbol{\phi}_{i+1}, &
(\rho u)^{\mathrm{HLLC}}_{i+\frac12}<0.
\end{cases}
\label{eq:Phi_up}
\end{equation}

The interface velocity consistent with the mass flux is defined as:
\begin{equation}
u^{\mathrm{HLLC}}_{i+\frac12}
\equiv
\frac{(\rho u)^{\mathrm{HLLC}}_{i+\frac12}}
{\rho^{\mathrm{up}}_{i+\frac12}}.
\label{eq:u_cons_def}
\end{equation}

First-order explicit time-stepping is adopted for time integration, where the time step is jointly constrained by the characteristic advection time and the local relaxation times:
\begin{equation}
\Delta t^n
=
\min\left(
C_{\mathrm{CFL}}
\frac{\Delta x}{\max\left(|u_i^n|+c_{W,i}^n\right)},
C_{\mathrm{src}}\tau_{M,\min}^n,
C_{\mathrm{src}}\tau_{T,\min}^n,
\right),
\label{eq:time_step}
\end{equation}
where $C_{\mathrm{CFL}}$ is the CFL number for the advection term and $C_{\mathrm{src}}$ is the source term CFL number, both taken as 0.5 in the calculations of this paper. The minimum values of the local relaxation times are defined as:
\begin{equation}
\begin{aligned}
\tau_{M,\min}^n
&=\min\left(\tau_{M,e,i}^n,\tau_{M,c,i}^n\right),\\
\tau_{T,\min}^n
&=\min\tau_{T,i}^n,
\end{aligned}
\label{eq:relaxation_time_step}
\end{equation}

The discretization presented above does not represent the sole numerical realization of the HRRF method. Since the configurations considered in this work involve moderate computational scales, explicit time-stepping serves as a simple and practical choice. When solving multidimensional problems with stiff relaxation times, implicit or exponential source-term integration and high-order primitive variable reconstruction can be incorporated to enhance computational efficiency and accuracy. Such numerical techniques have been widely investigated in the field of multiphase flow algorithms; their further extension to the HRRF framework is not elaborated upon here.

\section{MATLAB Program for the Oscillation-Free Homogeneous Relaxation Real Fluid Method}
This paper provides a one-dimensional MATLAB program for the oscillation-free Homogeneous Relaxation Real Fluid method as supplementary material to readers, with the URL https://github.com/boss-monkey/HRRF. Detailed comments are provided inside each function for reader reference.

\end{appendix}

\end{document}